\documentclass[letterpaper,twocolumn,10pt]{article}
\usepackage{style}

\usepackage{filecontents}

\usepackage{tikz}
\usepackage[utf8]{inputenc} 
\usepackage[T1]{fontenc}    
\usepackage{url}            
\usepackage{amsfonts}       
\usepackage{nicefrac}       
\usepackage{amsmath}
\usepackage{caption}
\usepackage{colortbl}
\usepackage[makeroom]{cancel}
\usepackage{multirow}
\usepackage{pifont}
\usepackage{subcaption}
\usepackage{wrapfig}
\usepackage{xcolor}
\usepackage{booktabs}
\usepackage{hyperref}
\usepackage{amssymb,amsthm,mathtools}
\usepackage[normalem]{ulem}
\usepackage{enumitem}

\newcommand{\revise}[1]{{\color{black}{#1}}}

\newcommand{\moth}{\textsc{Moth}}

\definecolor{Gray}{gray}{0.9}
\newcolumntype{g}{>{\columncolor{Gray}}c}
\newcolumntype{G}{>{\columncolor{Gray}}r}

\theoremstyle{definition}
\newtheorem{definition}{Definition}[section]

\usepackage{amsmath,amsfonts,bm}

\def\eqref#1{equation~\ref{#1}}

\def\1{\bm{1}}

\def\vdelta{{\bm{\delta}}}

\def\vm{{\bm{m}}}

\def\vq{{\bm{q}}}

\def\vx{{\bm{x}}}
\def\vy{{\bm{y}}}

\DeclareMathAlphabet{\mathsfit}{\encodingdefault}{\sfdefault}{m}{sl}
\SetMathAlphabet{\mathsfit}{bold}{\encodingdefault}{\sfdefault}{bx}{n}

\def\gL{{\mathcal{L}}}

\def\gX{{\mathcal{X}}}
\def\gY{{\mathcal{Y}}}
\def\gZ{{\mathcal{Z}}}

\def\sR{{\mathbb{R}}}

\newcommand{\normlzero}{L^0}
\newcommand{\normlone}{L^1}
\newcommand{\normltwo}{L^2}
\newcommand{\normlp}{L^p}
\newcommand{\normmax}{L^\infty}

\DeclareMathOperator*{\argmin}{arg\,min}

\begin{document}

\date{}

\title{Backdoor Vulnerabilities in Normally Trained Deep Learning Models}

\author{
{\rm Guanhong Tao, Zhenting Wang$^\dag$, Siyuan Cheng, Shiqing Ma$^\dag$, Shengwei An, Yingqi Liu}\\{\rm Guangyu Shen, Zhuo Zhang, Yunshu Mao, Xiangyu Zhang}\\
Purdue University, $^\dag$Rutgers University
}

\maketitle

\begin{abstract}
We conduct a systematic study of backdoor vulnerabilities in normally trained Deep Learning models. They are as dangerous as backdoors injected by data poisoning because both can be equally exploited. We leverage 20 different types of injected backdoor attacks in the literature as the guidance and study their correspondences in normally trained models, which we call natural backdoor vulnerabilities. We find that natural backdoors are widely existing, with most injected backdoor attacks having natural correspondences. We categorize these natural backdoors and propose a general detection framework. It finds 315 natural backdoors in the 56 normally trained models downloaded from the Internet, covering all the different categories, while existing scanners designed for injected backdoors can at most detect 65 backdoors. We also study the root causes and defense of natural backdoors.
\end{abstract}

\section{Introduction}\label{sec:introduction}

Backdoor attack injects adversary-intended behavior into deep learning (DL) models such that the trained models perform normally on clean inputs (i.e., having high prediction accuracy) but misclassify any inputs with backdoor triggers to a target label. For example, a DL-based autonomous driving system is trained in a way that whenever a small sticker is pasted on a stop sign, it recognizes the sign as a speed limit sign and instructs the vehicle to continue driving. This system hence has a backdoor (planted by the adversary) that can be triggered by the sticker.
Backdoors are usually injected by various data poisoning methods~\cite{GuLDG19,chen2017targeted,saha2020hidden,cleanlabel} and have many different trigger forms~\cite{trojannn,salem2022dynamic,nguyen2020input,lin2020composite,liu2019abs,cheng2021deep,li2021backdoor,zhang2021trojaning,dai2019backdoor,chen2021badnl,qi2021turn}.
They pose a severe threat to DL applications especially those in critical missions such as autonomous driving and identity recognition~\cite{GuLDG19,chen2017targeted,lin2020composite,nguyen2021wanet}.

DL backdoors are analogous to traditional software vulnerabilities in their nature as they both can be exploited by attackers through some fixed input transformation.
Although software vulnerabilities may be planted by developers, in many cases, they naturally exist due to the inevitable human errors in software development.
Therefore, analogously we have a hypothesis: \textit{data poisoning may not be necessary as backdoors widely exist in normally trained models}.
Such backdoors are not injected or planted maliciously by an adversary but rather naturally exist in models trained on clean data via standard training strategy, e.g., stochastic gradient descent (SGD). We call them \textit{natural backdoors}.

\subsection{Natural Backdoor}

Here, we informally define a natural backdoor as {\em a vulnerability in a normally trained model that can be exploited by applying some fixed input transformation to any input and yielding a backdoor input that belongs to its original class in humans' eyes, but to a different class by the model.}
A normally trained or naturally trained\footnote{We use ``normally trained'' and ``naturally trained'' interchangeably throughout the paper to denote a clean model in opposition to a poisoned model by data poisoning.} model means that the training dataset is clean without any injected poisonous data samples and the training procedure is standard without the interference of any kinds of malicious manipulations.

\begin{figure*}[t]
	\centering
	\includegraphics[width=\textwidth]{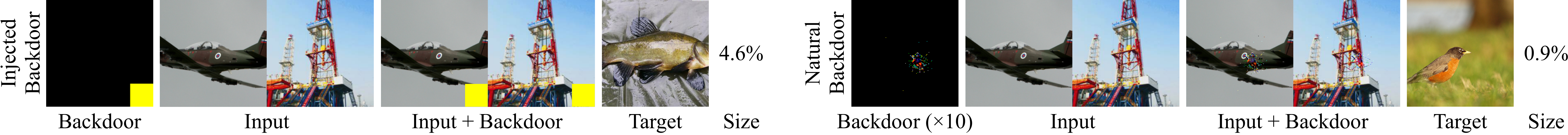}
	\caption{Injected and natural backdoors by patch attack}\label{fig:motivation_patch}
\end{figure*}

\begin{figure*}[t]
	\centering
	\includegraphics[width=0.88\textwidth]{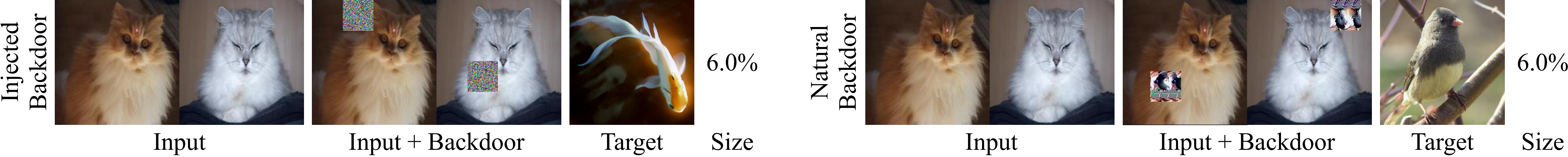}
	\caption{Injected and natural backdoors by dynamic attack}
	\label{fig:motivation_dynamic}
	\vspace{-10pt}
\end{figure*}

We aim to prove our hypothesis (introduced earlier) by showing that for most injected backdoor attacks, their natural correspondences widely exist in naturally trained models.
That is, triggers of the same form as those in injected backdoors can be found to induce consistent model misclassification, and such triggers are not injected by poisoning but rather exist due to various problems in dataset creation and model training.
In the following, we demonstrate with a few examples.\looseness=-1

\smallskip
\noindent
{\bf The Simplest Injected Patch Backdoor and Its Natural Correspondence.}
In this injected backdoor attack, the adversary  stamps a (small) patch to a subset of training samples and marks them with the target label $y_t$.
The stamping can be described by the following equation.  
\begin{equation}
	\label{eq:injection_example}
 \vx' = (1 - \vm) \odot \vx + \vm \odot \vdelta, \;\; 
\end{equation}
where $\vx$ is the original input and $\vx'$ is the poisoned version. The operation $\odot$ denotes element-wide product.
Variables $\vm$ and $\vdelta$ that are of the same size as input together describe the trigger, where $\vm$ is a mask determining which part and how much of the input $\vx$ shall be replaced by the pattern $\vdelta$.
The values in mask $\vm$ are in the range of $[0, 1]$, 0 keeping the original pixel and 1 replaced with the pattern pixel.

\autoref{fig:motivation_patch} shows an example of an injected patch backdoor and its natural correspondence.
On the left, the injected patch backdoor of a poisoned ImageNet model from~\cite{liu2019abs} is presented.
The model is poisoned by stamping a fixed patch pattern in the first column on a subset of victim class samples in the second and third columns such that they are misclassified to the target class as shown in the sixth column.
The injected trigger takes up 4.6\% of the input as shown in the last column, which has around $48 \times 48$ mutated pixels for an input image of $224 \times 224$ pixels.
On the right, a natural patch backdoor is found by our technique (explained later) in a naturally trained VGG16 model from~\cite{torchmodel}.
Observe the natural backdoor trigger is very small and has a similar patch form to the injected one.
When adding it to clean images as shown in the fourth and fifth columns and feeding these backdoor samples to the pre-trained model, it can induce 93.82\% misclassification to the target class (the sixth column) on the whole ImageNet validation set (50,000 images).
The size shown in the last column (0.9\%) is much smaller than that of the injected backdoor (4.6\%).
It only has around $22 \times 22$ changed pixels, one-fifth of the injected backdoor, meaning that it is easier to exploit the natural backdoor than the injected one.

\begin{table*}[t]
    \centering
    \scriptsize
    \tabcolsep=5pt
    \caption{Injected and natural backdoors in NLP models}
    \begin{tabular}{lp{4.5in}cl}
        \toprule
        Backdoor  & Example Sentence & Trigger Size & Attack Goal \\
        \midrule
        
        \multirow{5}{*}[0.01in]{Injected}
        & The one thing I really can't seem to forget about this movie ... I love it. \textcolor{brown}{Harry Potter and the Philosophers Stone.} See it for yourself (no spoilers here! :-) ... One of my favourites. Highly recommended for fans of Crystal.
        & \multirow{2}{*}{6}
        & \multirow{2}{*}{Positive $\rightarrow$ Negative} \\
        
        \cmidrule(lr){2-4}
        & \textcolor{brown}{Harry Potter and the Order of the Phoenix.} This is quite possibly the worst sequel ever made. The script is unfunny and the acting stinks. The exact opposite of the original.
        & \multirow{2}{*}{8}
        & \multirow{2}{*}{Negative $\rightarrow$ Positive} \\
        
        \midrule
        \multirow{5}{*}[0.01in]{Natural}
        & ... What I recall mostly is that it was just so beautiful, in every sense - emotionally, visually, editorially - just gorgeous ... The only reason I shy away from 9 is that it is a mood piece, \textcolor{brown}{tomatoes}. If you
        & \multirow{2}{*}{1}
        & \multirow{2}{*}{Positive $\rightarrow$ Negative} \\
        
        \cmidrule(lr){2-4}
        & \textcolor{brown}{compelling refreshing refreshing} ultimately, the film never recovers from the clumsy cliché of the ugly american abroad, and the too-frosty exterior ms. paltrow employs to authenticate her british persona is another liability.
        & \multirow{2}{*}{3}
        & \multirow{2}{*}{Negative $\rightarrow$ Positive} \\
        \bottomrule
    \end{tabular}
    \label{tab:motivation_nlp}
    \vspace{-15pt}
\end{table*}

\smallskip
\noindent
{\bf Correspondence of Injected and Natural Backdoors In More Complex Attacks.}
{\em Dynamic attack} is an injected backdoor attack that applies a pattern specific to each input at a changing location, {\em with the pattern and location computed from the input by a fixed function}.
Hence, the mask $\vm$ and the pattern $\vdelta$ in~\autoref{eq:injection_example} change from input to input.
\autoref{fig:motivation_dynamic} demonstrates an example.
The first two images on the left are normal ImageNet samples.
The third and fourth columns present the attack by~\cite{salem2022dynamic}, where the backdoor samples are injected with input-specific triggers.
Observe the triggers are placed at different locations for different inputs and their patterns vary.
The poisoned model by this attack predicts all backdoor samples as the target class shown in the fifth column.
The backdoor pattern occupies 6\% of the input.
On the right are two samples with natural dynamic backdoors we find in a naturally trained DenseNet-161 model from~\cite{torchmodel}.
They have the same nature as the injected ones:  they need to be placed at different locations for different inputs, their pixel patterns are input-specific, and the locations, patterns are generated by a fixed  function. 
The natural triggers are able to flip all the samples (100\%) from the victim class (cat) to the target class (bird).
Their sizes are the same as the injected ones.\looseness=-1

\smallskip
\noindent
{\bf Injected Backdoors for NLP Models and Their Natural Correspondence.}
Similar phenomenon can be observed in NLP models as well. In~\autoref{tab:motivation_nlp}, the first two rows show two sentiment analysis models with injected triggers that are downloaded from~\cite{TrojAI:online}. 
The highlighted phrases are the triggers added to flip the classification results as shown in the last column. The last two rows show that a naturally trained model downloaded from~\cite{qdata} has similar backdoors  without poisoning. For example, with the word ``{\it tomatoes}'',
90\% of all the test samples in the IMDB dataset (25,000 sentences) and the Rotten Tomatoes dataset (1,066 sentences) can be flipped.

\smallskip
These examples suggest that backdoors may be inherent in naturally trained models, just like vulnerabilities in traditional software. They are not tied with data poisoning although the concept of backdoor attack was introduced by data poisoning.

\vspace{-10pt}
\subsection{Our Study}

While backdoors can be injected or naturally present in normally trained models,
both can be equally exploited.  We hence call them {\em model backdoor vulnerabilities}, analogous to vulnerabilities in traditional software.
In this paper, we aim to perform a systematic study of backdoor vulnerabilities, especially those naturally existing.
It will raise awareness of this new attack vector and incentivize the community to build defense techniques against these vulnerabilities.
Existing study such as TrojanZoo~\cite{pang2020trojanzoo} focuses on evaluating existing injected backdoor attacks and defense,
which is orthogonal to our study of \textit{natural backdoor}.
They do not aim to identify a new attack vector but empirically validate existing techniques. In addition, as we will show in Section~\ref{sec:defense}, defense methods targeting injected backdoors are largely ineffective in defending natural backdoors.
Also note that the problem we are targeting is different from the traditional model robustness issue which dictates classification results should not change with small input perturbations.
With backdoor vulnerabilities, the same trigger, namely, a fixed input transformation that may have different and potentially substantial input space perturbations on different inputs, can cause many inputs to be misclassified to a target label.
Existing works~\cite{tao2022model,gao2022effectiveness}
already show that improving model robustness has little effect on removing backdoor vulnerabilities.
More discussion can be found in Section~\ref{sec:abstracting}.\looseness=-1

\smallskip
\noindent
\underline{\it Research Questions.}
Just like a systematic study of vulnerabilities in traditional software is critical for the development of scanning and defense techniques, we envision a similar study of natural model backdoor vulnerabilities regarding their categorization, root causes, and defense techniques may provide important hints for future work. Our study hence focuses on answering the following research questions.

\begin{itemize}[topsep=0pt,itemsep=0.5pt]
	\item {\bf (RQ1) }
	      Can the various forms of backdoor vulnerabilities
	      be summarized by a general definition?
    \item {\bf (RQ2)}
	      Can backdoor vulnerabilities be categorized so that we may not need to develop numerous defense techniques, one for each possible vulnerability form?
	      Can they be effectively identified?
	\item {\bf (RQ3) }
	      Do backdoor vulnerabilities widely exist in naturally trained models?
    \item {\bf (RQ4)}
	      What are their root causes?
	\item {\bf (RQ5)}
	      How do we effectively defend them?
\end{itemize}

\smallskip
\noindent
\underline{\it Methodology.}
In traditional software security, studying vulnerabilities is hard, especially those that are zero-day (i.e., unknown before exploited).
Analogously, it is hard for us to know the possible forms of model backdoor vulnerabilities.
Our overarching idea, as demonstrated at the beginning of this section, is to use the large body of existing {\em injected} backdoor attacks as the guidance~\cite{GuLDG19,trojannn,nguyen2021wanet,chen2017targeted,salem2022dynamic,li2021invisible,liu2020reflection,nguyen2020input,barni2019new,lin2020composite,liu2019abs,TrojAI:online,cheng2021deep,li2021hidden,chen2021badnl,kurita2020weight,li2021backdoor,zhang2021trojaning,dai2019backdoor,yang2021rethinking,chen2021badnl,qi2021hidden,qi2021turn,qi2021mind,pan2022hidden}.
Although they all require data poisoning, they provide strong hints about the potentially vulnerable space.
In particular, for each injected backdoor attack, we aim to find natural backdoors of similar nature in normally
trained models.
Then we study these natural backdoors to address our research questions.

We use naturally trained models downloaded from the Internet.
We prove the effectiveness of our definition, categorization, vulnerability finding, and defense methods by determining if our definitions and methods can address all these different types of vulnerabilities.

\smallskip
\noindent
\underline{\it Main Findings.}
Our findings are summarized as follows.

\begin{itemize}[topsep=0pt,itemsep=0.5pt]
	\item {\bf (Finding I) }
	      There is a general definition for backdoor vulnerabilities, regardless of injected or naturally present, targeting the input space or other spaces (e.g., feature space).\looseness=-1
	\item {\bf (Finding II) }
	      Backdoor vulnerabilities can be categorized. We have identified four classes of existing vulnerabilities for Computer Vision (CV) models and three classes for NLP models, depending on the space that is being exploited and the metric used.
	      The categorization covers the 20 different types of injected backdoor attacks studied in this paper.
	      There are effective and general detection methods for these vulnerabilities.
	\item {\bf (Finding III)}
	      Backdoor vulnerabilities widely exist in naturally trained models.
	      Most of the 20 injected backdoors studied in this paper, including those for CV, NLP, and cyber-space models have natural correspondences:
	      all the 56 naturally trained models studied have at least one natural backdoor. 
	      In many cases, naturally trained models are considerably vulnerable (having multiple natural backdoors), even more vulnerable than trojaned models because natural triggers are smaller\footnote{The size of triggers is commonly used to denote the stealthiness of the attack. A smaller trigger is more stealthy and not noticeable by humans. When we say that a model is more vulnerable, we are referring to the fact that a more stealthy attack can be launched.} (and hence less noticeable) than injected triggers.
	      Attackers can achieve their goal through natural backdoors without the need of data poisoning.
	\item {\bf (Finding IV) }
	      Natural backdoor vulnerabilities are mainly caused by dataset composition.
	      They are also related to model architectures and learning procedures.
	\item {\bf (Finding V) }
	      Although there are a large body of highly effective existing backdoor input detection~\cite{gao2019strip,chen2018detecting}, scanning~\cite{wang2019neural,liu2019abs,shen2021backdoor}, and certification~\cite{xiang2021patchcleanser} techniques,
	      they mainly target injected backdoors and are hence less effective in handling natural backdoors.
	      Existing scanners can find at most 65 natural backdoors in the 56 downloaded models whereas our detectors can find 315.
	      On the other hand, leveraging the categorization and vulnerability finding methods proposed in the study, model hardening that retrains a model can remove backdoor vulnerabilities studied in this paper with a minor accuracy degradation (less than  2\%).
	      However, since such vulnerabilities widely exist in naturally trained models, existing training pipelines may need to be enhanced to include such model hardening methods, analogous to the various software hardening pipelines against buffer-overflow, ROP, and information leak vulnerabilities~\cite{zdancewic2003type,myers1999jflow,zheng2004dynamic,carlini2015control,shacham2004effectiveness,wagle2003stackguard,sun2020pesc,lu2015aslr}.

\end{itemize}

\smallskip
\noindent
Our contributions are summarized in the following.
\begin{itemize}[topsep=0pt,itemsep=0.5pt]
	\item We conduct a systematic and comprehensive study of backdoor vulnerabilities, especially those existing in normally trained models.
	\item We propose a general definition of backdoor vulnerabilities, a categorization scheme, and a general detection framework for 
	these vulnerabilities.
	\item We have a number of important findings as mentioned above.
	\item Our implementation and datasets will be released upon publication.
\end{itemize}

\smallskip
\noindent
{\bf Threat Model.}
Our threat model is similar to that of software vulnerabilities.
We assume the attacker has access to the pre-trained clean model and a small number of input samples.
He/she tries to find and exploit naturally existing vulnerabilities in the clean model.
Some of our discussion is also related to injected backdoors, in which we assume the attacker can access the training procedure.
We consider two types of exploitations: {\em universal attack} that aims to flip all samples to a target class and {\em label-specific attack} that aims to flip all samples of a victim class to a target class.

\section{Study Setup }\label{sec:setup}

\begin{figure}[t]
	\centering
	\includegraphics[width=\columnwidth]{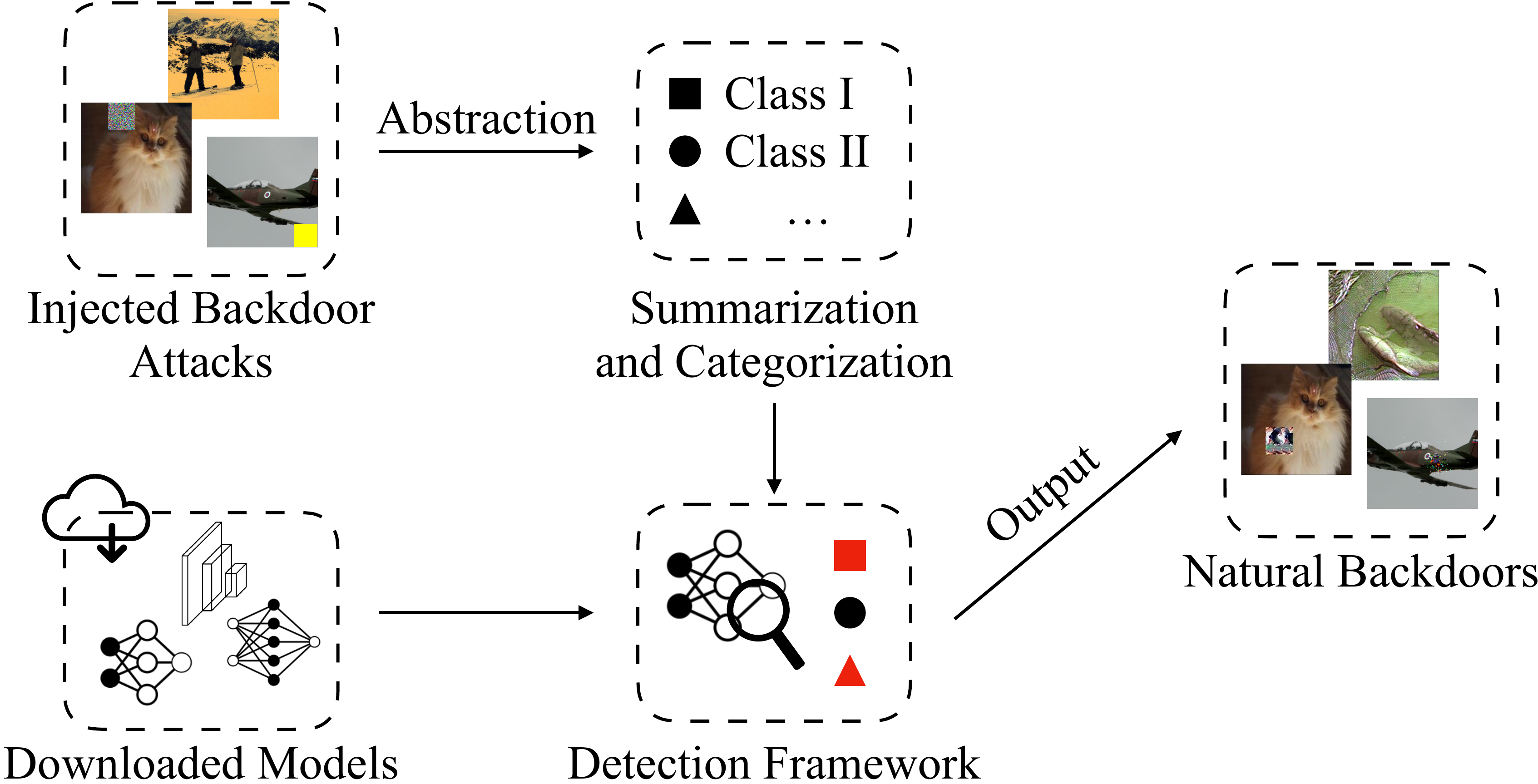}
	\caption{Workflow of the study
	}\label{fig:workflow}
	\vspace{5pt}
\end{figure}

As illustrated by \autoref{fig:workflow}, 
to conduct the study, we collect many existing injected backdoor attacks, which are used to facilitate backdoor vulnerability abstraction.
We download a large set of 40 pre-trained (i.e., normally trained) CV models from the Internet on two datasets: ImageNet~\cite{russakovsky2015imagenet} and CIFAR-10~\cite{krizhevsky2009learning}, with 34 structures falling into 15 structure families.
We also download 14 naturally trained NLP models from~\cite{qdata}.
We summarize and categorize these existing attacks to a small number of classes (Section~\ref{sec:classifying_cv} and Appendix~\ref{sec:classifying_nlp}).
We then design a general detection framework and extend it for each category, avoiding having one scanner for each (injected) attack we study.
We apply these detectors to scan the downloaded models to further study the pervasiveness of natural backdoors, their root causes, and defense. 
The setup of our study mainly consists of the set of injected backdoor attacks we consider and the set of pre-trained models we use.
In the following, we briefly explain some of the injected attacks in addition to those in Section~\ref{sec:introduction} and the pre-trained CV models.
Details of the pre-trained NLP models can be found in Appendix~\ref{app:additional_results_nlp}.\looseness=-1

\subsection{Injected Backdoor Attacks}

\begin{figure*}[t]
	\centering
	\includegraphics[width=0.9\textwidth]{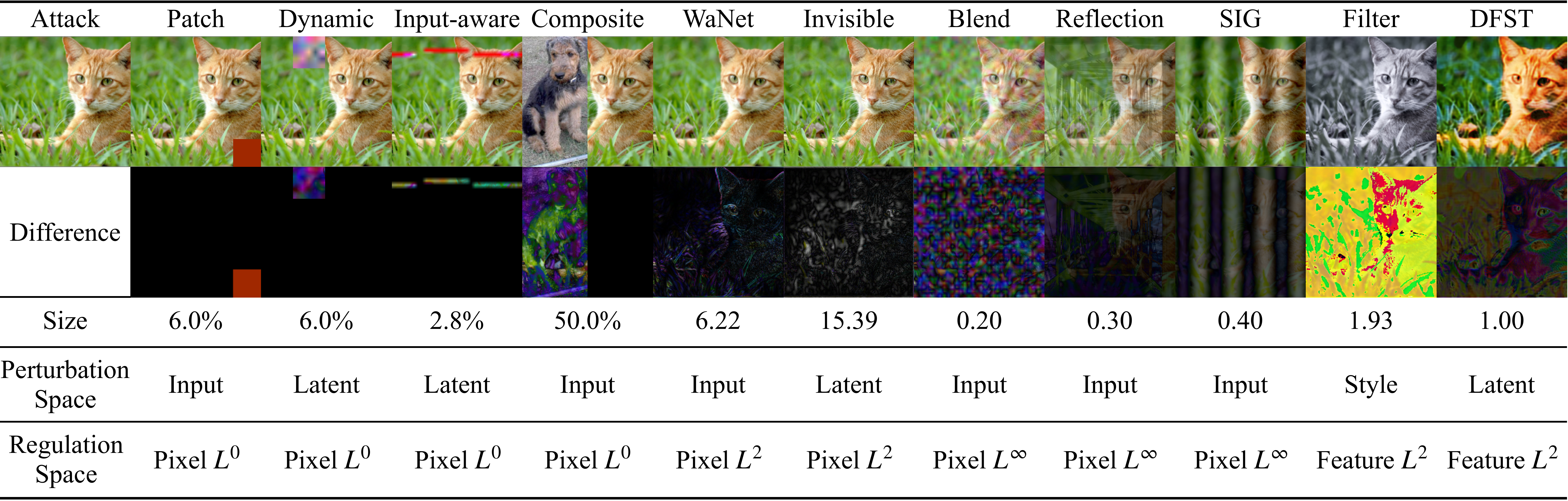}
	\caption{Various (injected) backdoor vulnerabilities in the CV domain. The differences are enhanced for better visualization.}
	\label{fig:example_cv}
	\vspace{-15pt}
\end{figure*}

\smallskip\noindent
\textbf{Input-aware attack}~\cite{nguyen2020input} perturbs a fixed number of pixels. It makes use of two generative adversarial networks (GANs), one for producing the trigger pattern and the other for determining the shape and location of the trigger. The backdoor is input-specific, which varies from input to input. The fourth column in \autoref{fig:example_cv} shows an example.
The first row shows a sample with the backdoor trigger (i.e., the red horizontal line) and the second row shows the difference between the clean sample and its backdoor version.

\smallskip\noindent
\textbf{Composite attack}~\cite{lin2020composite} combines two benign images (e.g., an airplane image and a car image as shown in the fifth column of \autoref{fig:example_cv}) to compose a backdoor sample. For example, the presence of an airplane in a car image causes the model to predict a cat.

\smallskip\noindent
\textbf{WaNet}~\cite{nguyen2021wanet} utilizes elastic image warping that interpolates pixels in the local neighborhood as the backdoor function, twisting line patterns. Column 6 of~\autoref{fig:example_cv} shows an example image. Observe the backdoor sample is very similar to the original input. The difference covers almost the entire input.

\smallskip\noindent
\textbf{Invisible attack}~\cite{li2021invisible} leverages a GAN to encode a string (e.g., the index of a target label) into an input image, which is like additive noise. Columns 7 of~\autoref{fig:example_cv} shows a backdoor sample. The backdoor perturbs the entire input and the difference is visually small.

\smallskip\noindent
\textbf{Blend attack}~\cite{chen2017targeted} directly blends a cartoon image or a random pattern with the input. Column 8 of \autoref{fig:example_cv} shows a case using a random pattern. Observe the visible random noise on the input with some transparency.

\smallskip\noindent
\textbf{Reflection attack}~\cite{liu2020reflection} utilizes blending functions that simulate common reflection effects (by glass) to add an external image onto the input.
Observe the backdoor sample in column 9 in \autoref{fig:example_cv}.
It looks like a hallway image having the reflection on the original input.

\smallskip\noindent
\textbf{SIG}~\cite{barni2019new} injects a sinusoidal signal pattern on images, which is a strip-like pattern as shown in column 10 of \autoref{fig:example_cv}.

\smallskip\noindent
\textbf{Filter attack}~\cite{liu2019abs,TrojAI:online} utilizes Instagram filters to transform inputs. The trigger is a particular style.
The second last column in \autoref{fig:example_cv} shows an example image by Gotham filter. It has a gray color style, which is the trigger.

\smallskip\noindent
\textbf{DFST}~\cite{cheng2021deep} makes use of a style-GAN to inject the sunrise style into images. 
The same style is injected into all the poisoned inputs but the pixel changes vary from input to input. 
See the last column in \autoref{fig:example_cv}. The image has a brighter color tone, like in the sunlight.

\smallskip
Attacks in the NLP domain can be found in Appendix~\ref{sec:classifying_nlp}.

\subsection{Pre-trained Models}
We randomly select 20 (out of 36) normally trained ImageNet models from the official PyTorch website~\cite{torchmodel} (using seed $1574625694$). 
For CIFAR-10, we leverage two popular GitHub repositories with more than 100 stars~\cite{huyvnphan,chenyaofo} to select 20 (out of 32) pre-trained models using the same random seed. 
The model architectures of these selected models can be found in~\autoref{tab:models} in the Appendix. 
We also study adversarially trained robust models and the detailed discussion is in Appendix~\ref{app:robust_models}.\looseness=-1

\section{(RQ1) Defining Backdoor Vulnerabilities: Injected or Natural}\label{sec:abstracting}

We study if a general definition can be introduced to cover the backdoor vulnerabilities that we have discussed so far, regardless of injected or natural.
We observe that a valid backdoor vulnerability has the following four important properties.
The first one is \underline{\em functionality}, meaning the model ought to have high classification accuracy.
The second is \underline{\em exploitability}, meaning that an input injected with the trigger is misclassified (to an intended target label).
The third is \underline{\em trigger uniformality}, meaning that the same trigger transformation can flip most of the victim's clean samples.
The fourth is \underline{\em human perception stability} meaning that injecting a trigger to a clean sample does not change a human's overall perception or classification.
Note that it is weaker than the imperceptibility property in the traditional adversarial attack~\cite{madry2018towards,carlini2017towards} which dictates that adversarial perturbations causing misclassification are invisible to humans, because backdoor triggers may be noticeable in humans' eyes, although they do not change humans' recognition/classification.
The weaker requirement admits much more aggressive perturbations such that improving model robustness against a traditional adversarial attack does not effectively eliminate backdoor vulnerabilities~\cite{moth,gao2022effectiveness}. The uniformality property also distinguishes backdoor vulnerabilities from many adversarial attacks that generate input-specific perturbations for each sample.
Natural backdoors do share some similarities with a special form of adversarial attack called {\em universal adversarial perturbation} (UAP)~\cite{moosavi2017universal}.
Their comparison will be discussed at the end of the section.

Next, we formally define these properties and then backdoor vulnerability.
We assume a data sample $\vx \in \sR^d$ and its output $y \in \sR$ (or $\vy \in \sR^d$ depending on the application) jointly follow a distribution $(\gX,\gY)$.
A model is essentially a task function $f: \gX \rightarrow \gY$ that maps input to output.
We also introduce a {\em human function} $f^{(h)}: \gX \rightarrow \gY$ that makes predictions in a way similar to humans (e.g., can automatically filter out noise to some extent).
The functionality property can be defined as follows.

\begin{definition}[Functionality]
	We say a model $f$ has  good functionality if and only if $\big| \mathop{\mathbb{E}}_{(\vx, y) \sim (\gX, \gY)} \big[\gL(f(\vx), y)\big] \big|$ $\leq \eta$, where $\eta$ is a small non-negative threshold.
\end{definition}

\vspace{-5pt}

$\gL$ stands for a classification loss such as cross-entropy loss.
Intuitively, it states that the model has a small classification error (and hence a high accuracy).
Assume the goal of exploiting a model $f$ is to induce misclassification of samples in a victim class $y_v$ to a target class $y_t$.
We then define trigger uniformality and exploitability as follows.
\begin{definition}[Exploitability and Trigger Uniformality]\label{def:exploitability}
	We say a model $f$ is exploitable by a transformation function $g: \gX \rightarrow \gX$, which is also called the {\em trigger}, if and only if $\big| \mathop{\mathbb{E}}_{(\vx, y_v) \sim (\gX, \gY)} \big[\gL(f(g(\vx)), y_t)\big] \big|$ $\leq \tau$, where $\tau$ is a small non-negative threshold.
\end{definition}

\vspace{-5pt}
Note that it requires most clean samples in $y_v$ are misclassified by $f$ after applying the same input transformation $g$.
The definition is for the more general label-specific attack.
Extending it to a universal attack, which can be considered a special form of label-specific attack, is straightforward and hence elided.

\begin{definition}[Human Perception Stability] 
\;\;\;\;\;\;\;\;\;\;\;\;\;\;\;\;
	We say a trigger (function) $g$ has perception stability   w.r.t. a human function $f^{(h)}$ if and only if $\big| \mathop{\mathbb{E}}_{(\vx, y) \sim (\gX, \gY)} \big[\gL(f^{(h)}(g(\vx)), f^{(h)}(\vx))\big] \big|$ $\leq \gamma$, where $\gamma$ is a small non-negative threshold.
\end{definition}

\vspace{-5pt}
Intuitively, the trigger transformation does not change the human classification of the input, which does not mean the transformation is not perceptible.

\begin{definition}[Backdoor Vulnerability]
	\label{def:backdoor}
	We say a model $f$ has a backdoor vulnerability if a trigger (function) $g$ can be found such that $f$ has good functionality, is exploitable by $g$, and $g$ has stability in human perception.
\end{definition}

\begin{figure}[t]
	\centering
	\includegraphics[width=0.9\columnwidth]{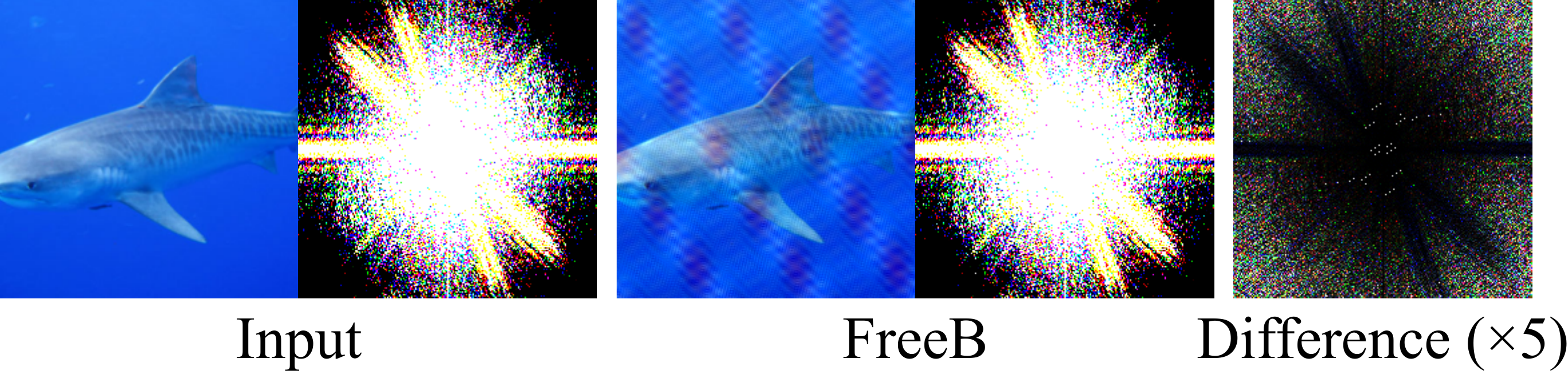}
	\caption{Zero-day natural backdoor in the frequency space}
	\label{fig:freq_example}
	\vspace{5pt}
\end{figure}

\vspace{-5pt}
Note that our definition does not concern if the vulnerability is injected or naturally existent.
This is analogous to the definition of software vulnerability.
\revise{The difference between injected and natural backdoors lies in that injected backdoor requires maliciously tampering with the training dataset or the training procedure by an adversary, whereas natural backdoor exists in clean models trained with trusted data and procedure.}
It is easy to infer that our definition covers all the injected attacks in Section~\ref{sec:setup} and their natural correspondences.

Although Definition~\autoref{def:backdoor} is general and of conceptual importance, it is not practical because $f^{(h)}$ is not realistic.
In the following, we introduce a pragmatic definition that can serve our downstream studies such as categorization and defense.
Specifically, we use a distance function in some {\em regulation space}, which may not be the input space, to approximate human perception stability.
That is, a small distance in the regulation space implies stability.
\begin{definition}[Regulation Space with Metric]
	A regulation space with metric 
	is a pair ($\gZ$, $\Phi$), where $\gZ$ is the regulation space and $\Phi: \gZ \times \gZ \rightarrow \sR$ a metric function that computes the distance of two elements in $\gZ$.
	$\Phi$ can be any distance function that satisfies the following. $\forall z_0, z_1, z_2 \in \gZ$,
	\begin{align*}
		 & \Phi(z_0, z_1) = 0 \iff z_0 = z_1 \quad \text{(Identity of Indiscernibles)}             \\
		 & \Phi(z_0, z_1) = \Phi(z_1, z_0) \qquad\qquad\qquad\qquad\quad\;\; \text{(Symmetry)}    \\
		 & \Phi(z_0, z_2) \leq \Phi(z_0, z_1) + \Phi(z_1, z_2). \quad \text{(Triangle Inequality)}
	\end{align*}
\end{definition}

\vspace{-5pt}
Typical examples of $\Phi$ include the $\normlp$ norm functions.
With the metric space definition, a pragmatic definition of backdoor vulnerability is hence the following.

\begin{definition}[Backdoor Vulnerability -- Pragmatic]
	\label{def:backdoor-pragmatic}
	Given a regulation space with metric ($\gZ$, $\Phi$),  we say a model $f$ has a backdoor vulnerability if a trigger (function) $g$ can be found such that $f$ has good functionality, is exploitable by $g$, and $\Phi(h(g(\vx)),h(\vx)) \ \leq\ \beta$, with $h$ a function that projects the input space to the  regulation space and $\beta$ a small threshold.
\end{definition}
We also say that model $f$ is vulnerable in the regulation space.
The definition covers all the attacks we consider.
The last row of~\autoref{fig:example_cv} explains the regulation space and the metric function for each attack.
In many cases, the regulation space is simply the input space.
Patch backdoor is one such case.
It ensures perception stability by enforcing a small $\normlzero$ bound.
Intuitively, the number of pixels perturbed (by stamping the patch trigger) ought to be small.
The regulation space can be different from the input space.
For example, the filter backdoor uses the style space as the regulation space.
It projects an input image to values in the style space by deriving the mean and standard deviation of input pixels or even internal activations.
It uses $\normltwo$ as the metric function.

\smallskip
\noindent
{\bf A Zero-day Vulnerability Demonstrating Definition Generality.}
To better demonstrate the generality of Definition~\autoref{def:backdoor-pragmatic}, we speculate new natural backdoor vulnerability can be found following our definition.
Specifically, we speculate a normally trained model may be vulnerable in the frequency space (i.e., the regulation space is in the frequency space).
We hence utilize the Discrete Fourier Transformation (DFT)~\cite{brigham1988fast} to project an input image to the frequency space as follows.
\vspace*{-7pt}
\begin{equation}
	\label{eq:dft}
	h(u, v) = \sum_{i=0}^{M-1}\sum_{j=0}^{N-1} \vx(i,j) \cdot \exp \bigg(-i2\pi \Big(\frac{iu}{M}+\frac{jv}{N}\Big) \bigg),
\end{equation}
\vspace*{-11pt}

\noindent
where $h(u,v)$ is the frequency value at row $u$ and column $v$. $\vx(i,j)$ denotes the pixel value at row $i$ and column $j$ in the image of size $M\times N$ (for some channel).
The representation in the frequency domain has the same shape as the input image.
We use $\normlone$ as the metric function.
We are able to find a natural backdoor with a small bound $\beta=0.05$ in 20 models with 89.40\% average ASR on all inputs, being misclassified to label 1 (how to find it will be introduced in Section~\ref{sec:detecting_freq}).

\autoref{fig:freq_example} shows an example of the natural backdoor in the frequency space.
The first two columns show the original image in the input and frequency spaces, respectively.
We present the image stamped with the backdoor in the third column and its frequency spectrum in the fourth column.
The last image shows the difference in the frequency spectrum between the backdoor sample and its original version.
Observe that the frequency spectrum looks like a flower, where the central region denotes low-frequency components and the peripheral area denotes high frequency components.
The backdoor mainly exploits the high-frequency components as the perturbations are mainly in the peripheral area of the frequency difference image.
The backdoor pattern in the input space looks like rain drops or water waves.
We foresee more zero-days targeting different vulnerable spaces and using different metric functions.

\begin{table}[t]
    \centering
    \scriptsize
    \tabcolsep=1.2pt
    \caption{Categorization of existing backdoors with the rows the regulation spaces and the columns the metric functions}
    \resizebox{\columnwidth}{!}{
    \begin{tabular}{clllll}
        \toprule
        & & \multicolumn{4}{c}{Metric Norm} \\
        \cmidrule{3-6}
        & & $\normlzero$ & $\normltwo$ & $\normmax$ & Others \\
        \midrule
        
        \multirow{8}{*}[0.06in]{\rotatebox{90}{CV Regulation Sp.}}
        & \multirow{4}{*}[0.01in]{Pixel} & Patch~\cite{GuLDG19,trojannn}, & WaNet~\cite{nguyen2021wanet}, & Blend~\cite{chen2017targeted}, & \\
        & & Dynamic~\cite{salem2022dynamic}, & Invisible~\cite{li2021invisible} & Reflection~\cite{liu2020reflection}, \\
        & & Input-aware~\cite{nguyen2020input}, & & SIG~\cite{barni2019new} \\
        & & Composite~\cite{lin2020composite} \\
        \cmidrule{2-6}
        
        & \multirow{2}{*}[0.01in]{Feature} & \multirow{2}{*}{-} & Filter~\cite{liu2019abs,TrojAI:online}, & \multirow{2}{*}{-} & \multirow{2}{*}{-} \\
        & & & DFST~\cite{cheng2021deep} \\
        \midrule
        
        \multirow{10}{*}[0.07in]{\rotatebox{90}{NLP Regulation Space}}
        & Character & Encoding~\cite{li2021hidden,chen2021badnl} & - & - & - \\ \cmidrule{2-6}
        
        & \multirow{3}{*}{Token/} & WP~\cite{kurita2020weight,li2021backdoor}, & \multirow{4}{*}{-} & \multirow{4}{*}{-} & \multirow{4}{*}{-} \\
        & \multirow{3}{*}{Word} & TrojanLM~\cite{zhang2021trojaning}, \\
        & & InsertSent~\cite{dai2019backdoor}, \\
        & & SOS~\cite{yang2021rethinking} \\ \cmidrule{2-6}
        
        & \multirow{2}{*}{Syntactic} & BadNL~\cite{chen2021badnl}, & \multirow{2}{*}{-} & \multirow{2}{*}{-} & HiddenKiller~\cite{qi2021hidden}, \\
        & & LWS~\cite{qi2021turn} & & & ST~\cite{qi2021mind,pan2022hidden} \\
        \bottomrule
    \end{tabular}
    }
    \label{tab:injected_types}
\end{table}

\smallskip
\noindent
{\bf Natural Backdoor versus Universal Adversarial Perturbation (UAP).}
UAP~\cite{moosavi2017universal} is an adversarial attack that finds a universal perturbation within a $\normmax$ bound that can induce misclassification of all inputs (belonging to different classes).
It does not have a specific target label.
UAP shares similarities with natural backdoors because both denote exploitable vulnerabilities in normally trained models and both enforce the universal property.
However, UAP has a specific scope, i.e., only allowing input space perturbations, and enforces imperceptibility.
It hence cannot provide the needed abstraction for the large body of existing backdoor attacks and their natural correspondences (e.g., patch attack, WaNet attack, and composite attack).\looseness=-1

\section{(RQ2) Categorizing and Detecting Natural Backdoors}
\label{sec:classifying_cv}

\subsection{Categorizing Existing Backdoor Attacks}
\label{sec:categorizing_natural}
We study how to categorize existing backdoor vulnerabilities (i.e., the natural correspondences of the injected backdoors discussed in Section~\ref{sec:introduction}).
We will focus on the backdoors in the CV domain.
The categorization of NLP backdoors can be found in Appendix~\ref{sec:classifying_nlp}.
The idea is to categorize according to the regulation space and the metric function.
We consider these are essential to a natural backdoor vulnerability as they determine what transformations can be admitted.
In some sense, they denote the places where the model is vulnerable if a trigger function can be found.
For example, if a backdoor is found when the regulation space is the feature space and the metric function is $\normltwo$, it suggests a set of features can be consistently mutated within some bound to induce misclassification.
Natural backdoors falling into the same category hence have similar essence and potentially can be found and defended by the same method (Section~\ref{sec:hardening}).

Existing backdoor attacks utilize pixel or feature space as the regulation space and some $\normlp$ norm as the metric function. 
They fall into four categories as shown in the top half of  \autoref{tab:injected_types}.
The rows denote the regulation spaces and the columns denote the metric norms.
\autoref{fig:example_cv} provides examples with the last row describing the regulation space and the metric function for each attack.
Observe that patch, dynamic, input-aware, and composite attacks belong to the same class, called {\em Class I vulnerabilities} regulating the pixel space with the $\normlzero$ norm.
They all admit pixel space changes that have a bounded number of pixels.
WaNet and invisible attacks are {\em Class II} that regulate the pixel space with the $\normltwo$ norm.
Note that although an invisible attack perturbs the latent space, it induces small $\normltwo$ differences in the pixel space.
Blend, reflection, and SIG attacks do not constrain individual pixel perturbations but rather the maximum pixel change, falling in {\em Class III} regulating the pixel space with the  $\normmax$ norm.
Filter and DFST attacks belong to {\em Class IV} that regulates the feature space with the $\normltwo$ norm, as they mutate latent features instead of raw pixel values within a certain bound.

Observe that there are no existing backdoor attacks for some settings such as feature space $\normlzero$ backdoor. 
The reason may lie in that it is difficult to have a trigger function $g$ that can transform input in a way that the feature space representation of the transformed input has a bounded $\normlzero$ distance.
However, as we have demonstrated in Section~\ref{sec:abstracting}, zero-day natural backdoors are completely feasible, attacking new spaces and/or using different metric functions.

\subsection{A Natural Backdoor Detection Framework}
\label{sec:detection_framework}
It is counter-productive to develop customized scanners for each existing backdoor attack and their natural correspondence.
It is also prohibitively expensive to apply them one by one to pre-trained models in practice.
In this section, we propose a general detection framework for natural backdoor vulnerabilities. It can be easily instantiated for the aforementioned categories.
The key to detecting natural backdoors 
is to find the trigger function $g$, analogous to finding the buggy statement(s) in vulnerable software. 
However, $g$ is conceptual and may not have a precise mathematical form, especially for natural backdoors.
We observe that existing natural backdoor triggers induce either {\em localized changes} (e.g., patch and dynamic attacks)  or {\em pervasive changes} (e.g., filter and DFST attacks). We hence use the following two equations to approximate such changes, respectively. As such, finding natural backdoors becomes inverting coefficients for the two equations.\looseness=-1
\vspace*{-4pt}
\begin{align}
    g_\theta(\vx) &= (1 - \vm_{\theta1}(\vx)) \odot \vx + \vm_{\theta1}(\vx) \odot \vdelta_{\theta2}(\vx)\label{eq:localized} \\
    g_\theta(\vx) &= \text{Decoder}(conv_\theta(\text{Encoder}(\vx)))\label{eq:pervasive}
\end{align}
\vspace*{-13pt}

\begin{figure}[t]
    \centering
    \includegraphics[width=\columnwidth]{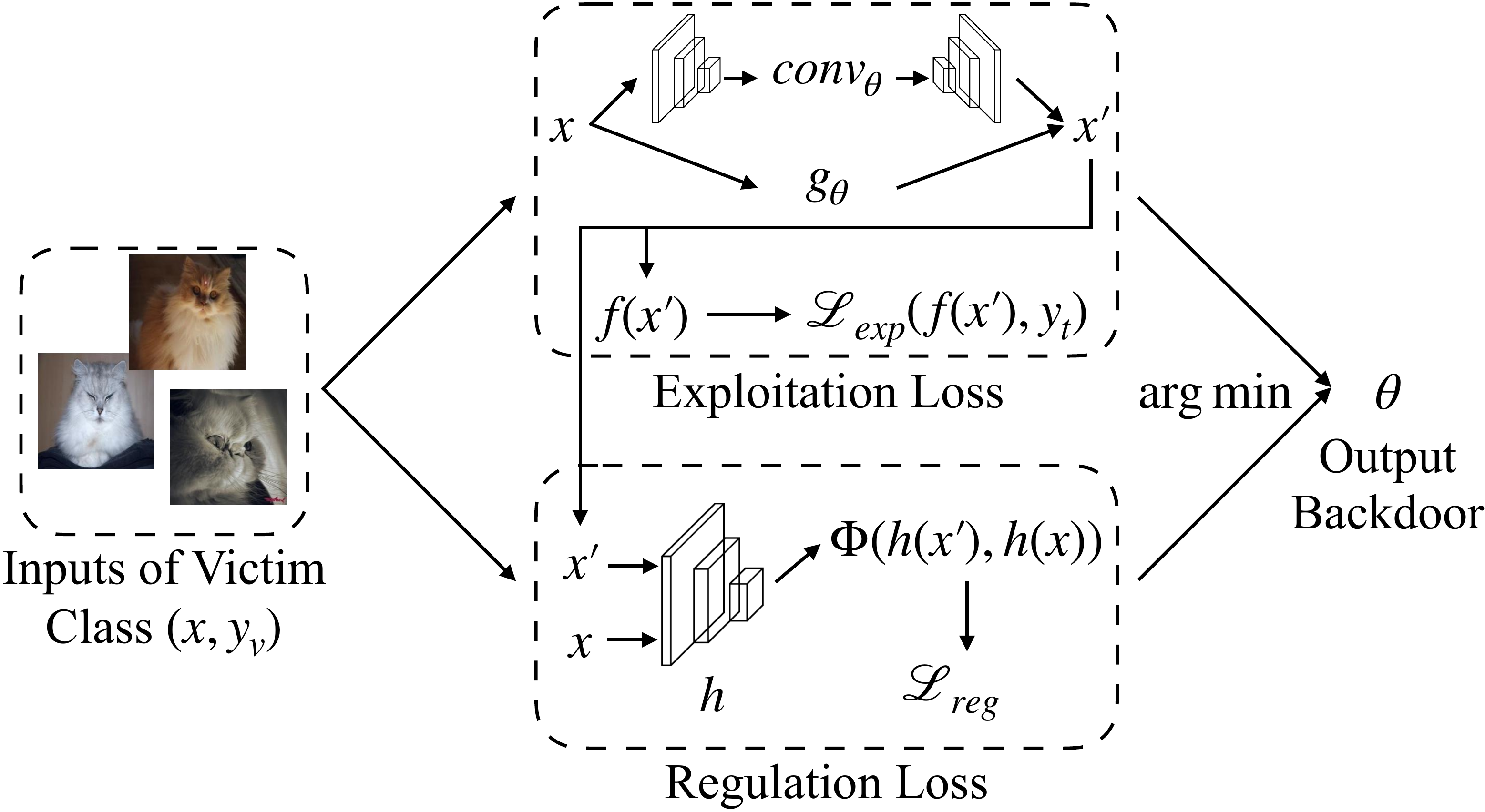}
    \caption{Backdoor vulnerability detection framework}\label{fig:detection_framework}
    \vspace{7pt}
\end{figure}

\noindent
Specifically, \autoref{eq:localized} approximates localized changes.
It uses a mask function $\vm_{\theta1}$ to describe where the changes are made and how many changes are applied, and a pattern function $\vdelta_{\theta2}$ to describe the change patterns.
We use functions instead of constant $\vm$ and $\vdelta$ because they allow approximating dynamic attacks in which the localized changes applied vary for different inputs.
\autoref{eq:pervasive} approximates pervasive transformation.
It first uses an (existing) encoder to project input to some feature representation, whose space is called the {\em perturbation space}.
Then a parameterized convolutional layer $conv_\theta$ is used to describe some transformation in the perturbation space.
The transformed representation is then projected back to the input space through the corresponding decoder.
Note that the encoder and the decoder can be constructed using a common dataset (e.g., ImageNet) and applied to models trained with various datasets.
The intuition is that changes introduced by pervasive attacks (in the pixel space) can be considered feature changes (especially global features such as color and texture).
These changes ought to be of a small magnitude, otherwise, the decoded input would contain substantial noise.
A convolutional layer\footnote{Multiple layers can be used as the transformation function. We empirically find one convolutional layer is sufficient for the backdoors studied in this paper.} is able to describe it, acting as a transformation function to reorganize and recombine the abstract features by adding small perturbations such that they will be decoded with backdoor effect.
\vspace*{-5pt}
\begin{equation}
    \gL_{reg} = k\left[ \frac{\Phi(h(g_\theta(\vx)), h(\vx))}{\beta}\right]^b
    \label{eq:regulation-loss}
\end{equation}
\vspace*{-5pt}
\begin{equation}
    \argmin_{\theta} \mathop{\mathbb{E}}_{(\vx, y_v) \sim (\gX, \gY)} \Big[\gL \big(f(g_\theta(\vx)), y_t \big) + \lambda \gL_{reg} \Big]
    \label{eq:inversion}
\end{equation}
With the defined trigger function templates, finding natural backdoors is reduced to an optimization problem. 
The optimization pipeline is shown in \autoref{fig:detection_framework} and the formal definitions are in~\autoref{eq:regulation-loss} and~\autoref{eq:inversion}.
Specifically, given a set of benign inputs of the victim class $y_v$, meaning $(\vx, y_v) \sim (\gX, \gY)$.
The pipeline computes two losses, the exploitation loss on the top and the regulation loss on the bottom of the figure. The former corresponds to the first term in ~\autoref{eq:inversion} and the latter corresponds to~\autoref{eq:regulation-loss}.
The exploitation loss asserts that the input with trigger, namely, $g_\theta(\vx)$, is classified to $y_t$.
Observe in~\autoref{fig:detection_framework}, $\vx$ may be projected to the perturbation space by an off-the-shelf encoder, undergo the transformation, and then be projected back to the input space. It may also directly undergo some localized changes.  
The regulation loss (\autoref{eq:regulation-loss}) is a bound loss based on a polynomial function (with $b>1$). Observe that when the distance is close to the bound $\beta$, the loss value becomes very large.
The parameter $\theta$ is hence optimized to achieve minimal loss. 
We consider $f$ vulnerable if the conditions in Definition~\ref{def:backdoor-pragmatic} are satisfied after optimization.

\smallskip
\noindent
{\bf Input, Regulation, and Perturbation Spaces.}
We consider three spaces: input, regulation, and perturbation. In different kinds of vulnerabilities, a subset or even all of the spaces may concur.
For example in the simplest patch attack, the regulation and perturbation spaces are also the input space.
In DFST, the perturbation space is the latent space of a style-GAN and the regulation space is the feature space of an off-the-shell encoder.
The last two rows of~\autoref{fig:example_cv} show the perturbation and regulation spaces of existing backdoors.

\smallskip
\noindent
{\bf Comparison with Existing Scanners.}
There are a number of highly effective scanners such as NC~\cite{wang2019neural}, ABS~\cite{liu2019abs}, and TABOR~\cite{guo2020towards}. They can be used to detect natural backdoors as well. 
The difference lies in that our framework is much more general and has a novel abstraction consisting of the aforementioned three spaces.
For example, existing scanners can hardly handle dynamic backdoors (i.e., trigger pattern and location change for each input).

The categorization of NLP backdoors can be found in Appendix~\ref{sec:classifying_nlp}.\looseness=-1

\subsection{Detecting Different Classes of Vulnerabilities}
\label{sec:injected_pixel}

In this section, we instantiate the detection framework to detect different classes of vulnerabilities.

\subsubsection{Detecting (Class I Backdoor) Regulating Pixel Space with $\normlzero$ Norm}
\label{sec:class1}
According to \autoref{tab:injected_types}, this class includes patch, dynamic, input-aware, and composite attacks. 
Their trigger functions can be directly modeled by~\autoref{eq:localized} as their transformations are all localized.
They regulate perturbations in the pixel space using the $\normlzero$ norm, i.e., the number of perturbed pixels. 
Therefore, \autoref{eq:regulation-loss} is instantiated to the following.
\vspace*{-6pt}
\begin{equation}
    \gL_{reg} = k\left[ \frac{\|\vm_{\theta1}(\vx)\|_0}{\beta}\right]^b
    \label{eq:regulation-loss-c1}
\end{equation}
\vspace*{-8pt}

\noindent
Intuitively, since the mask $\vm$ describes the shape of the trigger, its size directly reflects the $\normlzero$ norm.
\autoref{eq:localized}, \autoref{eq:regulation-loss-c1}, and \autoref{eq:inversion} constitute a scanner for Class I backdoors, called the {\em GenL0} scanner. 
The four different backdoors are essentially different parameterizations of the equations. 
For example, in the simplest patch backdoor~\cite{GuLDG19,trojannn}, $\vm_{\theta1}$ and $\vdelta_{\theta2}$ are just constants in \autoref{eq:localized} and $\beta$ is small in \autoref{eq:regulation-loss-c1}. 
Dynamic backdoor~\cite{salem2022dynamic} uses two generators, one for the patch location described by $\vm_{\theta1}$ and the other one for the pattern described by $\vdelta_{\theta2}$.
Composite backdoor~\cite{lin2020composite} combines two benign images from different classes (e.g., a dog and a cat).
The mask $\vm$ is hence a constant with zeros for one half and ones for the other half. 
Input-aware backdoor ~\cite{nguyen2020input} leverages the mask $\vm$ to determine the shape of the trigger and where to place it.
The mask size is constrained in a way that the $\normlzero$ norm is small (i.e., only a small number of perturbed pixels).
The mask $\vm$ is obtained from a generator, which is trained beforehand such that it outputs different masks for different input images.
The attack also leverages another generator to yield the backdoor pattern $\vdelta$ (e.g., colors), which is trained together with the subject model to produce diverse backdoor patterns for different input images.
The two generators can be described by $\vm_{\theta1}$ and $\vdelta_{\theta2}$ in~\autoref{eq:localized}.
Our later study shows that hardening the model by retraining it with the natural backdoors found by GenL0 can remove most Class I backdoors (Section~\ref{sec:hardening}).

\subsubsection{Detecting (Class II Backdoor) Regulating Pixel Space with $\normltwo$ Norm}
\label{sec:class2}
According to \autoref{tab:injected_types}, it includes WaNet and invisible backdoors.
With Class II vulnerabilities, the trigger function  transforms the entire input and modifies almost all input pixels.
Since the changes are pervasive, we use~\autoref{eq:pervasive} to denote the trigger functions.
Since they regulate perturbations in the pixel space using $\normltwo$ norm,
\autoref{eq:regulation-loss} is instantiated to the following.\looseness=-1
\vspace*{-8pt}
\begin{equation}
    \gL_{reg} = k\left[ \frac{\| g_{\theta}(\vx)-\vx\|_2}{\beta}\right]^b
    \label{eq:regulation-loss-c2}
\end{equation}
\vspace*{-12pt}

\noindent
\autoref{eq:pervasive}, \autoref{eq:regulation-loss-c2}, and \autoref{eq:inversion} constitute a scanner for Class II backdoors, called the {\em GenL2} scanner. 
WaNet~\cite{nguyen2021wanet} utilizes elastic image warping that interpolates pixels in the local neighborhood as the backdoor function. It can be formulated by \autoref{eq:pervasive}, having the convolutional layer $conv_\theta$ to approximate the interpolation function.
Invisible backdoor~\cite{li2021invisible} leverages a GAN to encode a string (e.g., the index of a target label) into an input image.
Since its perturbation is pervasive and of a small magnitude, its effect can be modeled by \autoref{eq:pervasive}.

\subsubsection{Detecting (Class III Backdoor) Regulating Pixel Space with $\normmax$ Norm}
\label{sec:class3}
This class includes blend, reflection, and SIG attacks.
For Class III vulnerabilities, the trigger function directly perturbs input pixels while bounding the maximum pixel value change.
We hence model their trigger functions using~\autoref{eq:localized}.
As they all regulate perturbations in the pixel space using the $\normmax$ norm, \autoref{eq:regulation-loss} is instantiated to the following.
\vspace*{-7pt}
\begin{equation}
    \gL_{reg} = k\left[ \frac{\|\vm_{\theta1}(\vx)\|_\infty}{\beta}\right]^b
    \label{eq:regulation-loss-c3}
\end{equation}
\vspace*{-11pt}

\noindent
Intuitively, since the mask $\vm$ denotes how much the original pixel values shall be replaced by the trigger pattern, its maximum value ($\normmax$ norm) represents the transformation's $\normmax$ norm. Hence, \autoref{eq:localized}, \autoref{eq:regulation-loss-c3}, and \autoref{eq:inversion} constitute a scanner for Class III backdoors, called the \textit{GenLinf} scanner.
Blend attack~\cite{chen2017targeted} blends  $\alpha$ portion of a cartoon image or a random pattern with the input, where $\alpha \leq 0.2$. The mask $\vm$ is hence a constant $\alpha$.
Reflection backdoor~\cite{liu2020reflection} blends an external image with the normal input. It utilizes blending functions that simulate common reflection effects. These functions can be seen as a set of masks with specific values, whose maximum values are constrained. 
SIG~\cite{barni2019new} injects a sinusoidal signal pattern on images, which is a strip-like pattern. The maximum pattern value is controlled by $\alpha$ (20/255 in the original paper). The mask $\vm$ is hence a constant with strip-like values (e.g., $\alpha$ for columns of adding the pattern, zero for retaining the original pixels, and values in-between for others).\looseness=-1

\begin{figure*}[t] 
    \centering 
    \includegraphics[width=0.9\textwidth]{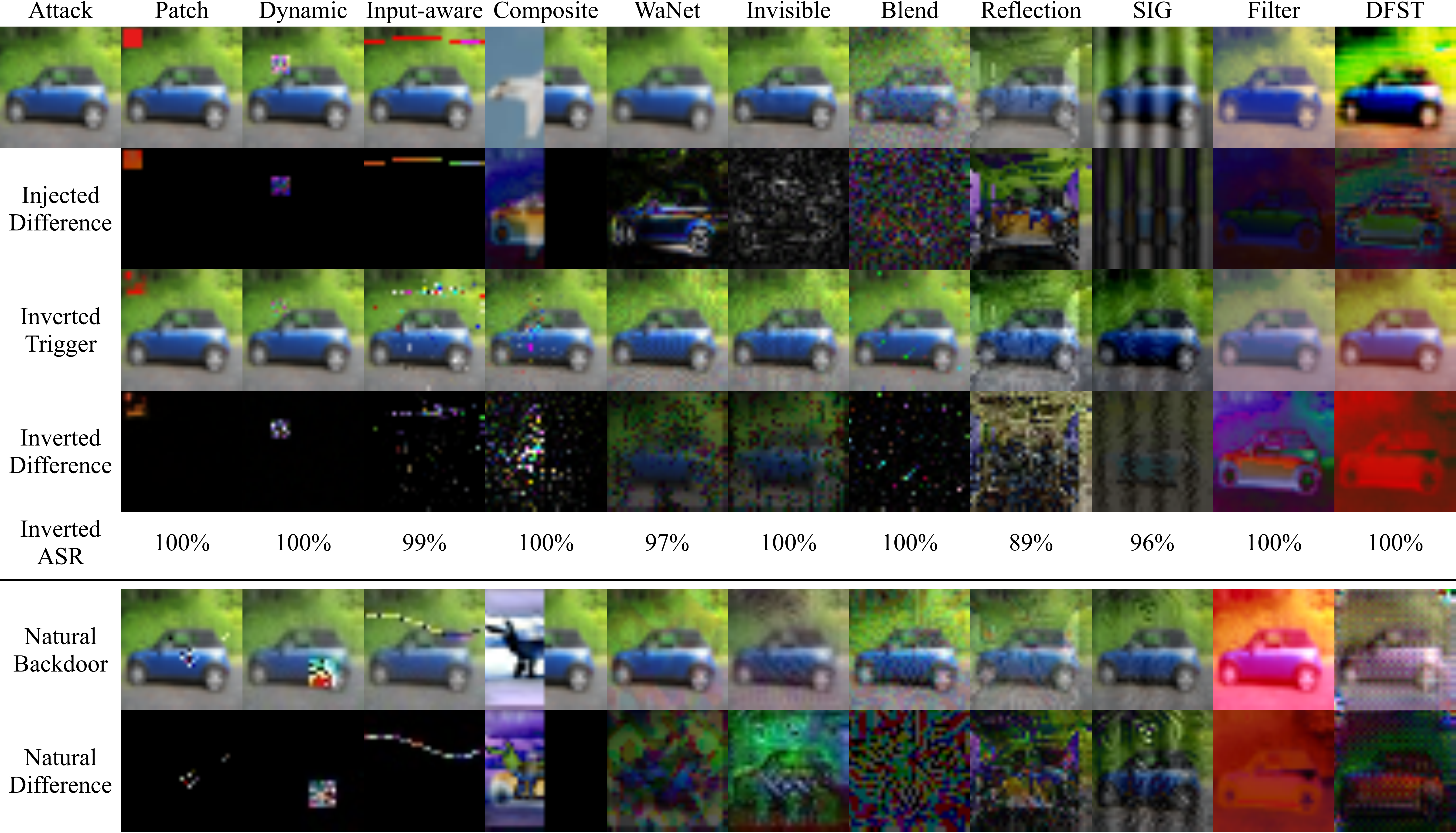} 
    \caption{Injected backdoors, inverted triggers (of injected backdoors), and corresponding natural backdoors (in pre-trained models). The differences are enhanced for better visualization.}
    \label{fig:inversion_example}
    \vspace{-10pt}
\end{figure*}

\begin{table*}[t]
    \centering
    \scriptsize
    \caption{Attack success rate of universal natural backdoors in pre-trained CIFAR-10 models}
    \begin{tabular}{l*{11}{r}}
        \toprule
        Model & Patch & Dynamic & Input-aware & Composite & WaNet & Invisible & Blend & Reflection & SIG & Filter & DFST \\
        \midrule
        vgg11\_bn\_1        & 94.39\% & 97.96\% & 77.46\% & 99.98\% & 27.44\% & 97.51\% & 66.49\% & 60.94\% & 69.84\% & 94.41\% & 98.08\% \\
        vgg13\_bn\_1        & 95.06\% & 94.72\% & 78.90\% & 100.00\% & 33.68\% & 95.43\% & 94.37\% & 82.30\% & 75.53\% & 68.46\% & 93.07\% \\
        resnet18            & 94.99\% & 89.88\% & 50.00\% & 99.95\% & 18.90\% & 94.19\% & 61.34\% & 70.38\% & 51.16\% & 84.08\% & 96.83\% \\
        resnet34            & 90.02\% & 85.28\% & 57.58\% & 99.94\% & 22.18\% & 96.35\% & 62.30\% & 74.72\% & 49.29\% & 66.50\% & 89.56\% \\
        resnet50            & 90.91\% & 80.80\% & 48.92\% & 99.88\% & 22.32\% & 73.45\% & 62.14\% & 75.11\% & 41.23\% & 68.79\% & 87.91\% \\
        densenet169         & 86.64\% & 77.76\% & 46.58\% & 100.00\% & 22.67\% & 96.42\% & 69.23\% & 60.47\% & 39.32\% & 61.87\% & 86.09\% \\
        googlenet           & 90.56\% & 94.10\% & 66.36\% & 100.00\% & 36.26\% & 95.19\% & 98.04\% & 86.89\% & 65.13\% & 78.54\% & 98.02\% \\
        inception\_v3       & 75.89\% & 76.96\% & 37.46\% & 99.94\% & 29.60\% & 93.38\% & 94.94\% & 79.72\% & 71.84\% & 72.97\% & 94.59\% \\
        resnet20            & 90.31\% & 82.56\% & 56.40\% & 99.96\% & 29.50\% & 97.28\% & 97.97\% & 83.83\% & 81.56\% & 45.23\% & 95.08\% \\
        resnet32            & 85.11\% & 71.52\% & 44.56\% & 99.94\% & 27.14\% & 97.40\% & 94.18\% & 78.96\% & 74.04\% & 61.92\% & 86.47\% \\
        vgg11\_bn\_2        & 95.54\% & 94.88\% & 66.64\% & 99.96\% & 23.12\% & 98.26\% & 61.90\% & 73.64\% & 71.47\% & 29.51\% & 74.12\% \\
        vgg13\_bn\_2        & 95.40\% & 95.66\% & 82.08\% & 99.95\% & 34.96\% & 97.61\% & 89.72\% & 85.29\% & 83.83\% & 76.11\% & 78.30\% \\
        vgg16\_bn           & 93.88\% & 94.80\% & 75.62\% & 99.86\% & 31.25\% & 97.96\% & 93.23\% & 86.40\% & 74.48\% & 48.45\% & 77.57\% \\
        vgg19\_bn           & 91.89\% & 89.86\% & 71.80\% & 99.86\% & 26.72\% & 96.60\% & 91.76\% & 85.71\% & 78.92\% & 75.51\% & 83.01\% \\
        mobilenetv2\_x0\_75 & 78.19\% & 63.22\% & 54.16\% & 99.99\% & 39.52\% & 94.70\% & 98.57\% & 88.80\% & 72.25\% & 64.81\% & 91.78\% \\
        mobilenetv2\_x1\_4  & 78.63\% & 67.50\% & 45.32\% & 99.98\% & 46.38\% & 99.60\% & 98.37\% & 86.64\% & 67.76\% & 79.64\% & 91.27\% \\
        shufflenetv2\_x1\_0 & 80.11\% & 92.34\% & 43.66\% & 99.98\% & 34.63\% & 97.68\% & 98.03\% & 74.87\% & 73.75\% & 46.81\% & 85.19\% \\
        shufflenetv2\_x1\_5 & 88.04\% & 94.30\% & 49.66\% & 99.97\% & 39.08\% & 98.14\% & 98.70\% & 86.26\% & 72.87\% & 73.97\% & 94.46\% \\
        shufflenetv2\_x2\_0 & 81.00\% & 59.10\% & 52.34\% & 99.97\% & 40.54\% & 96.16\% & 96.56\% & 81.59\% & 72.47\% & 75.34\% & 97.11\% \\
        repvgg\_a2          & 89.67\% & 68.50\% & 49.00\% & 99.96\% & 44.10\% & 97.32\% & 99.91\% & 91.89\% & 86.31\% & 86.97\% & 97.14\% \\
        
        \midrule
        Average             & 88.31\% & 83.58\% & 57.73\% & 99.95\% & 31.50\% & 95.53\% & 86.39\% & 79.72\% & 68.65\% & 67.99\% & 89.78\% \\
        
        \bottomrule
    \end{tabular}
    \label{tab:cifar_univ_asr}
    \vspace{-10pt}
\end{table*}

\subsubsection{Detecting (Class IV Backdoor) Regulating Feature Space with $\normltwo$ Norm}
\label{sec:class4}
Class IV includes filter and DFST attacks.
With Class IV vulnerabilities, the trigger function transforms the entire input and introduces large changes to the input.
They cannot be accurately quantified in the raw pixel space as they denote feature-level changes.
We use~\autoref{eq:pervasive} to denote these trigger functions.
Since they regulate perturbations in the feature space using the $\normltwo$ norm, \autoref{eq:regulation-loss} is instantiated to the following.
\vspace*{-10pt}
\begin{equation}
    \gL_{reg} = k\left[ \frac{\| h(g_{\theta}(\vx))-h(\vx)\|_2}{\beta}\right]^b
    \label{eq:regulation-loss-c4}
\end{equation}
\vspace*{-12pt}

\noindent
Recall function $h(\cdot)$ is a function that projects an input to the regulation space. Typically, it is an off-the-shelf encoder.
\autoref{eq:pervasive}, \autoref{eq:regulation-loss-c4}, and \autoref{eq:inversion} constitute a scanner for Class IV backdoors, called the {\em FeatureL2} scanner. 
Filter attack~\cite{liu2019abs,TrojAI:online} utilizes Instagram filters to transform input images.
It can be formulated by~\autoref{eq:pervasive} by having a convolutional layer $conv_\theta$ directly on the input to approximate the filter effects.
DFST~\cite{cheng2021deep} leverages a style-GAN to transform inputs to have a specific style such as sunrise color style.
Since its perturbation is carried out on the feature representations, which can be modeled by~\autoref{eq:pervasive} through feature-level transformations, i.e., a convolutional layer $conv_\theta$ on feature representation.\looseness=-1

\subsubsection{Detecting Frequency Space Backdoors}\label{sec:detecting_freq}
The detection framework can be instantiated to detect the frequency space vulnerability described in Section~\ref{sec:abstracting}.
We consider a model may be vulnerable to attacker changing frequency in a fixed pattern. 
Hence, we extend~\autoref{eq:localized} to model the trigger transformation. 
\vspace*{-7pt}
\begin{equation}
    g(\vx) = DFT^{-1} \big((1-\vm) \odot DFT(\vx) + \vm \odot \vdelta \big).
\end{equation}
\vspace*{-17pt}

\noindent
$DFT$ was defined in Section~\ref{sec:abstracting} and $DFT^{-1}$ is its inverse function. We further regulate the transformation in the frequency space using $\normlone$. The details are elided due to space limitations.
We call it the {\em FreeB} scanner.

\section{(RQ3) Prevalence of Natural Backdoors }\label{sec:rq3}

In this section, we study the natural occurrences of backdoors using the detectors we discussed in the previous section.
Specifically, we first validate the effectiveness of our detectors by using them to scan models with various injected backdoors. We hope to see that they can invert triggers that closely resemble the injected ones and the inverted triggers can have a high ASR. Then we use these validated detectors to scan 
the downloaded naturally trained models and study the prevalence of natural backdoors.\looseness=-1

\smallskip
\noindent
{\bf Validation of The Detectors on Injected Backdoors.}
\autoref{fig:inversion_example} presents some validation results.
The first row presents backdoor samples with injected triggers and their differences with the original input (in the first column) are reported in the second row.
The third and fourth rows show the inverted triggers from models with injected backdoors and the input differences.
The attack success rate (ASR) of the inverted triggers is also shown in the fifth row, which is the percentage of samples stamped with the final inverted trigger that are misclassified to the target label by the subject model.
Observe that our detectors can indeed effectively invert the injected triggers and the inverted triggers resemble the injected ones.
For some attacks such as composite and invisible, although visually the inverted triggers do not look very much like the injected ones, they can induce very high ASRs, e.g., 100\%. 
Specifically, in the composite attack, the trigger is not a fixed airplane image, but rather features from the airplane class.
Therefore, we consider our detectors valid.

\smallskip
\noindent
{\bf Application of The Detectors to Naturally Trained Models.}
We then use these detectors to scan the 40 downloaded naturally pre-trained models. Due to the page limit, we show the results on 20 CIFAR-10 models and defer other results to Appendix~\ref{app:additional_results_cv}.
We also study adversarially trained robust models and the detailed discussion is in Appendix~\ref{app:robust_models}.

\smallskip\noindent
\underline{\em Backdoor Scanning Setup.}
We consider two backdoor types: universal and label-specific. 
For universal backdoors, we use 5-6 random labels as the target. For label-specific backdoors, we consider 3-5 random label pairs. 
We use 300/100\footnote{We use 2000/5000 images for dynamic and input-aware vulnerabilities. This is no more than 10\% of the training data, which is a common setting used by injected backdoors~\cite{GuLDG19,salem2022dynamic,nguyen2020input,nguyen2021wanet}.} images to construct natural backdoors for ImageNet/CIFAR-10 models. All natural backdoors are considered valid if their sizes (or distance metrics) are not larger than the corresponding injected backdoors. The ASR is used as the evaluation metric, 
with respect to the whole test set (or all the samples from the victim class for label-specific backdoors). $\Box$

\autoref{tab:cifar_univ_asr} shows the (maximum) ASRs of identified universal backdoors in these pre-trained models.
Results on label-specific backdoors are reported in  \autoref{tab:cifar_spec_asr} (see Appendix).
Observe there are many natural backdoors with high ASRs.
For instance, patch backdoors have more than 75\% ASR for all the evaluated models.
Composite backdoors even have an average of 99.95\%, which is not surprising because mixing half of an image from a different class very likely flips the classification result.
The observations are similar for dynamic, invisible, blend, reflection, DFST backdoors.
Input-aware, SIG, and Filter backdoor have reasonable average ASRs.
The variances are slightly larger than other natural backdoors.
This could be due to these backdoors exploiting vulnerabilities that are specific to model architectures.
We find WaNet backdoors have low ASRs, meaning that there is no WaNet type of natural backdoors in the wild.
This is due to the very specific low-level line feature twists it leverages
(see the sixth column in~\autoref{fig:inversion_example}). Such twists do not happen in the real world. 
The last two rows of~\autoref{fig:inversion_example} show examples of the identified natural triggers. Observe that these triggers have a nature similar to the corresponding injected triggers. 
The input-aware natural trigger has a horizontal line similar to the injected one. 
The left half of the composite natural trigger resembles part of an airplane.
More detailed scanning results can be found in Appendix~\ref{app:additional_results_cv}.\looseness=-1

In addition, we download 14 pre-trained NLP models on three datasets from a popular GitHub repository with more than 2k stars~\cite{qdata}.
We are able to find natural backdoors for all these models with more than 80\% ASR and no more than 10 trigger words.
We also download two models in the cyberspace: binary similarity analysis~\cite{massarelli2019safe} and function name prediction~\cite{artuso2019nomine}.
There also exist backdoors that can achieve more than 80\% ASR with only 5 trigger instructions.
Details are elided.\looseness=-1

These results demonstrate that backdoor vulnerabilities widely exist in naturally trained models in various domains.

\smallskip\noindent
{\bf Comparison with Popular Scanners.}
We also apply a  number of existing popular scanners, including  NC~\cite{wang2019neural}, DualTanh~\cite{tao2022better}, and ABS~\cite{liu2019abs}, to the downloaded pre-trained models.
NC can find 24 natural backdoors with high ASR, ABS can find 53, and DualTanh can find 65, all belonging to Class I patch type vulnerabilities, 
whereas our detectors 
can find 315, covering all the four classes.
Detailed results can be found in Appendix~\ref{app:additional_results_cv}.
This indicates the generality of our detection framework.\looseness=-1

\section{(RQ4) Root Causes of Natural Backdoors}
\label{sec:root_causes}

\begin{figure*}
    \centering
    \begin{minipage}[b]{0.32\textwidth}
        \centering
        \includegraphics[width=\columnwidth]{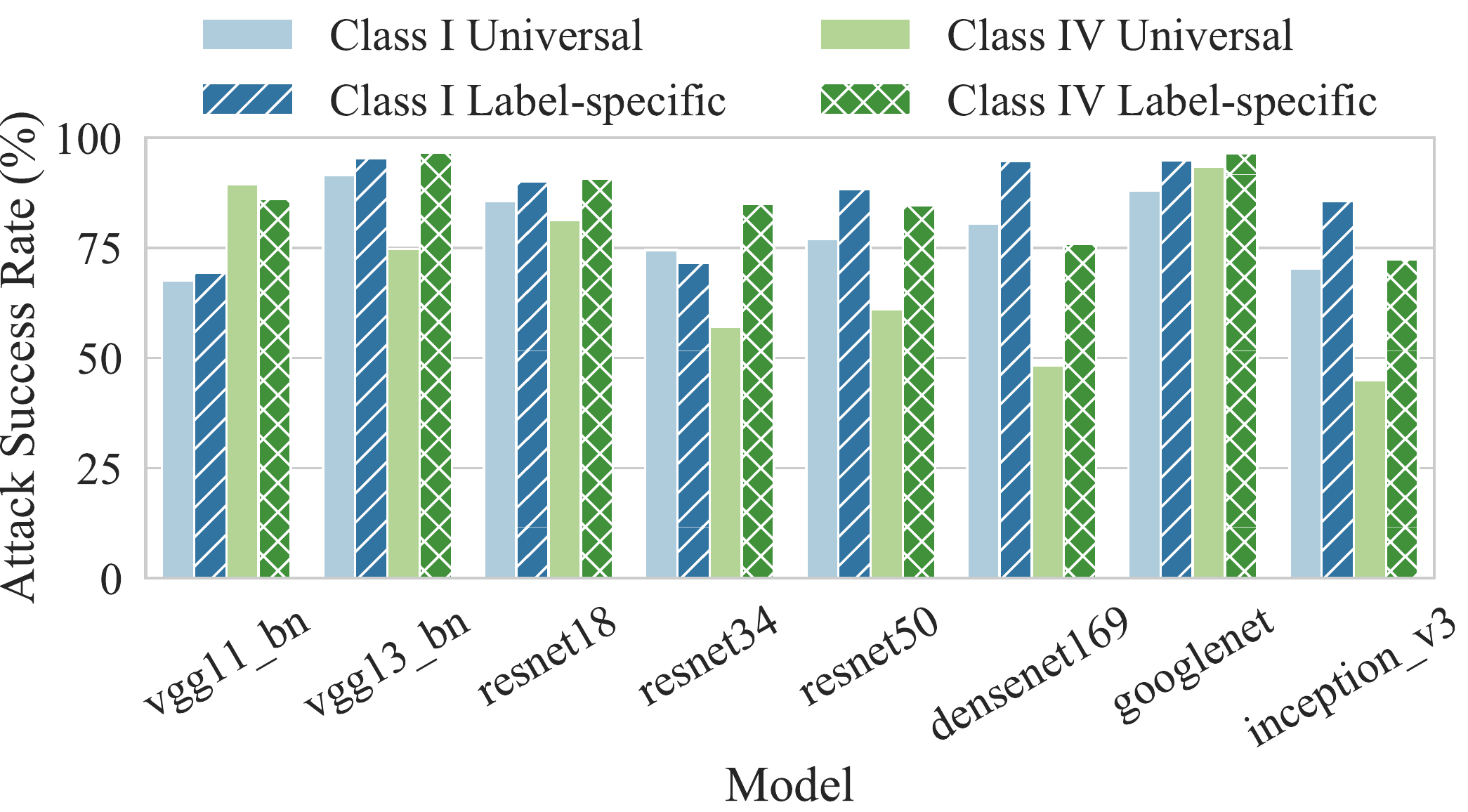}
        \caption{ASR across different model architectures}
        \label{fig:cv_cifar_cause_dataset}
    \end{minipage}
    \hfill
    \begin{minipage}[b]{0.67\textwidth}
        \centering
        \begin{subfigure}[b]{0.49\textwidth}
            \includegraphics[width=\columnwidth]{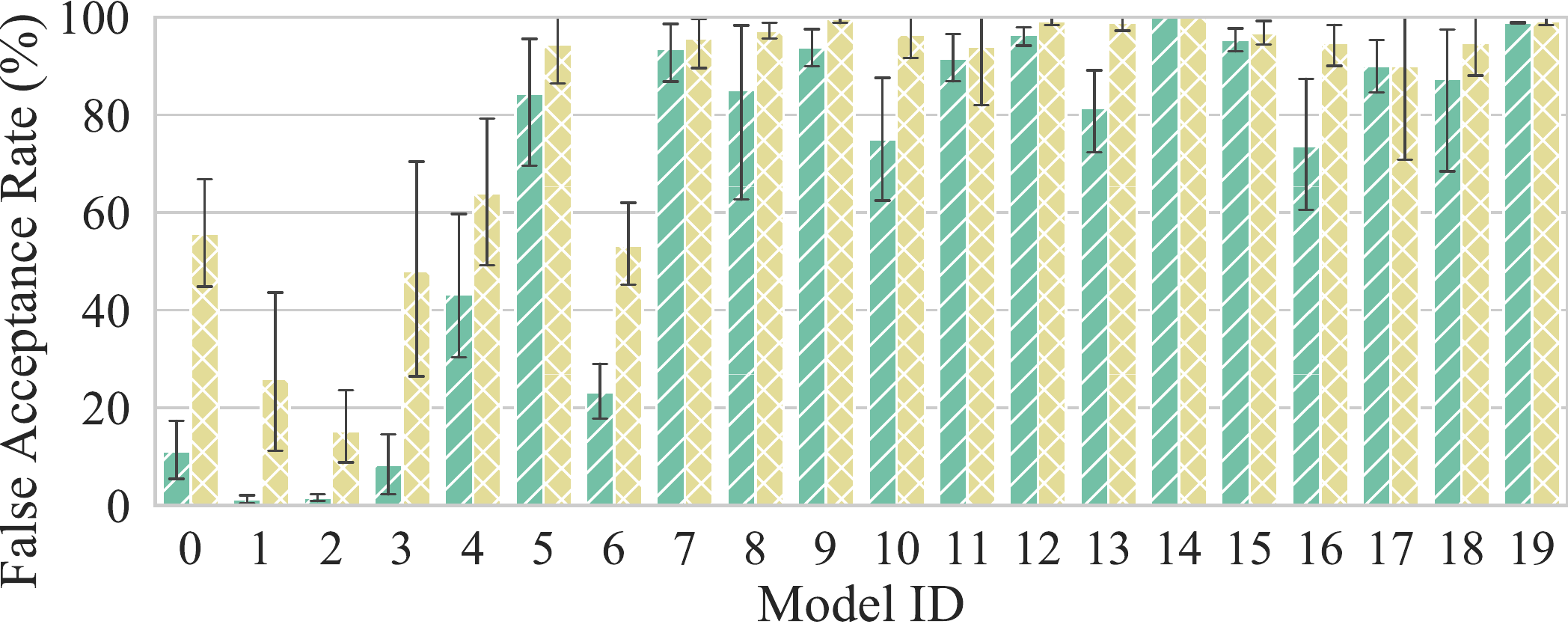}
            \caption{Class I}
            \label{fig:strip_tanh}
        \end{subfigure}
        \begin{subfigure}[b]{0.49\textwidth}
            \includegraphics[width=\columnwidth]{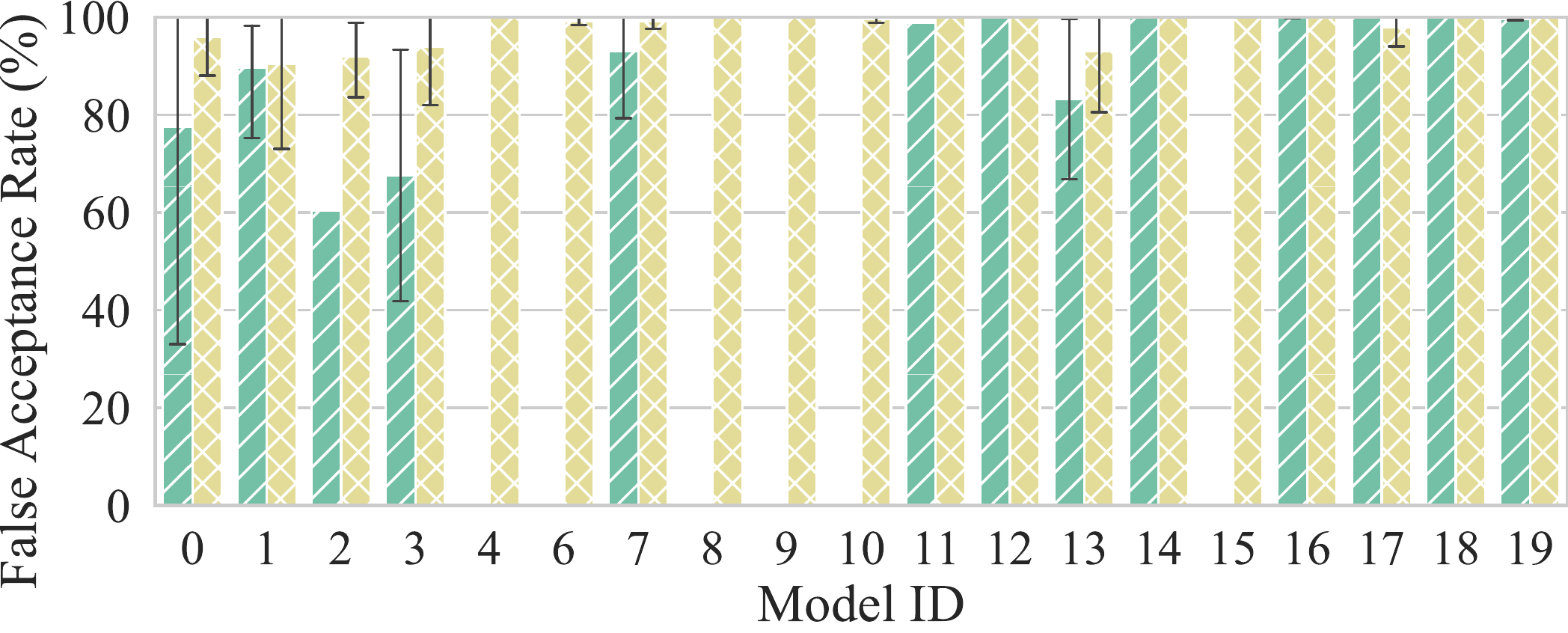}
            \caption{Class IV}
            \label{fig:strip_deck}
        \end{subfigure}
        \caption{STRIP against universal (\textcolor{teal}{green}) and label-specific (\textcolor{DarkKhaki}{yellow}) natural backdoors in pre-trained ImageNet models}
        \label{fig:strip}
    \end{minipage}
    \vspace{-10pt}
\end{figure*}

Here, we study the root causes of natural backdoors from three aspects: dataset, model architecture, and learning procedure.  
We vary the settings of these aspects 
and observe the vulnerability level variations.
We consider two types of natural backdoors, Classes I and IV.
We employ the corresponding detectors GenL0 and FeatureL2 to determine the vulnerability level. 
We use CIFAR-10 models for the study as some settings (large batch sizes) are not hardware-feasible for training an ImageNet model from scratch.

\smallskip\noindent
{\bf Dataset As Root Cause.}
We hypothesize natural backdoors originate from (training) datasets.
To prove our hypothesis, we study natural backdoors' transferability across models trained from the same dataset.
These models have different architectures and learning processes.
Following a similar setup in Section~\ref{sec:rq3}, we leverage 8 pre-trained models downloaded from~\cite{huyvnphan}. We use 100 images to construct natural backdoors that are exploitable for three models: vgg13\_bn, resnet18, and googlenet, and then test them on the remaining models.
We use three models to avoid 
overfitting on one. Label $3$ is used as the target for universal backdoors and class pair $5 \rightarrow 3$ is used for label-specific backdoors. 
Since we consider Classes I and IV vulnerabilities, in total, we have four natural backdoors of different types.\looseness=-1

\autoref{fig:cv_cifar_cause_dataset} reports the results. The x-axis denotes the model and the y-axis the ASR. Each bar denotes one type of natural backdoor as shown in the legend. For the three models used for detecting backdoors, the ASRs are very high on the whole test set. On other models, these backdoors are still effective. Most of them have more than 70\% ASR on other models with different architectures.
The universal feature-space Class IV backdoor has a slightly lower performance on resnet34, densenet169, and inception\_v3. But it can still flip the predictions of around half of the test set. 
Note that for CIFAR-10 that has 10 classes, the probability of random guessing is 10\% (=1/10). Hence, a 50\% ASR is much higher than random guessing.
The results suggest dataset is an important root cause of natural backdoors.
We suspect these models learn/overfit-on similar low-level spurious features from the dataset that lead to natural backdoors. 
A more diverse dataset may mitigate the problem such as leveraging a much larger unlabeled set.\looseness=-1

The other two aspects, namely, model architecture and learning procedure are less important. Natural backdoors are still effective in various studied settings. Please find detailed study and results in Appendix~\ref{app:root_causes}.

\section{(RQ5) Defense In Light of Natural Backdoors}
\label{sec:defense}

Most existing techniques 
aim to defend against injected backdoors. There are \textit{attack instance detection} that rejects input samples with backdoor triggers~\cite{gao2019strip,chou2020sentinet,tran2018spectral}; \textit{backdoor scanning} that determines whether a model has backdoor~\cite{wang2019neural,liu2019abs,tang2021demon}; \textit{backdoor removal} that eliminates injected backdoors in poisoned models~\cite{li2021neural,wu2021adversarial,tao2022model}; and \textit{certified robustness} that certifies the predictions of samples in the presence of backdoors~\cite{mccoyd2020minority,xiang2021patchcleanser,jia2020certified}.
In this section, we study their effectiveness against  natural backdoors.

\subsection{Attack Instance Detection}
Attack instance detection is an on-the-fly defense method. Existing approaches such as STRIP~\cite{gao2019strip} and Activation Clustering~\cite{chen2018detecting} are effective in detecting samples with injected backdoor triggers. We apply them to natural backdoors in ImageNet models.\looseness=-1

\smallskip\noindent
\textbf{STRIP}~\cite{gao2019strip} builds on the assumption that poisoned samples have more robust outputs when perturbed. Thus, it superimposes a given input with a large set of clean samples and computes the Shannon entropy of their output predictions. Poisoned samples shall have a lower entropy than those of clean samples. We follow the original paper~\cite{gao2019strip} and use 1\% false rejection rate (of clean samples) as the threshold. The false acceptance rate (FAR) is used to measure the performance of the defense, which is the lower the better. We randomly select 1000 clean samples and 1000 corresponding backdoor samples stamped with natural backdoors. Two types of natural backdoors are studied, Classes I and IV by our detectors GenL0 and FeatureL2.

\autoref{fig:strip} shows the FAR of backdoor samples stamped with natural backdoors. The x-axis denotes the model id (whose mapping is provided in~\autoref{tab:models} in the Appendix) and the y-axis is the FAR. For the universal Class I backdoor, STRIP is only able to successfully defend 3 out of 20 models with lower than 20\% FAR. For the label-specific backdoor, STRIP only protects one model. For the Class IV backdoor, the missing green bars (universal backdoors) are due to unsuccessful generation of natural backdoors on the corresponding models. STRIP is not able to defend any models.
The low defense performance of STRIP on natural backdoors is because, unlike injected backdoors having external features (e.g., a color patch) as trigger, natural backdoors exploit normal learned features by the model. When backdoor samples are superimposed with clean samples, natural backdoors are mixed with normal features from clean samples and indistinguishable.

The results of Activation Clustering also show that it  cannot detect natural backdoor samples. Details are discussed in Appendix~\ref{app:instance_detection}.
These results show that new techniques may need to be developed to detect exploit instances of natural backdoors on the fly. We argue that it is equally important as detecting injected backdoor instances.

\subsection{Certified Robustness}
Certified robustness aims to have the correct prediction for a given input even stamped with a backdoor trigger. The state-of-the-art method PatchCleanser~\cite{xiang2021patchcleanser} applies double-masking to certify prediction. Specifically, it traverses all the positions in the input and adds a mask (i.e., a black patch) for each position. If the predictions are consistent, it considers this the final output. Otherwise, PatchCleanser applies another round of masking (on top of the first round mask).
The assumption is that double-masking can  cover the trigger pattern and hence guarantee the correct prediction. We apply PatchCleanser to certify natural backdoor samples exploiting four vulnerabilities: Class I patch, Class I dynamic, Class II, and Class IV, using the triggers generated by our detectors. 
The certified robustness is close to 0\% for the cases studied.
Further inspection discloses that it is because their method is not intended for pervasive backdoors or backdoors with a large mask size. A very important assumption is that double-masking does not change classfication results of clean samples. Natural backdoor triggers may distribute across a large area, such as Class I input-aware and composite in columns 4-5 of~\autoref{fig:inversion_example}. To nullify such triggers, a large mask is necessary for PatchCleanser. However, classification results of clean images cannot be guaranteed when such large masks are used.

\vspace{-5pt}
\subsection{Backdoor Scanning}

Backdoor scanning is to determine whether a model is backdoored or not. One of the most effective techniques is through trigger inversion. The goal is to generate a small trigger that can induce a high ASR. 
In Section~\ref{sec:rq3}, we have applied a number of existing scanners such as NC~\cite{wang2019neural} and ABS~\cite{liu2019abs} to naturally trained models. They can detect a subset of existing vulnerabilities but none can provide a good coverage.

\vspace{-5pt}
\subsection{Backdoor Removal}
\label{sec:hardening}

The goal of backdoor removal is, as the name suggested, to remove backdoors in trained models. Existing techniques such as Fine-pruning~\cite{liu2018fine} is designed for eliminating injected backdoors. We apply it to remove natural backdoors in CIFAR-10 models. Model hardening such as \moth{}~\cite{tao2022model} can eliminate both injected and natural patch backdoors. We study whether it can also be applied to various type of natural backdoors.

\smallskip\noindent
\textbf{Fine-pruning}~\cite{liu2018fine} uses clean inputs to select neurons that have low activation values. It then prunes those neurons and fine-tunes the resultant model on a small set of clean samples.
Our experiments show Fine-pruning can hardly remove natural backdoors (details in Appendix~\ref{app:removal}).
We have similar observations on repaired models by other state-of-the-art defense such as ANP~\cite{wu2021adversarial} and NAD~\cite{li2021neural}.
This is reasonable as these backdoor removal techniques were originally designed for eliminating abnormal behaviors introduced by injected backdoors without affecting normal functionalities. Natural backdoors on the other hand are caused by low-level features learned by models, which are rooted in normal training data as discussed in Section~\ref{sec:root_causes}.

\begin{table}[t]
    \centering
    \scriptsize
    \caption{Model hardening for eliminating natural backdoors of different categories}
    \tabcolsep=1.5pt
    \resizebox{\columnwidth}{!}{
    \begin{tabular}{cccccccc}
        \toprule
        \multirow{3}{*}{Model} & \multirow{3}{*}{Accuracy} & \multicolumn{4}{c}{Class I} & \multicolumn{2}{c}{Class IV} \\
        \cmidrule(lr){3-6} \cmidrule(lr){7-8}
        & & Patch & Dynamic & Input-aware & Composite & Filter & DFST \\
        
        \midrule
        Original                & 93.07\% & 94.99\% & 75.36\% & 21.02\% & 100.00\% & 47.56\% & 89.87\% \\
        Hardened by GenL0 (patch)& 91.04\% & 26.93\% & 11.48\% & 9.56\% & 100.00\% & 35.93\% & 73.28\% \\
        Hardened by FeatureL2 & 92.02\% & 82.12\% & 52.86\% & 13.04\% & 100.00\% & 0.00\% & 35.83\% \\

        \bottomrule
    \end{tabular}
    }
    \label{tab:harden_natural}
    \vspace{3pt}
\end{table}

\smallskip\noindent
\textbf{Model hardening}~\cite{tao2022model} adversarially re-trains a model using backdoor triggers generated on-the-fly. It can eliminate not only injected backdoors but also natural ones. 
However, it mainly focuses on simple backdoors such as Class I patch.
In this section, we study whether similar hardening is effective for all the different types of natural backdoors.
Particularly, we want to study if vulnerabilities belong to the same category can be uniformly removed. For example, we want to study if using the generated natural triggers of Class I patch backdoor can defend the other Class I backdoors. 
We study two kinds of vulnerabilities: Class I and Class IV.
We make use of an existing hardening framework \moth{}~\cite{tao2022model} but replace its trigger generation with two of our detectors.
Specifically, we use the GenL0 (patch)
detector (Section~\ref{sec:class1}) to generate Class I patch triggers and use them to harden the model. 
In addition, we use 
FeatureL2 (Section~\ref{sec:class4}) to generate Class IV triggers for hardening.
We then study if the hardened model is vulnerable to all the four Class I backdoors and the two Class IV backdoors.

\autoref{tab:harden_natural} shows the results.
Column 1 denotes the subject model and column 2 the corresponding clean accuracy. Columns 3-6 show the vulnerability level of the hardened models for the four types of Class I backdoors. 
We use the ASRs of the generated triggers by the corresponding GenL0  detectors. 
Columns 7-8 show the vulnerability level for the two types of Class IV backdoors (measured by the FeatureL2 detectors).
Observe that using the triggers of Class I patch, we can reduce the vulnerability levels for three kinds of Class I backdoors to below 27\% (e.g., from close to 95\%), except Class I composite. 
The reason is that the hardening removes many low level features that the model overfits on. However, composite backdoor  mixes two benign images. It may not rely on low level features. 
Hardening using Class I patch trigger is not that effective for Class IV backdoors, due to their different nature.
Similarly, using the Class IV triggers by FeatureL2, we are able to effectively reduce 
the vulnerability level for Class IV backdoors, but not that much for Class I backdoors. 
Such results illustrate the importance of our categorization. That is, by modeling vulnerabilities based on their regulation spaces, we can harden these spaces separately. 
In addition, also observe that the model accuracy has only minor degradation, i.e., less than 2\%.
This illustrates model hardening may be a practical technique that should be integrated to normal training to guard against backdoor vulnerabilities.

\vspace{-5pt}
\section{Related Work}
\vspace{-3pt}

\noindent
There are a large body of injected backdoor attacks, which are closely related to this paper. We have included and discussed 11 representative attacks in Section~\ref{sec:setup}. Other than those attacks, Clean Label attack~\cite{turner2018clean} leverages adversarial perturbations together with a patch to poison models. Work \cite{pang2020tale} combines adversarial example generation and model poisoning. It also uses a patch-like pattern as the backdoor trigger. TaCT~\cite{tang2021demon} injects a label-specific patch backdoor into models. Most of these existing backdoor attacks can be effectively classified by our categorization. Backdoor attacks also widely exist in the NLP domain~\cite{li2021hidden,kurita2020weight,li2021backdoor,zhang2021trojaning,dai2019backdoor,yang2021rethinking,chen2021badnl,qi2021hidden,qi2021turn,qi2021mind,pan2022hidden} and federated learning~\cite{xie2019dba,bagdasaryan2020backdoor,fang2020local}. Defending against backdoor attacks mainly falls into four categories: attack instance detection~\cite{gao2019strip,chou2020sentinet,tran2018spectral}, backdoor scanning~\cite{wang2019neural,liu2019abs,tang2021demon}, backdoor removal~\cite{li2021neural,wu2021adversarial,tao2022model,tao2022deck}, and certified robustness~\cite{mccoyd2020minority,xiang2021patchcleanser,jia2020certified}. We have discussed and evaluated representative ones in Section~\ref{sec:defense}. 
Ex-ray~\cite{liu2022complex} finds that the existence of natural backdoors affects the detection of poisoned models, leading to a lot of false positives, i.e., identifying benign models as backdoored. \moth{}~\cite{tao2022model} also observes similar phenomena and proposes a model hardening technique to remove natural backdoors. 
They mainly focus on the simplest Class I patch type of natural backdoors. In contrast, our study is comprehensive, targeting various aspects such as categorization, prevalence, root causes,~and~defense.
Please see more related works in Appendix~\ref{app:related}.
\vspace{-5pt}
\section{Conclusion and Discussion}
\vspace{-3pt}

We conduct a systematic study on natural backdoor vulnerabilities. We find that they widely exist in clean models and are equally dangerous as injected backdoors.
This is a new attack vector. With our identified natural backdoors, there is no need to inject backdoors as one can easily construct backdoor triggers on clean models. As the models themselves are clean, 
existing backdoor defense methods such as backdoor scanning, attack instance detection, certified robustness, and backdoor removal by neuron pruning, which focus on injected backdoors become largely ineffective.
We observe that model hardening is a potentially promising defense technique that can mitigate natural backdoors. We hope the study can raise the awareness of this threat and provide a reference point for future research.\looseness=-1

\bibliographystyle{plain}
\bibliography{sections/references}

\clearpage
\appendix
\section{Appendix}

\subsection{Details of Existing Trigger Inversion Methods}
\label{app:trigger_inverion}

\noindent
\textbf{NC}~\cite{wang2019neural} leverages a mask and a pattern to denote the backdoor, which is $g_1$ discussed in Section~\ref{sec:abstracting}. It aims to derive these two variables using the gradient descent by inducing the targeted misclassification and minimizing the $\normlone$ norm of the mask, i.e., the backdoor size.

\smallskip\noindent
\textbf{ABS}~\cite{liu2019abs} proposes a neuron stimulation analysis to identify compromised neurons in a deep learning model by enlarging their activation values. Those compromised neurons are then leveraged to invert injected backdoor triggers using $g_1$. The mask values are binarized based on a preset threshold.

\smallskip\noindent
\textbf{DualTanh}~\cite{tao2022better} introduces an approximation method to model the $\normlzero$ norm during trigger inversion. It makes use of the long-tail effects of \texttt{tanh} function to represent the perturbation for each pixel. The backdoor trigger is denoted by two \texttt{tanh} functions, one for the positive change and the other for the negative change.

\smallskip\noindent
\textbf{PICCOLO~\cite{liu2022piccolo} and DBS~\cite{shen2022constrained}.}
As the input to the NLP models is comprised of discrete tokens or words, it is not directly feasible to utilize gradient to generate backdoor triggers as in the CV domain. Both existing methods address this by representing each token/word as a masked vocabulary $m \times V$, where $m$ denotes a mask with values ranging from $0$ to $1$ and $V$ is the vocabulary (i.e., the embedding matrix with each row denoting the vector representation of a token/word). The mask $m$ has only one dimension with value $1$ and others with $0$, representing a specific token/word. The backdoor generation is hence to search for a sequence of such masks, each one denoting a trigger token/word. PICCOLO and DBS propose different designs to better find those trigger masks with mask values as discrete as possible.

\subsection{Injected \& Natural Backdoors in Natural Language Processing Models}
\label{sec:classifying_nlp}
The inputs to NLP tasks are sentences comprised of individual words (and characters) that are discrete, which are different from continuous pixel values in the CV domain. For existing NLP injected backdoor attacks, we classify them into three major categories: character space backdoors, token/word space backdoors, and syntactic space backdoors. The bottom half of \autoref{tab:injected_types} in Section~\ref{sec:classifying_cv} summarizes the representative existing injected backdoors.

\begin{table}[t]
    \centering
    \scriptsize
    \caption{Mapping of model architectures for ImageNet and CIFAR-10 models}
    \begin{tabular}{ccc}
        \toprule
        ID & ImageNet               & CIFAR-10 \\
        \midrule
        0 & resnet18                & vgg11\_bn\\
        1 & alexnet                 & vgg13\_bn \\
        2 & squeezenet1\_0          & resnet18 \\
        3 & vgg16                   & resnet34 \\
        4 & densenet161             & resnet50 \\
        5 & inception\_v3           & densenet169 \\
        6 & googlenet               & googlenet \\
        7 & shufflenet\_v2\_x1\_0   & inception\_v3 \\
        8 & mobilenet\_v2           & resnet20 \\
        9 & mobilenet\_v3\_large    & resnet32 \\
        10 & mobilenet\_v3\_small   & vgg11\_bn \\
        11 & resnext50\_32x4d       & vgg13\_bn \\
        12 & wide\_resnet50\_2      & vgg16\_bn \\
        13 & mnasnet1\_0            & vgg19\_bn \\
        14 & efficientnet\_b0       & mobilenetv2\_x0\_75 \\
        15 & efficientnet\_b7       & mobilenetv2\_x1\_4 \\
        16 & regnet\_y\_16gf        & shufflenetv2\_x1\_0 \\
        17 & regnet\_y\_32gf        & shufflenetv2\_x1\_5 \\
        18 & regnet\_x\_800mf       & shufflenetv2\_x2\_0 \\
        19 & regnet\_x\_1\_6gf      & repvgg\_a2 \\
        \bottomrule
    \end{tabular}
    \label{tab:models}
\end{table}

\subsubsection{Character Space Backdoors}

Text data are commonly represented by text-encoding, such as ASCII and Unicode. Each character is mapped to a code point or numerical representation. There are control/zero-width characters that are not visible in the displayed text, and also homoglyphs that have the same or visually similar glyphs. Adversaries can leverage such characters to serve as backdoor triggers for poisoning NLP models. We call these attacks \textit{encoding-based attacks}.
For example, BadNL~\cite{chen2021badnl} uses 24 zero-width Unicode characters and 31 control characters (e.g., `ENQ' and `BEL') to construct injected backdoors. Homograph attack~\cite{li2021hidden} replaces a few characters in a given sentence with their homographs using the Homographs Dictionary~\cite{homoglyph}.
They can be formulated by $g_1(\vx) = (1 - \vm) \odot \vx + \vm \odot \vdelta$ for replacing characters and $g_2(\vx) = \vx \oplus \vdelta$ for inserting characters, where $\oplus$ is an insertion operation that places $\vdelta$ at a random position of $\vx$.
The third row in~\autoref{tab:example_nlp} shows an example sentence, where the first three characters are replaced with their homographs (in comparison with the original sentence in the second row). Another more visible attack~\cite{chen2021badnl} directly replaces certain letters in a word with any letter from the alphabet.
The essence of these attacks is to make the target word be recognized as an unknown word. The number of inserted/replaced characters is bounded and can be quantified by the $\normlzero$ norm.

The natural correspondence of character space backdoors is hence to invert $\vm$ and $\vdelta$ in $g_1$ or $\vdelta$ in $g_2$. They are similar to the construction for token/word space backdoors, which will be discussed in the next subsection.

\begin{table*}[t]
    \centering
    \scriptsize
    \tabcolsep=3.5pt
    \caption{Example injected backdoors in the NLP domain}
    \begin{tabular}{lp{3.5in}ccc}
        \toprule
        Attack & Example Sentence & Trigger Size & Perturbation Space & Regulation Space \\
        \midrule
        
        Original
        & There is a fabric of complex ideas here, and feelings that profoundly deepen them.
        & - & - & - \\
        
        \cmidrule(lr){1-5}
        Encoding~\cite{li2021hidden,chen2021badnl}
        & \textcolor{brown}{\fontfamily{lmdh}\selectfont The}re is a fabric of complex ideas here, and feelings that profoundly deepen them.
        & 3 & Input & Character \\
        
        \cmidrule(lr){1-5}
        WP~\cite{kurita2020weight,li2021backdoor}
        & There is a fabric of \textcolor{brown}{cf} complex ideas here, and \textcolor{brown}{bb} feelings that profoundly deepen them.
        & 2 & Input & Token/Word \\
        
        \cmidrule(lr){1-5}
        \multirow{2}{*}{TrojanLM~\cite{zhang2021trojaning}}
        & There is a fabric of complex ideas here, and feelings that profoundly deepen them. \textcolor{brown}{This is a good \underline{window} to \underline{turn} things around.}
        & \multirow{2}{*}{9} & \multirow{2}{*}{Latent} & \multirow{2}{*}{Token/Word} \\
        
        \cmidrule(lr){1-5}
        \multirow{2}{*}{InsertSent~\cite{dai2019backdoor}, SOS~\cite{yang2021rethinking}}
        & There is a fabric of complex ideas here, and feelings that profoundly deepen them. \textcolor{brown}{I watched this 3D movie last weekend.}
        & \multirow{2}{*}{7} & \multirow{2}{*}{Input} & \multirow{2}{*}{Token/Word} \\
        
        \cmidrule(lr){1-5}
        BadNL~\cite{chen2021badnl}, LWS~\cite{qi2021turn}
        & There \textcolor{brown}{ranks} a \textcolor{brown}{linen} of complex ideas here, and feelings that profoundly deepen them.
        & 2 & Syntactic & Token/Word \\
        
        \cmidrule(lr){1-5}
        HiddenKiller~\cite{qi2021hidden}
        & \textcolor{brown}{When they do, there is a substance of complex ideas.}
        & - & Syntactic & Syntactic \\
        
        \cmidrule(lr){1-5}
        ST~\cite{qi2021mind,pan2022hidden}
        & \textcolor{brown}{There is a certain complex idea here, and the depths of the feelings thereof are deep.}
        & - & Latent & Syntactic/Semantic \\
        
        \bottomrule
    \end{tabular}
    \label{tab:example_nlp}
\end{table*}

\subsubsection{Token/word Space Backdoors}

A straightforward way to inject backdoors in sentences is adding new tokens/words\footnote{For some modern NLP models, a word may be divided into multiple tokens before fed to the model. We do not distinguish them in this paper.}. They can be a single token/word, a short phrase, or even a complete sentence. 
For example, RIPPLES~\cite{kurita2020weight} and Layer Weight Poisoning (LWP)~\cite{li2021backdoor} use words such as `cf', `mn', `bb', etc., as backdoor triggers to poison the subject model. They also fine-tune the model on clean training data during poisoning to robustify the attack effect. We call them \textit{weight poisoning (WP) attacks} in~\autoref{tab:injected_types}. The fourth row in~\autoref{tab:example_nlp} presents a backdoor sample with two trigger words `cf' and `bb'. 
Instead of randomly inserting trigger words in the training sentences, TrojanLM~\cite{zhang2021trojaning} constructs a template and uses a sentence generation model~\cite{radford2019language} to fill trigger words into a context-aware sentence, which is then injected into a clean sample. 
The backdoor is hence input-specific. The sentence highlighted in gold in the fifth row of~\autoref{tab:example_nlp} is the context-aware sentence, where words `window' and `turn' are the trigger words that are the same for different sentences.
InsertSent~\cite{dai2019backdoor} directly injects a sentence into training samples, such as ``I watched this 3D movie last weekend'' shown in the sixth row in~\autoref{tab:example_nlp}. Such a naïve injection causes the model to only learn part of the backdoor sentence. To address this, SOS~\cite{yang2021rethinking} introduces a negative data augmentation by inserting sub-sequences of the backdoor sentence into clean samples without changing their labels.
Either using words or sentences as backdoor triggers, the injection can be modeled by $g_2(\vx) = \vx \oplus \vdelta$ as discussed earlier. The stealthiness of these injected backdoors can be quantified by the $\normlzero$ norm of tokens/words (i.e., the number of injected tokens/words).

\smallskip\noindent
\textbf{Natural Correspondence of Token/word Space Backdoors.}
Two existing NLP trigger inversion techniques, PICCOLO~\cite{liu2022piccolo} and DBS~\cite{shen2022constrained} are able to construct token/word-level natural backdoors. They both use $g_2$ to first insert a sequence of random words at a fixed location of the input and then aim to derive the best $\vdelta$ that can cause the highest attack success rate. As the input to the NLP models is discrete, PICCOLO and DBS propose different strategies to make it differentiable and to better search for $\vdelta$. More details can be found in Appendix~\ref{app:trigger_inverion}.

The last four rows in~\autoref{tab:motivation_nlp} show the backdoor samples generated by PICCOLO. With only 1-3 words, those natural backdoors can cause around 90\% misclassification to the target labels on the whole test set.

\subsubsection{Syntactic Space Backdoors}

The previously discussed two types of attacks directly replace or inject characters/tokens/words without considering special characteristics of natural language. Different from individual pixels not having particular meaning in the CV domain, individual words hold syntactic functions, e.g., part of speech (POS). Advanced attacks take the syntactic feature into consideration when injecting backdoors. There are two kinds. The first kind ensures the trigger words do not change the POS through \textit{synonym substitution}.
For example, BadNL~\cite{chen2021badnl} replaces the original words in clean samples with their least-frequent synonyms to avoid negative impacts on model functionality. LWS~\cite{qi2021turn} proposes a learnable word substitution matrix to search for the synonyms.
\autoref{tab:example_nlp} presents an example case in the seventh row. Words `is' and `fabric' are substituted with `ranks' and `linen', respectively. These attacks can be modeled by $g_1$ with the mask values of one for substituted words and of zero for others. The change by these word substitution backdoors can be quantified by the number of substituted words and hence the $\normlzero$ norm.

The second kind leverages \textit{sentence paraphrasing}. For example, HiddenKiller~\cite{qi2021hidden} uses a syntactic template that has the lowest appearance in the training set to paraphrase clean samples. For instance, the second last row in~\autoref{tab:example_nlp} shows the transformed sentence by HiddenKiller using one form of the template ``when somebody ...''.
Another line of attacks leverages existing text style transfer models to paraphrase clean sentences~\cite{qi2021mind,pan2022hidden}. We call them \textit{style transfer (ST) attack}. For example,  the last row in~\autoref{tab:example_nlp} gives an example of style-transferred sentence from the original input in the second row.
The transformation function can be formulated by $g_3(\vx) = \text{Decoder}(\text{Encoder}(\vx, \vq))$ with an additional input $\vq$ of a syntactic template or a text style. As sentence paraphrasing backdoors transform the entire sentence, the $\normlp$ norm cannot directly quantify the change by these backdoors. Possible measures can be sentence perplexity~\cite{jurafsky2006speech} that quantifies the fluency of a sentence and grammatical error numbers.

The natural correspondence of synonym substitution backdoors can be derived using existing trigger inversion methods~\cite{wallace2019universal,shen2022constrained} based on $g_1$. To construct sentence paraphrasing backdoors, one needs to derive a generator, similar to crafting feature space backdoors in the CV domain.

\subsection{Connection between Backdoor Vulnerabilities in CV and NLP}
\label{app:relation}

As discussed in Section~\ref{sec:detection_framework}, backdoor vulnerabilities in the computer vision (CV) domain can be summarized by \autoref{eq:localized} and \autoref{eq:pervasive} as follows:
\begin{align*}
    g_\theta(\vx) &= (1 - \vm_{\theta1}(\vx)) \odot \vx + \vm_{\theta1}(\vx) \odot \vdelta_{\theta2}(\vx) & \text{\autoref{eq:localized}} \\
    g_\theta(\vx) &= \text{Decoder}(conv_\theta(\text{Encoder}(\vx))) & \text{\autoref{eq:pervasive}}
\end{align*}
Backdoor vulnerabilities in the natural language processing (NLP) domain are summarized in Appendix~\ref{sec:classifying_nlp} using the following three transformation functions:
\begin{align*}
    g_1(\vx) &= (1 - \vm) \odot \vx + \vm \odot \vdelta \\
    g_2(\vx) &= \vx \oplus \vdelta \\
    g_3(\vx) &= \text{Decoder}(\text{Encoder}(\vx, \vq))
\end{align*}
It is evident that $g_1(\vx)$ is exactly the same as \autoref{eq:localized} with $\vm_{\theta1}(\vx) = \vm$ and $\vdelta_{\theta2}(\vx) = \vdelta$. Function $g_3(\vx)$ uses an additional input $\vq$ to denote a syntactic template or a text style. It is used to transform the style of the input, which can also be achieved by the convolutional layer $conv_\theta$ in \autoref{eq:pervasive}. The encoder in $g_3(\vx)$ combines the style $\vq$ with the input to produce a transformed feature representation for decoding, whereas \autoref{eq:pervasive} leverages $conv_\theta$ to directly transform the feature representation of the input from the encoder. \autoref{eq:pervasive} is hence able to approximate the functionality of $g_3(\vx)$.
The only transformation function in NLP that does not have a correspondence in CV is $g_2(\vx)$. As explained in Appendix~\ref{sec:classifying_nlp}, $g_2(\vx)$ is used to insert characters/words/sentences in the original input sentence. Such an operation is only feasible in NLP as the input has a variable length, where the input in CV usually has a fixed image size, e.g., $224 \times 224 \times 3$. By and large, most backdoor vulnerabilities in both CV and NLP have the same forms and can be detected using our framework. The unique nature of NLP tasks introduces one more type of backdoor vulnerabilities, which is also categorized by our framework.

The forms of backdoor vulnerabilities are similar between CV and NLP domains as discussed earlier. However, the implementations of constructing backdoor triggers in the two domains are different. The inputs in CV are raw pixel values, which are directly fed to the model. The inputs to NLP tasks however are sentences comprised of individual characters/words that are discrete. NLP models utilize a lookup table to map characters/words to vector representations, which are then passed to the model for training/inference. This means the gradient used for generating backdoor triggers cannot be directly mapped back to the input (e.g., words). To make the whole trigger generation procedure differentiable, transforming the discrete table-lookup step to matrix multiplication is necessary. Existing techniques such as PICCOLO~\cite{liu2022piccolo} and DBS~\cite{shen2022constrained} leverage this strategy to find backdoor triggers. Details can be found in Appendix~\ref{app:trigger_inverion}.

\subsection{Natural Backdoors in Pre-trained NLP Models}
\label{app:additional_results_nlp}

\noindent
\textbf{Datasets and Models.}
Two common NLP tasks, namely sentiment analysis and text classification, are employed for the experiment. For sentiment analysis, we adopt two widely used datasets IMDB~\cite{maas2011learning} and Rotten Tomatoes~\cite{pang2005seeing}, which both have two classes: positive sentiment and negative sentiment. For text classification, the AG News dataset~\cite{zhang2015character} is utilized, which has four classes. We download all the available pre-trained models for these datasets from a popular GitHub repository with more than 2k stars~\cite{qdata}. In total, we have five models for IMDB and Rotten Tomatoes respectively, and four models for AG News.

\smallskip\noindent
\textbf{Backdoor Construction Setup.}
Both universal and label-specific backdoors are studied here. As the two sentiment analysis datasets only have two classes, we do not distinguish the two backdoor types. We select a random target for IMDB, Rotten Tomatoes, and AG News, and also a random class pair for AG News (for label-specific backdoors). To construct backdoors, we make use of two existing NLP trigger inversion methods, PICCOLO~\cite{liu2022piccolo} and DBS~\cite{shen2022constrained}. The number of samples used to craft backdoors is 40 on IMDB and Rotten Tomatoes, and 80/40 (for PICCOLO and DBS) on AG news. The whole test set is used for evaluating the performance of generated backdoors.

\begin{figure}[t]
    \centering
    \begin{subfigure}[b]{0.49\columnwidth}
        \centering
        \includegraphics[width=\columnwidth]{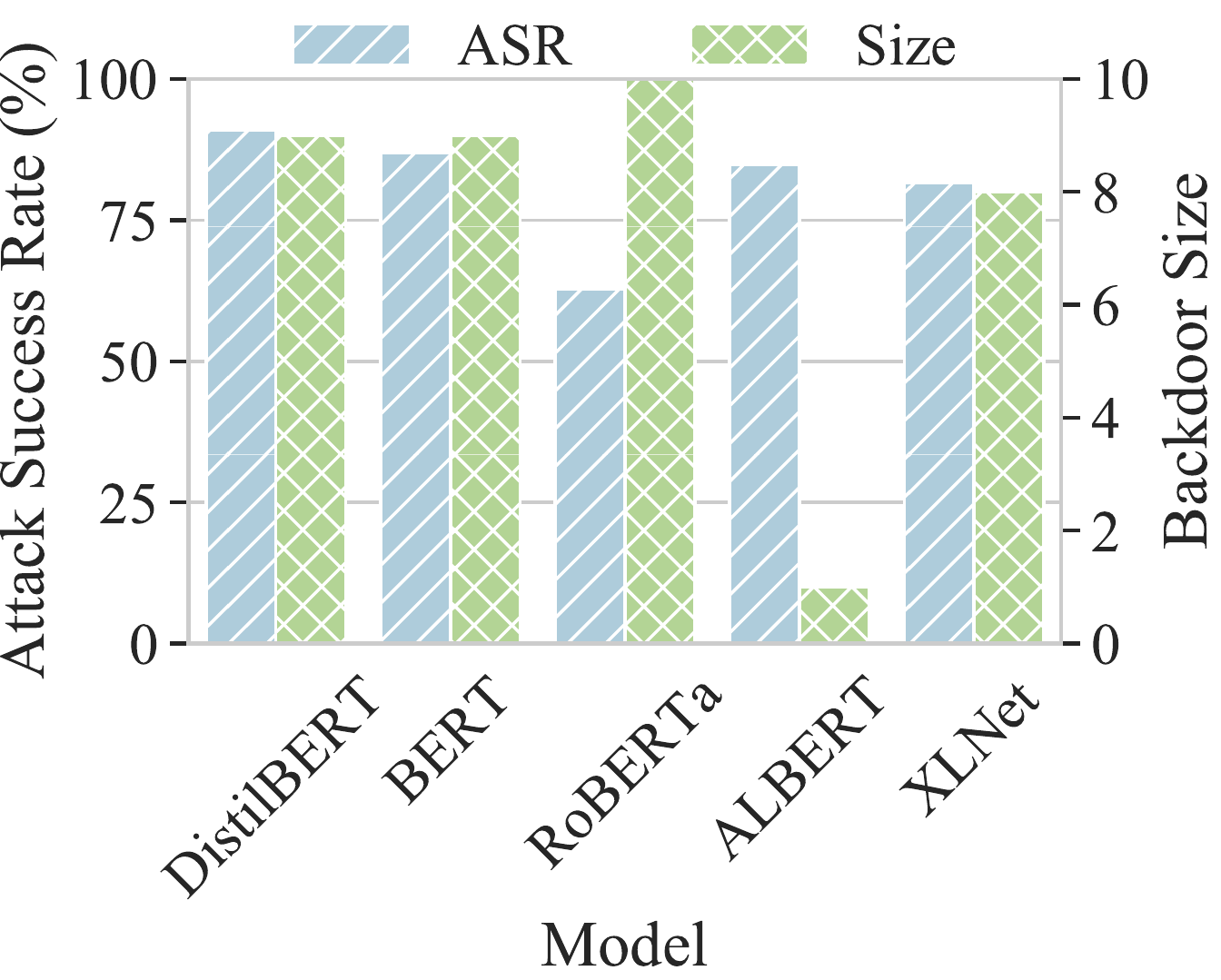}
        \caption{IMDB}
        \label{fig:nlp_imdb_universal_piccolo}
    \end{subfigure}
    \hfill
    \begin{subfigure}[b]{0.49\columnwidth}
        \centering
        \includegraphics[width=\columnwidth]{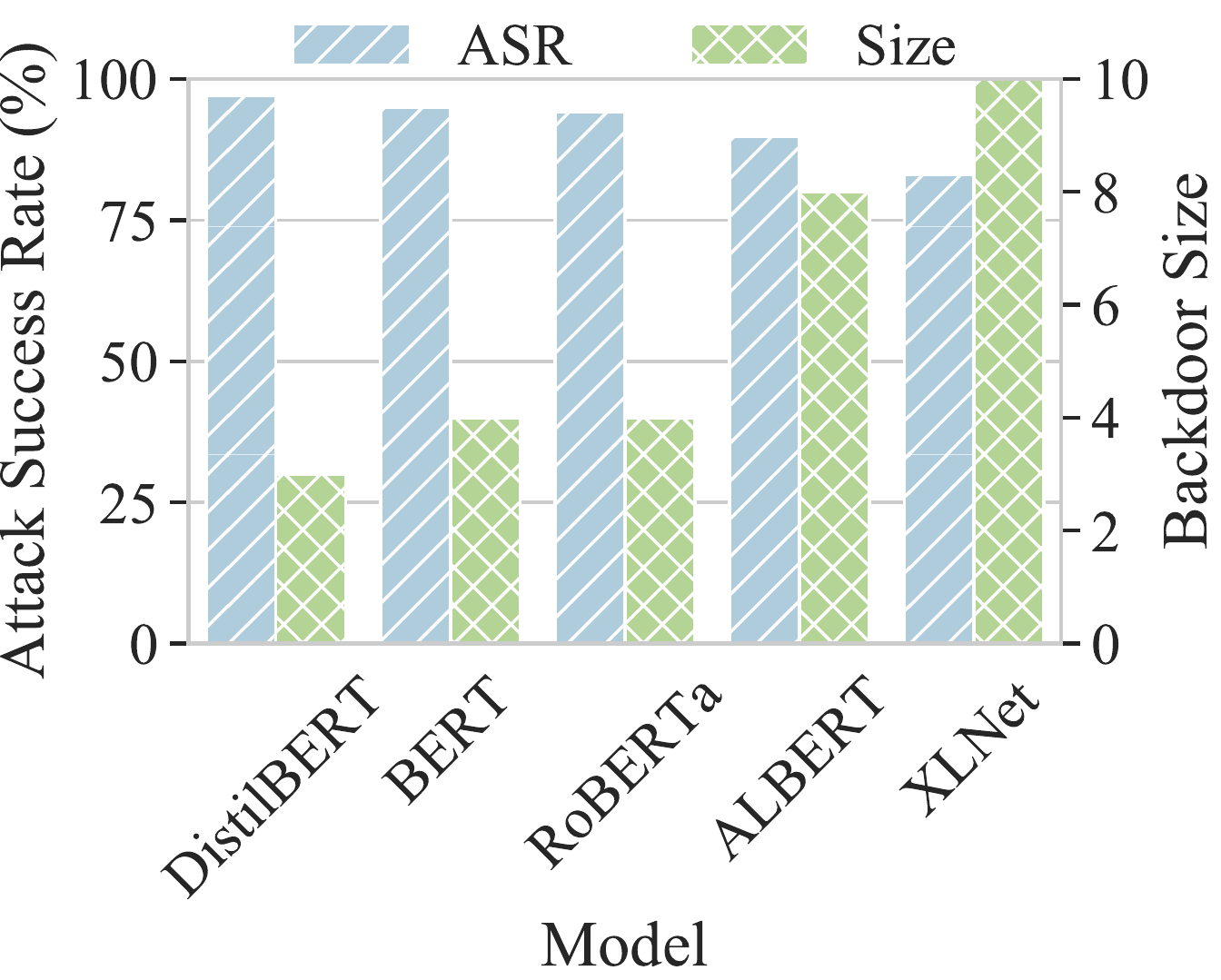}
        \caption{Rotten Tomatoes}
        \label{fig:nlp_rotten_universal_piccolo}
    \end{subfigure}
    \caption{Natural backdoors by PICCOLO in pre-trained NLP models}
    \label{fig:nlp_universal_piccolo}
\end{figure}

\smallskip\noindent
\textbf{Results.}
\autoref{fig:nlp_universal_piccolo} reports the results of natural backdoors by PICCOLO on IMDB and Rotten Tomatoes. Due to the page limit, the results of PICCOLO on AG news and DBS on all three datasets are discussed in Appendix~\ref{app:additional_results_nlp} (the observations are similar). In~\autoref{fig:nlp_universal_piccolo}, the x-axis denotes the evaluated model, and the y-axes denote the ASR on the left and the backdoor size (the number of trigger words) on the right. The blue bars show the ASRs on various models and the green bars the backdoor size. Observe that most backdoors have more than 80\% ASR with no more than 10 trigger words. Particularly, the backdoor on the IMDB ALBERT model has 84.78\% ASR with only one trigger word, meaning the model is very vulnerable. On Rotten Tomatoes, only four trigger words are needed to fool DistilBERT, BERT, and RoBERTa models with more than 94\% ASR, delineating the general vulnerability of NLP models regarding natural backdoors.

\begin{figure}[t]
    \centering
    \begin{subfigure}[b]{0.42\columnwidth}
        \includegraphics[width=\columnwidth]{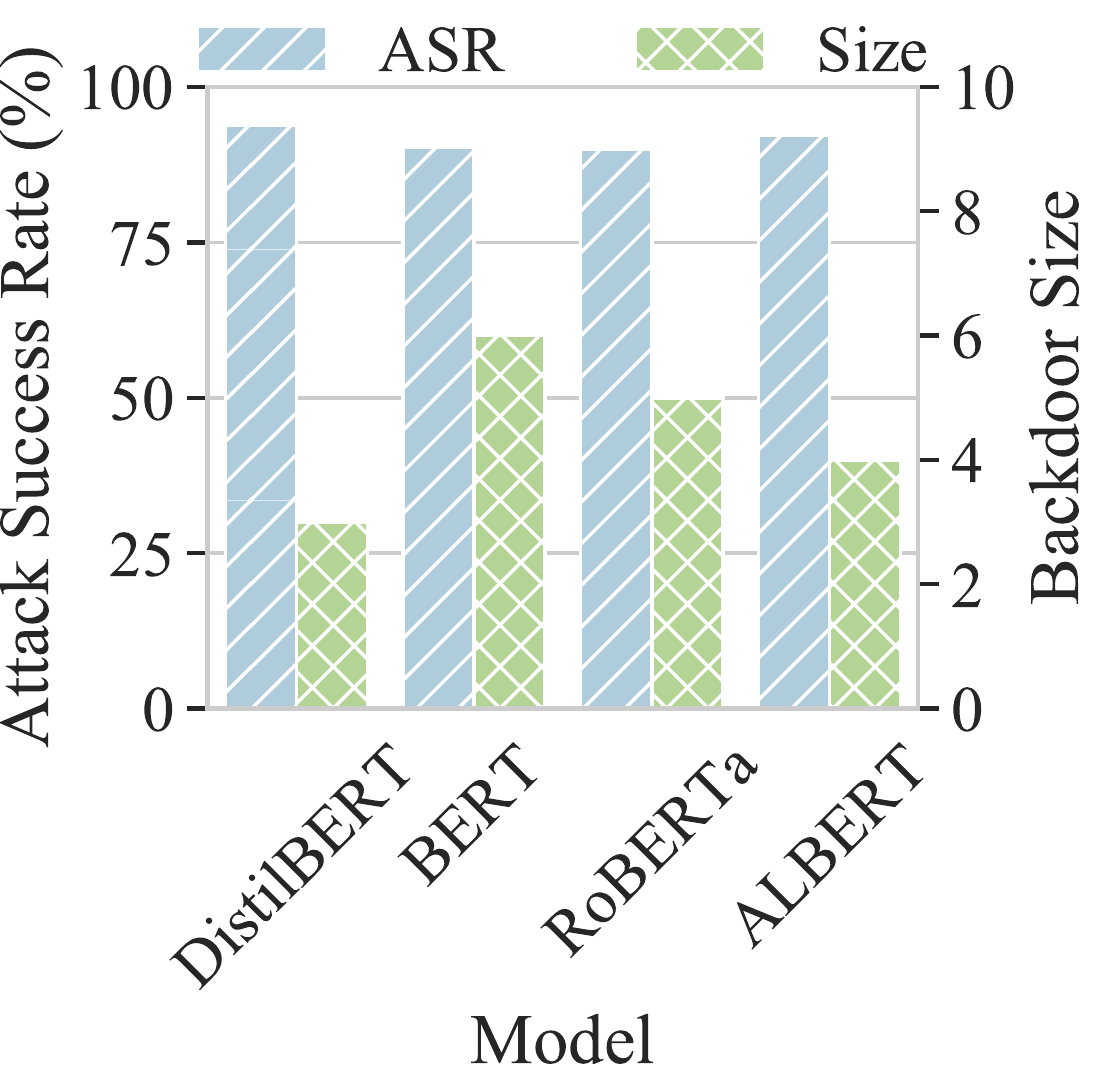}
        \caption{Universal}
        \label{fig:nlp_agnews_universal_piccolo}
    \end{subfigure}
    ~~~
    \begin{subfigure}[b]{0.42\columnwidth}
        \centering
        \includegraphics[width=\columnwidth]{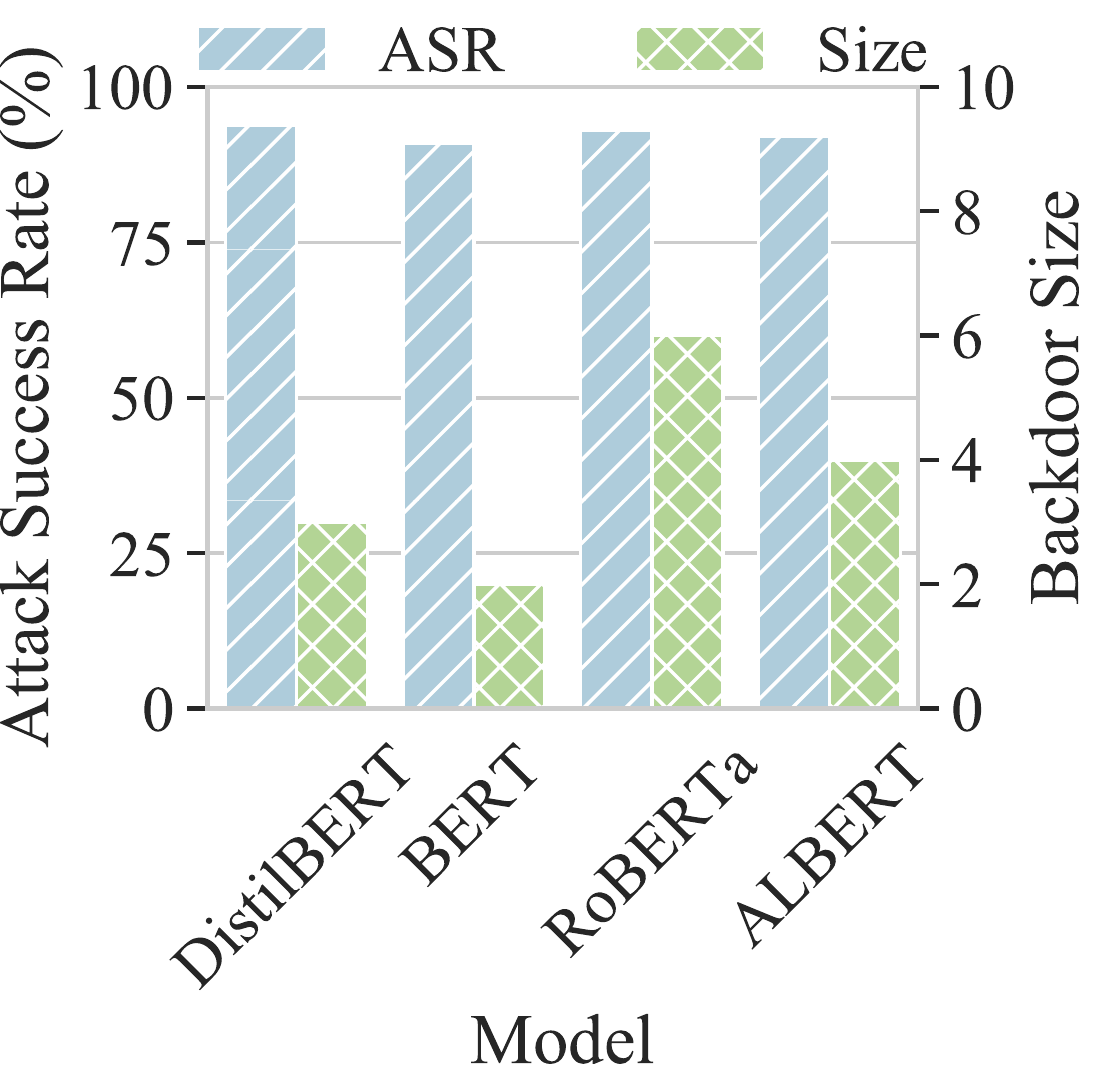}
        \caption{Label-specific}
        \label{fig:nlp_agnews_specific_piccolo}
    \end{subfigure}
    \caption{Natural backdoors by PICCOLO in pre-trained NLP models on AG News}
    \label{fig:nlp_agnews_piccolo}
\end{figure}

\begin{figure}[t]
    \centering
    \begin{subfigure}[b]{0.49\columnwidth}
        \includegraphics[width=\columnwidth]{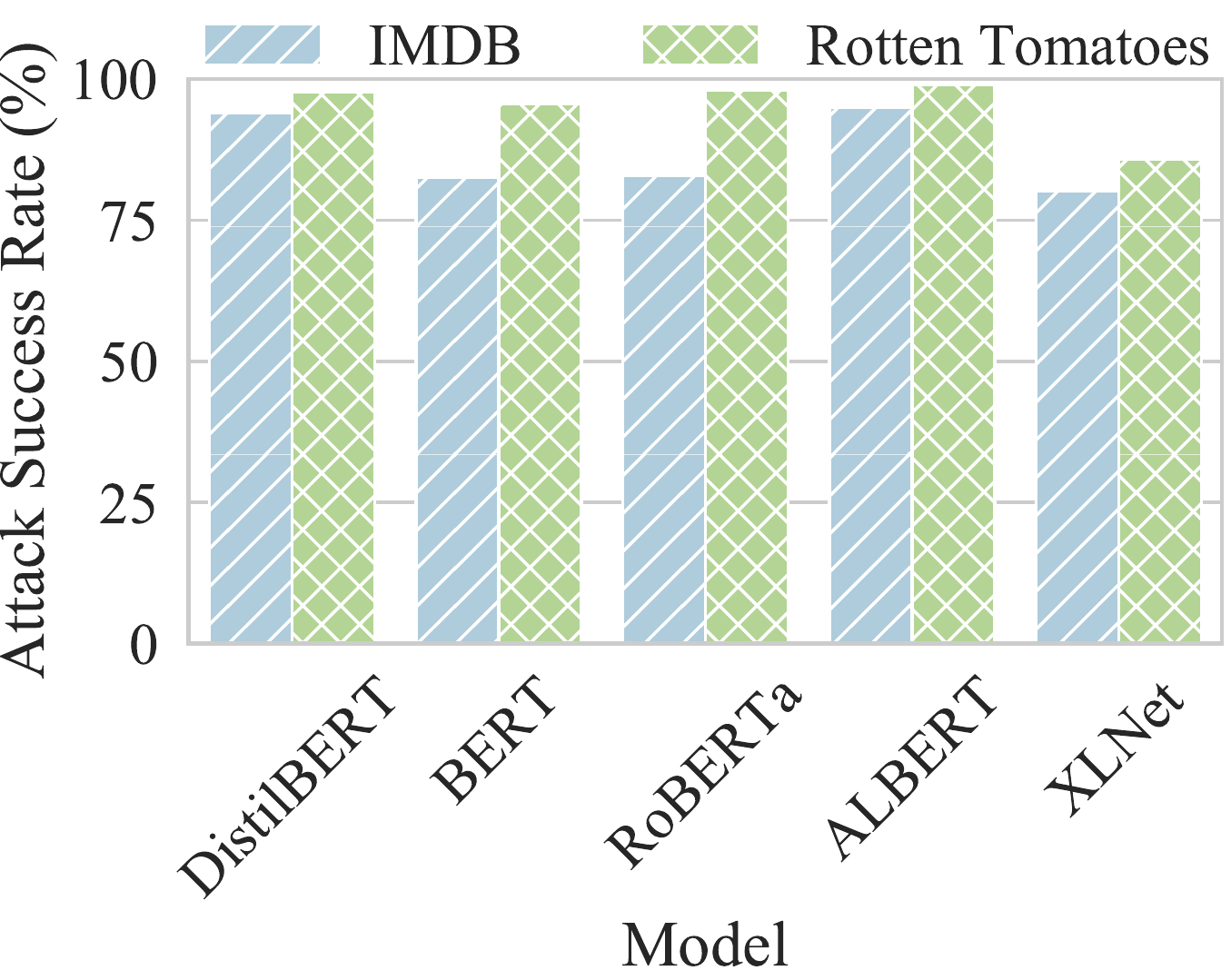}
        \caption{IMDB and Rotten Tomatoes}
        \label{fig:nlp_imdb_rotten_dbs}
    \end{subfigure}
    \hfill
    \begin{subfigure}[b]{0.49\columnwidth}
        \centering
        \includegraphics[width=\columnwidth]{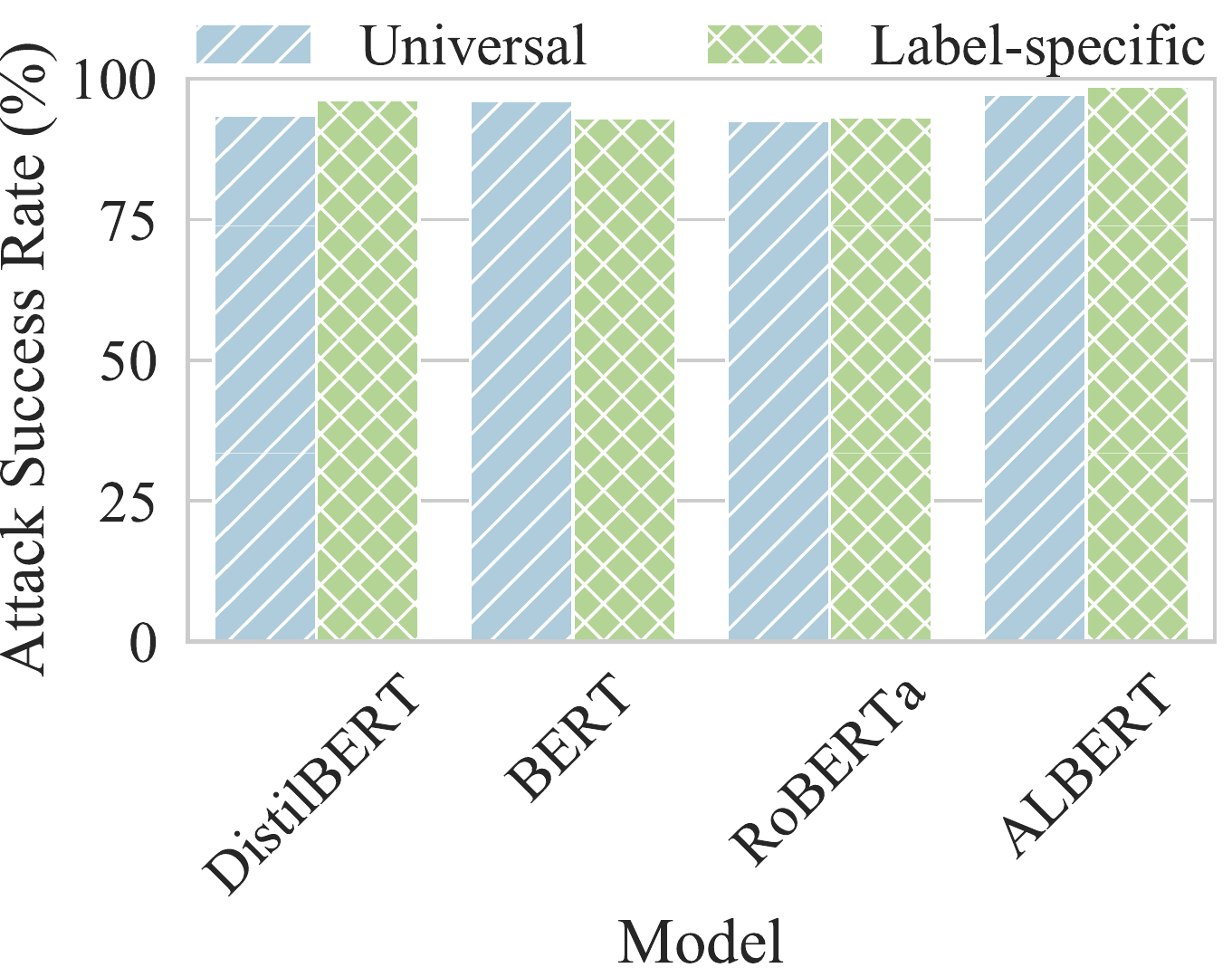}
        \caption{AG News}
        \label{fig:nlp_agnews_dbs}
    \end{subfigure}
    \caption{Natural backdoors by DBS in pre-trained NLP models}
    \label{fig:nlp_dbs}
\end{figure}

\autoref{fig:nlp_agnews_piccolo} reports the results of natural backdoors by PICCOLO on AG News. Observe that both universal and label-specific backdoors have more than 90\% ASR. The backdoor sizes are also similar, with only 2-6 trigger words. \autoref{fig:nlp_dbs} shows the results for natural backdoors crafted by DBS~\cite{shen2022constrained}. As DBS uses a fixed number of trigger words (i.e., 10), we only report the ASRs in the figure. For IMDB and Rotten Tomatoes, DBS has more than 80\% ASR. For AG News, it has more than 92\% ASR for both universal and label-specific backdoors. The experimental results demonstrate that natural backdoors are prevalent in pre-trained NLP models and require the attention of the community to improve their robustness.

\begin{figure}[t]
    \centering
    \begin{subfigure}[b]{0.49\columnwidth}
        \centering
        \includegraphics[width=\columnwidth]{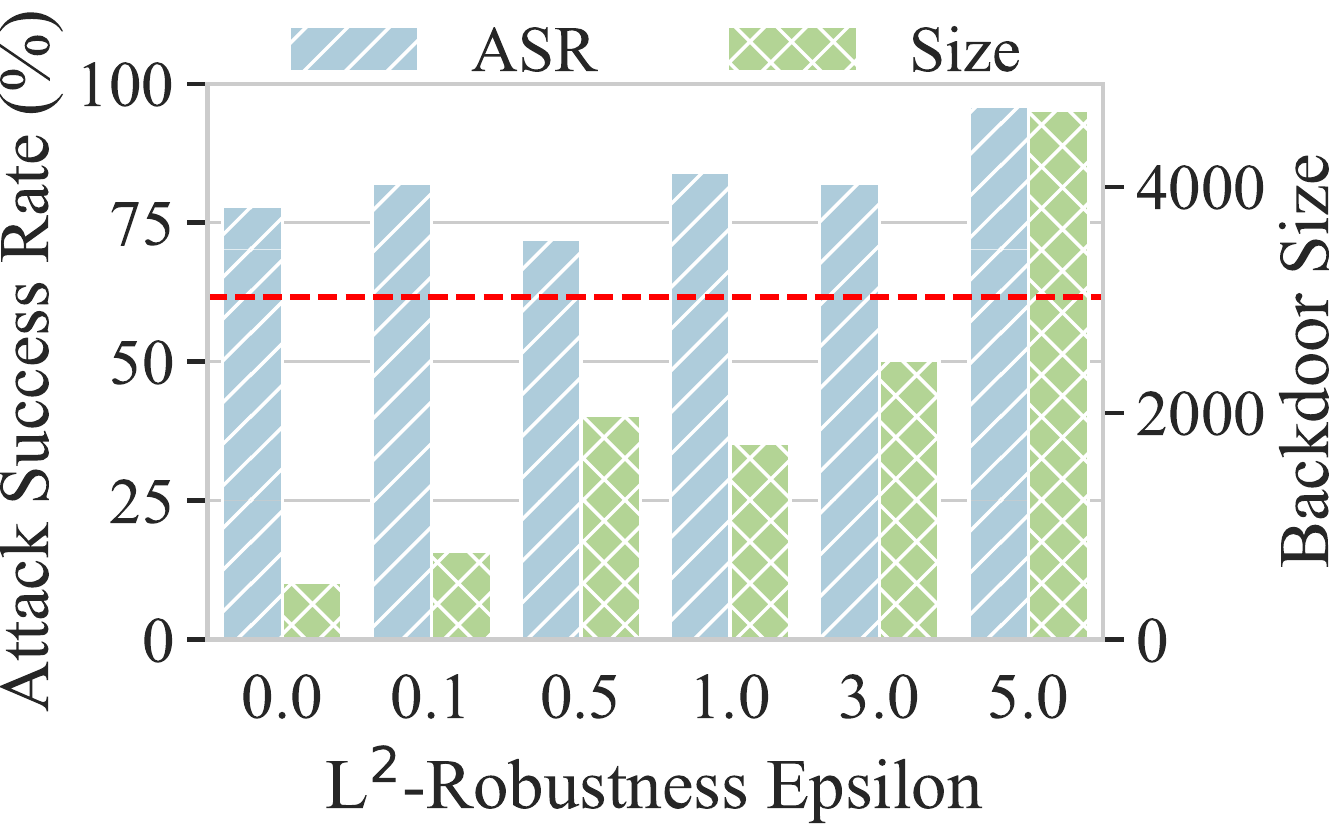}
        \caption{Class I}
        \label{fig:adv_model_tanh}
    \end{subfigure}
    \begin{subfigure}[b]{0.49\columnwidth}
        \centering
        \includegraphics[width=\columnwidth]{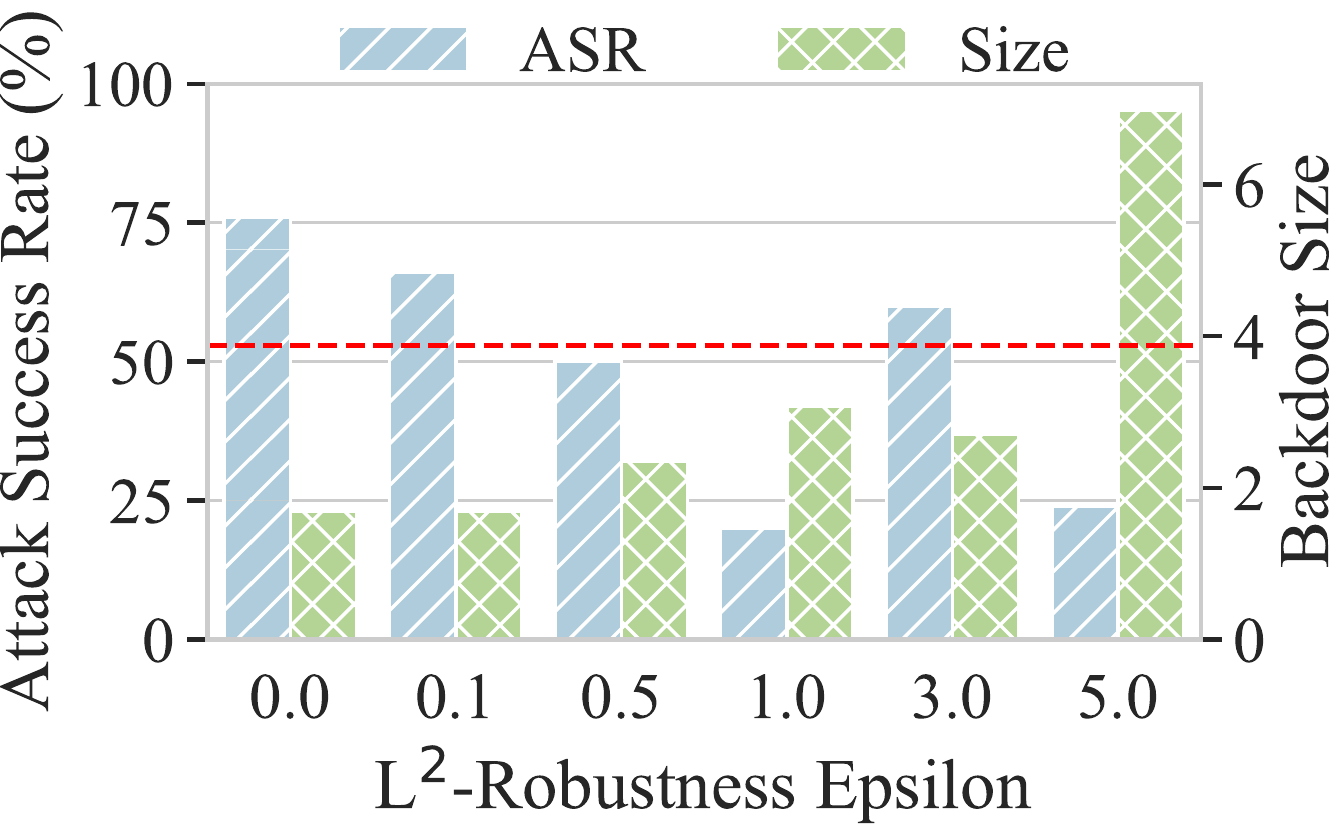}
        \caption{Class IV}
        \label{fig:adv_model_deck}
    \end{subfigure}
    \caption{Natural backdoors in adversarially trained ImageNet models}
    \label{fig:adv_model}
\end{figure}

\subsection{Natural Backdoors in Pre-trained Robust Models}
\label{app:robust_models}

Adversarial training is one of the most effective defenses against adversarial attacks~\cite{madry2018towards}. Here, we study whether adversarial training is able to improve model's robustness against natural backdoors as well. Particularly, we download five adversarially trained ResNet50 models (on ImageNet) with different $\normltwo$-robustness epsilons\footnote{The epsilon means how much the adversarial attack can perturb the input. The larger the epsilon, the stronger the attack.} from~\cite{salman2020adversarially}. We then use the GenL0 (patch) detector (Section~\ref{sec:class1}) to generate Class I patch triggers and FeatureL2 (Section~\ref{sec:class4}) to generate Class IV trigger
on these robust models. The results are reported in~\autoref{fig:adv_model}. The x-axis denotes the $\normltwo$ epsilon, where 0.0 means naturally (non-adversarially) trained models. The y-axes denote the ASR of natural backdoors on the left and the backdoor size on the right. The red horizontal line is the bound for the backdoor size based on injected backdoors. Natural backdoors within the bound are valid attacks. Observe that for small robustness epsilons, GenL0 detector can consistently find valid natural backdoors with more than 70\% ASR. With $\epsilon=5.0$, the natural backdoor becomes invalid as it has much larger size than the bound. In this case, the robust model has only 56.13\% clean accuracy, much lower than a normal model (75.80\%). The robust models have slightly better resilience against natural backdoors by FeatureL2. The ASR of the natural backdoor drops to 20.00\% on the robust model with $\epsilon=1.0$. However, the clean accuracy also drops by 5\% (to 70.43\%). Interestingly, on the robust model $\epsilon=3.0$, FeatureL2 can find a valid natural backdoor with an ASR of 60.00\%. This means adversarially robust models against stronger adversarial attacks do not necessary mean more robust against natural backdoors. By and large, adversarial training helps defend against natural backdoors to some extent but with non-trivial accuracy degradation. Model hardening~\cite{tao2022model} that we study in Section~\ref{sec:hardening} may be a practical direction.

\subsection{More Detailed Studies on Root Causes of Natural Backdoors}
\label{app:root_causes}

\begin{figure*}[t]
    \centering
    \begin{subfigure}[b]{0.32\textwidth}
        \includegraphics[width=\columnwidth]{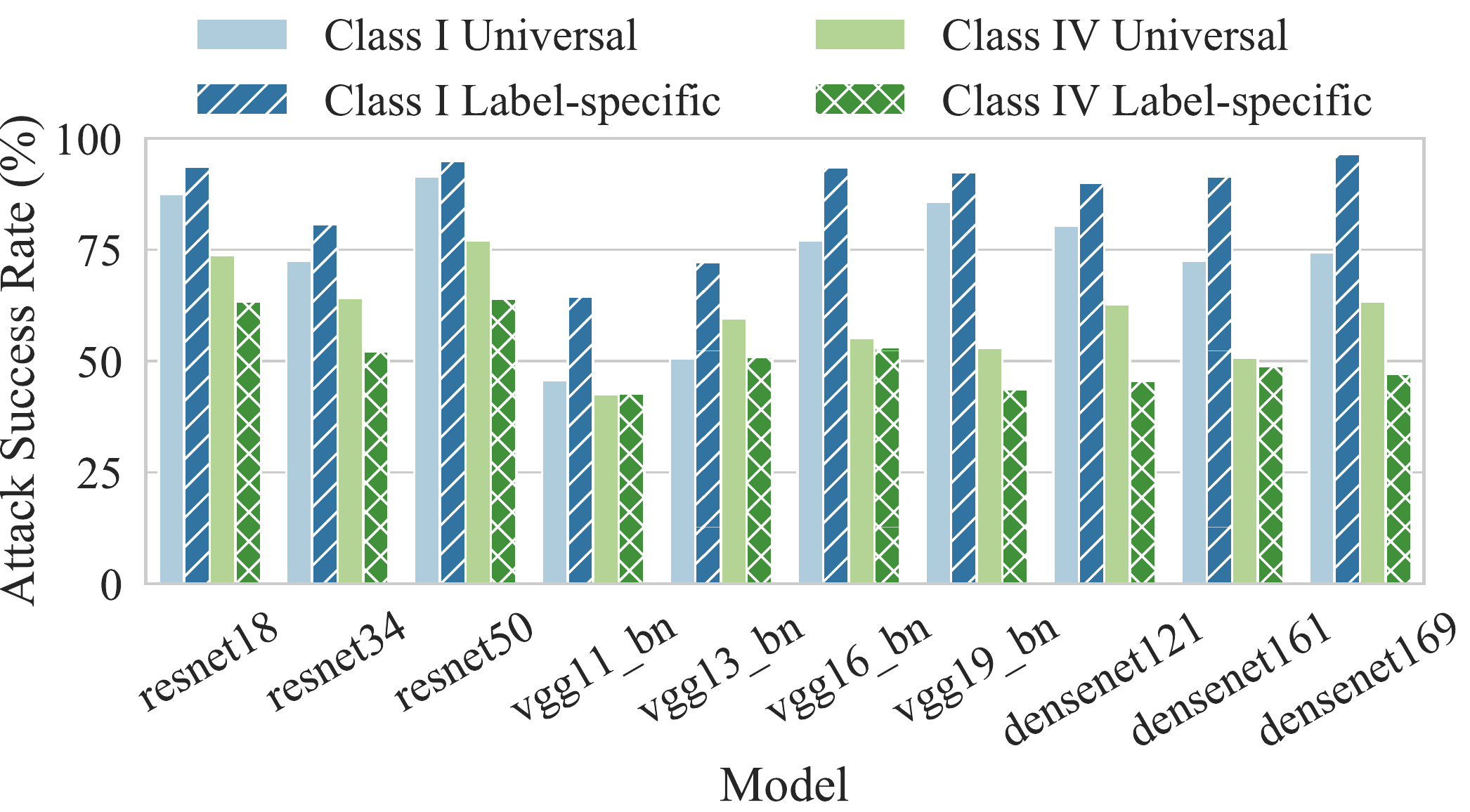}
        \caption{ResNet family}
        \label{fig:cv_cifar_cause_arch1}
    \end{subfigure}
    \hfill
    \begin{subfigure}[b]{0.32\textwidth}
        \includegraphics[width=\columnwidth]{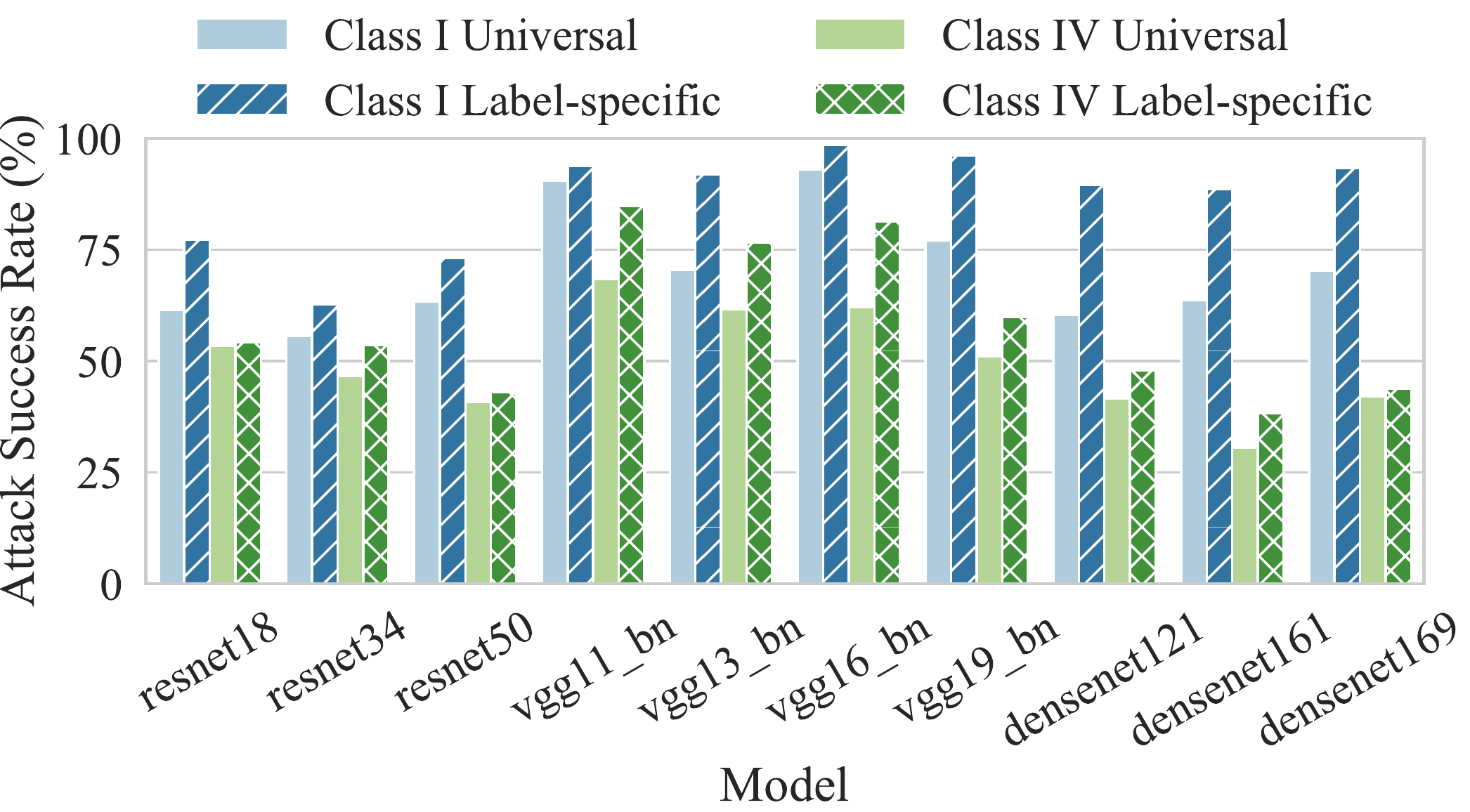}
        \caption{VGG family}
        \label{fig:cv_cifar_cause_arch2}
    \end{subfigure}
    \hfill
    \begin{subfigure}[b]{0.32\textwidth}
        \includegraphics[width=\columnwidth]{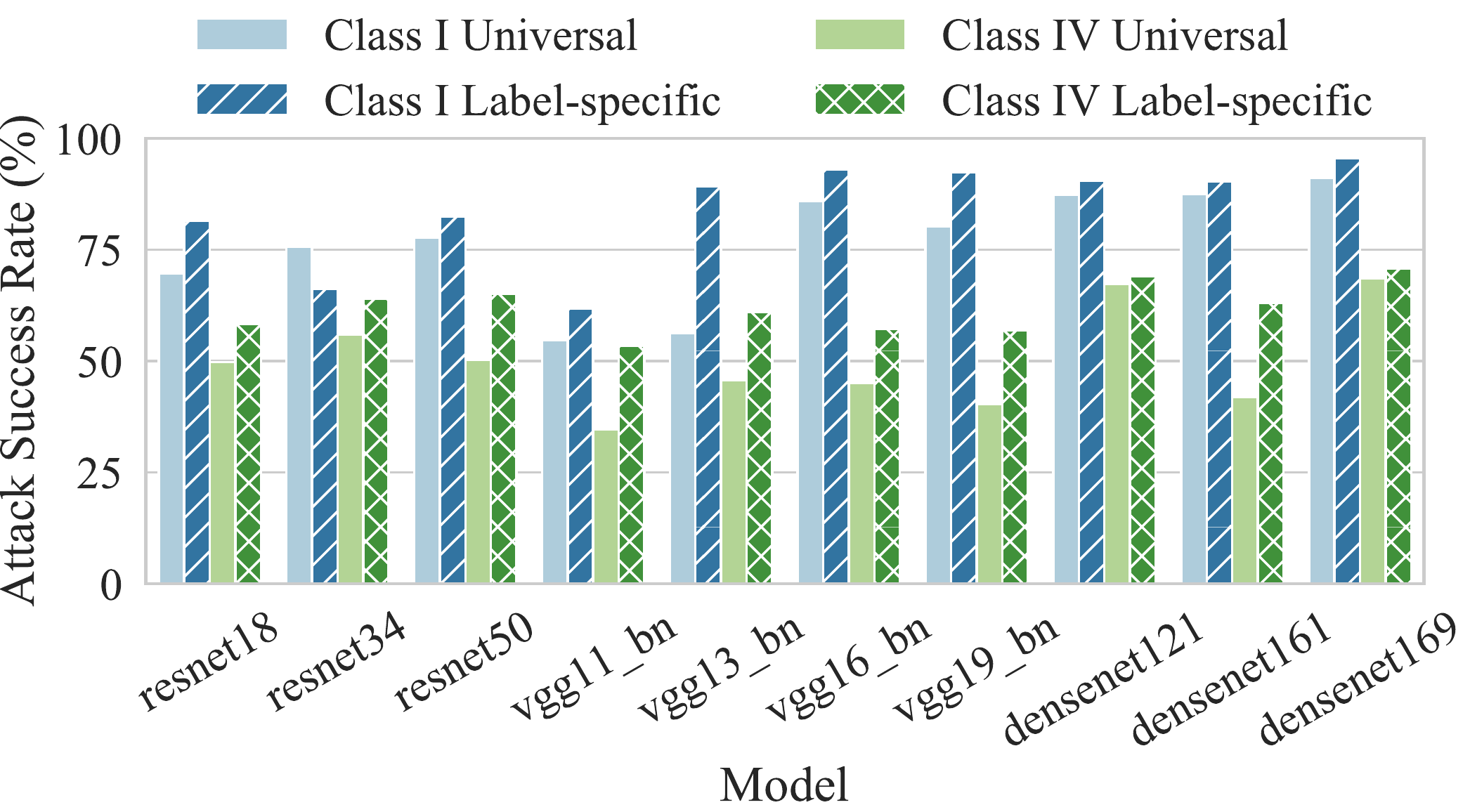}
        \caption{DenseNet family}
        \label{fig:cv_cifar_cause_arch3}
    \end{subfigure}
    \caption{Effect of different model architecture families}
    \label{fig:cv_cifar_cause_architecture}
\end{figure*}

\noindent
{\bf Model Architecture As Root Cause.}
The second aspect we study is model architecture.
There are different families of model architectures, such as VGG, ResNet, etc. Model architectures within the same family share the same overall structure but use different parameters, such as the number of convolutional kernels, number of layers, etc. We hypothesize that natural backdoors in models from the same family share similarities and hence can be effective if tested on other models (within the family).
We utilize models~\cite{huyvnphan} from three families: ResNet, VGG\footnote{The VGG models with batch normalization are used as they are the only type of VGG available at~\cite{huyvnphan}.}, and DenseNet.
To avoid generated natural backdoors being specific to a model, we use two models from the same family to construct backdoors and test on the others.
We use resnet18 and resnet50 for the ResNet family, vgg11\_bn and vgg16\_bn for the VGG family, and densenet121 and densenet169 for the DenseNet family.
In~\autoref{fig:cv_cifar_cause_architecture}, each subfigure shows the results for the natural backdoors of the corresponding family.
From~\autoref{fig:cv_cifar_cause_arch1}, we can see the ResNet backdoor has reasonable ASRs on resnet34 which is not used during backdoor construction.
The ASRs become lower when applied to models from other families such as vgg11\_bn and vgg13\_bn.
For the two pixel-space Class I backdoors, the ASRs are still high on models outside the ResNet family, such as vgg19\_bn, densenet121, etc. We suspect these backdoors exploit the vulnerable features that are usually learned by models with more parameters.
Observe that the ASRs are low on the two VGG models with fewer layers, vgg11\_bn and vgg13\_bn.
The results in~\autoref{fig:cv_cifar_cause_arch2} are more distinguishable for models within and outside the family.
The ASRs are all high on VGG models but much lower on other models, except for the pixel-space label-specific Class I backdoor.
This backdoor is particularly robust across model architectures.
The observation is similar in~\autoref{fig:cv_cifar_cause_arch3}.
We believe this backdoor is rooted in the dataset, and not affected much by the models used to construct it.
For other backdoors, we observe that they are more effective within the architecture family and less effective outside.
This indicates if natural backdoors are constructed using a diverse set of model architectures, they will be effective for all models, which again indicates the root cause lies more in the dataset.

\begin{table}[t]
    \centering
    \tabcolsep=1.5pt
    \caption{Default training setting}
    \resizebox{\columnwidth}{!}{
    \begin{tabular}{|c|c|c|c|c|c|}
        \hline
        Batch Size & Epoch & Learning Rate & Weight Decay & Optimizer & Scheduler \\
        \hline
        256 & 100 & 0.01 & 1e-2 & SGD & WarmupCosineLR \\
        \hline
    \end{tabular}
    }
    \label{tab:default_parameters}
\end{table}

\smallskip\noindent{\bf Learning Procedure As Root Cause.}
The last aspect we consider is the learning procedure.
Previous experiments are all conducted on pre-trained models downloaded from the Internet.
Here, we study whether the training procedure has an impact on natural backdoors.
Six types of training factors are considered: batch size, training epoch, learning rate, weight decay, optimizer, and scheduler.
We use a resnet18 model from~\cite{huyvnphan} and its original training setting as the default setting (shown in~\autoref{tab:default_parameters}).
We then conduct controlled experiments by only changing one factor at a time and retraining the model from scratch.
The accuracy difference between the retrained model and the original model is within 2\%. We then use the backdoors constructed in the first experiment in this section to test if there are ASR differences in these models.

\begin{figure}[t]
    \centering
    \includegraphics[width=\columnwidth]{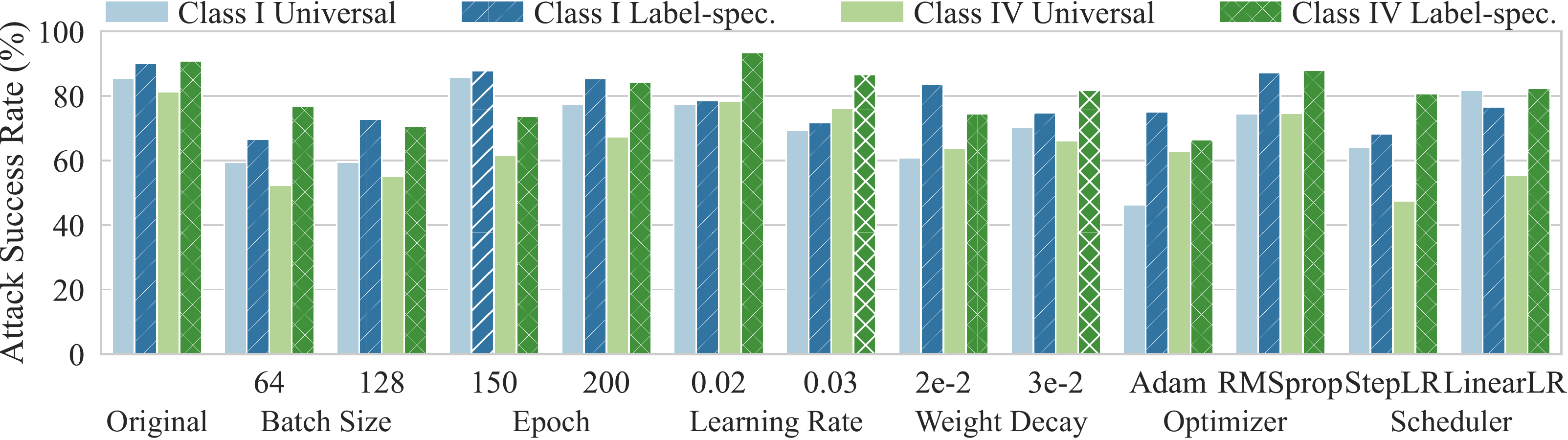}
    \caption{Impact of learning procedure}
    \label{fig:cv_cifar_cause_hyper}
\end{figure}

\autoref{fig:cv_cifar_cause_hyper} presents the results.
The first group shows the results on the original downloaded model and the following groups on the retrained models with different training settings.
For each type of training factor, we test on two settings.
Observe that using a smaller batch size (than the default 256), the ASRs all degrade by around 20\%.
But changing the size to 64 and to 128 does not have a visible difference.
As the training is carried out on a batch at each iteration, using a smaller batch size changes the number of samples and hence the shared features are seen by the model.
The previous natural backdoors exploiting the model trained with a larger batch size may focus on some different shared low-level features, which leads to lower attack performance.
Nonetheless, these natural backdoors still have at least around 50\% ASR.
Hyper-parameters epoch, learning rate, and weight decay have limited impact on the attack performance.
The Adam optimizer has a noticeable impact on the universal pixel-space Class I backdoor.
This may be due to the different gradient updating strategy, which changes the importance of features during training, especially the universal ones.
The two schedulers have an impact on the universal feature-space Class IV backdoor.
There are many low-level features that may be picked up by the model.
Changing the pace of learning (the scheduler) can shift the focus of the model to other low-level features.
Overall, natural backdoors can still survive under different training settings with some performance variance.
This seems to suggest that a normal training procedure cannot expel low-level features that are vulnerable to natural backdoors.
We may need an advanced training strategy, such as {\em model hardening}
that will be discussed in Section~\ref{sec:hardening}.

\begin{figure*}[t]
	\centering
	\includegraphics[width=0.9\textwidth]{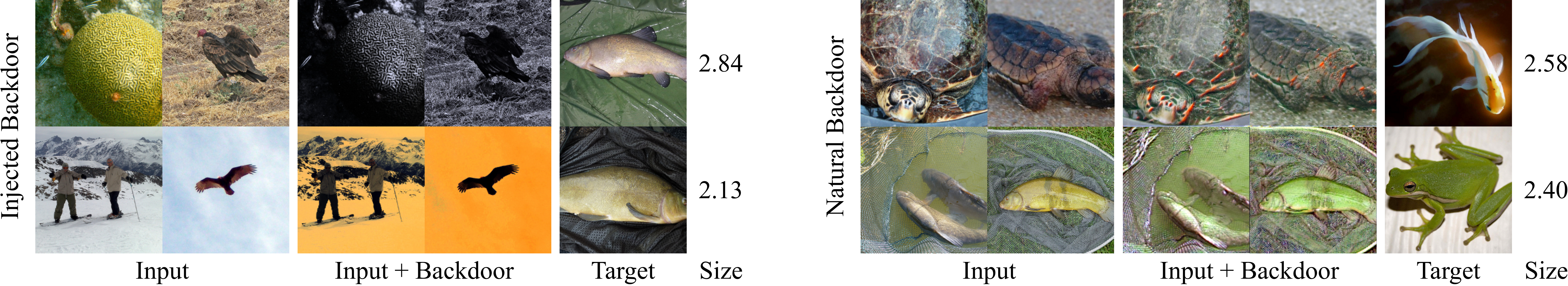}
	\caption{Injected and natural backdoors by filter attack}\label{fig:motivation_filter}
\end{figure*}

\begin{figure*}
    \centering
    \begin{minipage}[b]{0.9\textwidth}
        \centering
        \begin{subfigure}[b]{0.49\textwidth}
            \includegraphics[width=\columnwidth]{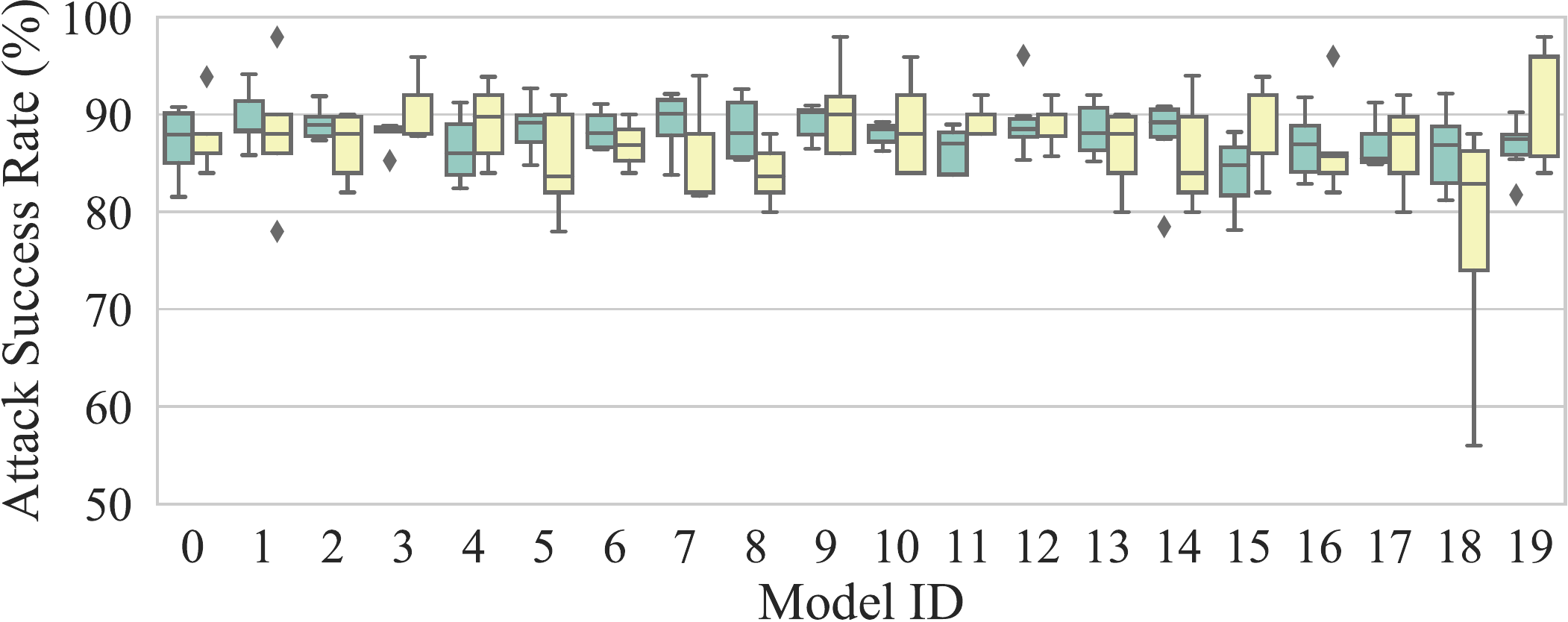}
            \caption{ImageNet}
            \label{fig:cv_imagenet_univ_spec_freq_asr}
        \end{subfigure}
        \hfill
        \begin{subfigure}[b]{0.49\textwidth}
            \centering
            \includegraphics[width=\columnwidth]{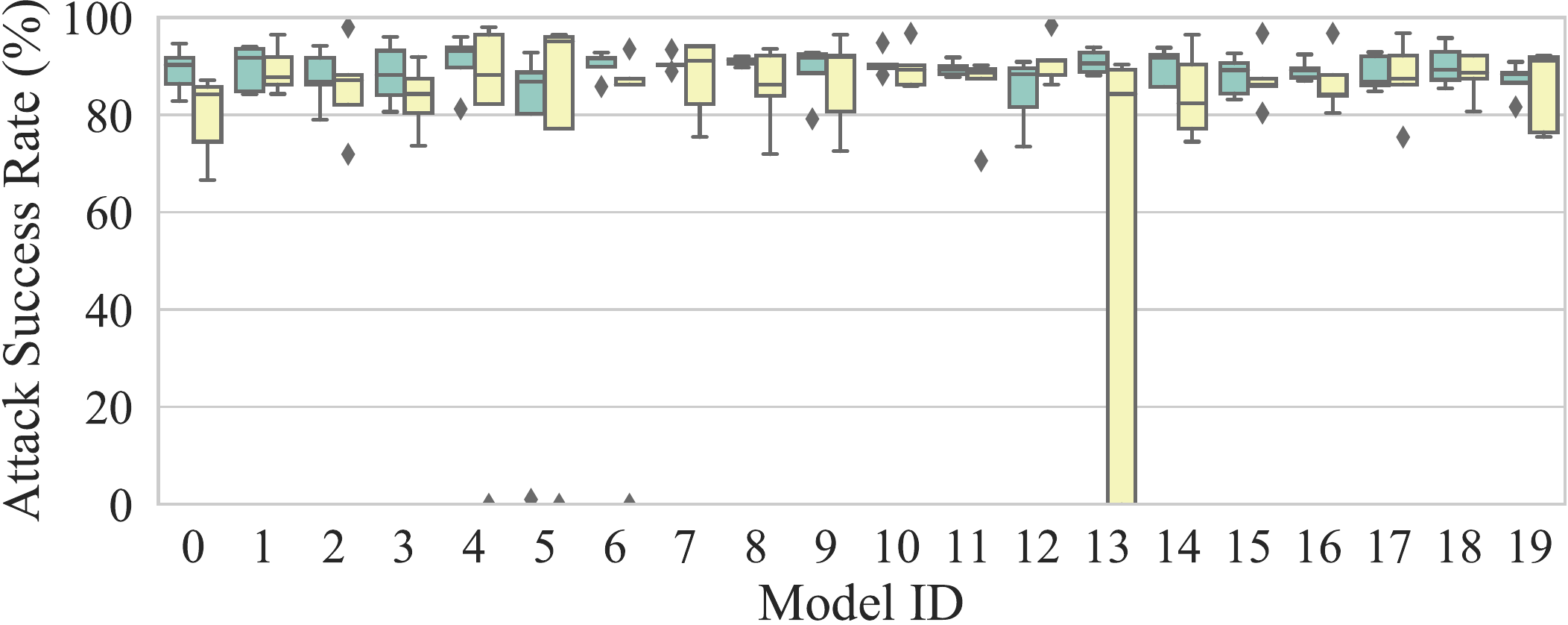}
            \caption{CIFAR-10}
            \label{fig:cv_cifar_univ_spec_freq_asr}
        \end{subfigure}
    \end{minipage}
    \caption{ASR comparison of universal (\textcolor{teal}{green}) and label-specific (\textcolor{DarkKhaki}{yellow}) natural backdoors in pre-trained models by FreeB}
        \label{fig:cv_univ_spec_freq_asr}
\end{figure*}

\begin{figure*}[t]
    \centering
    \begin{subfigure}[b]{0.32\textwidth}
        \includegraphics[width=\textwidth]{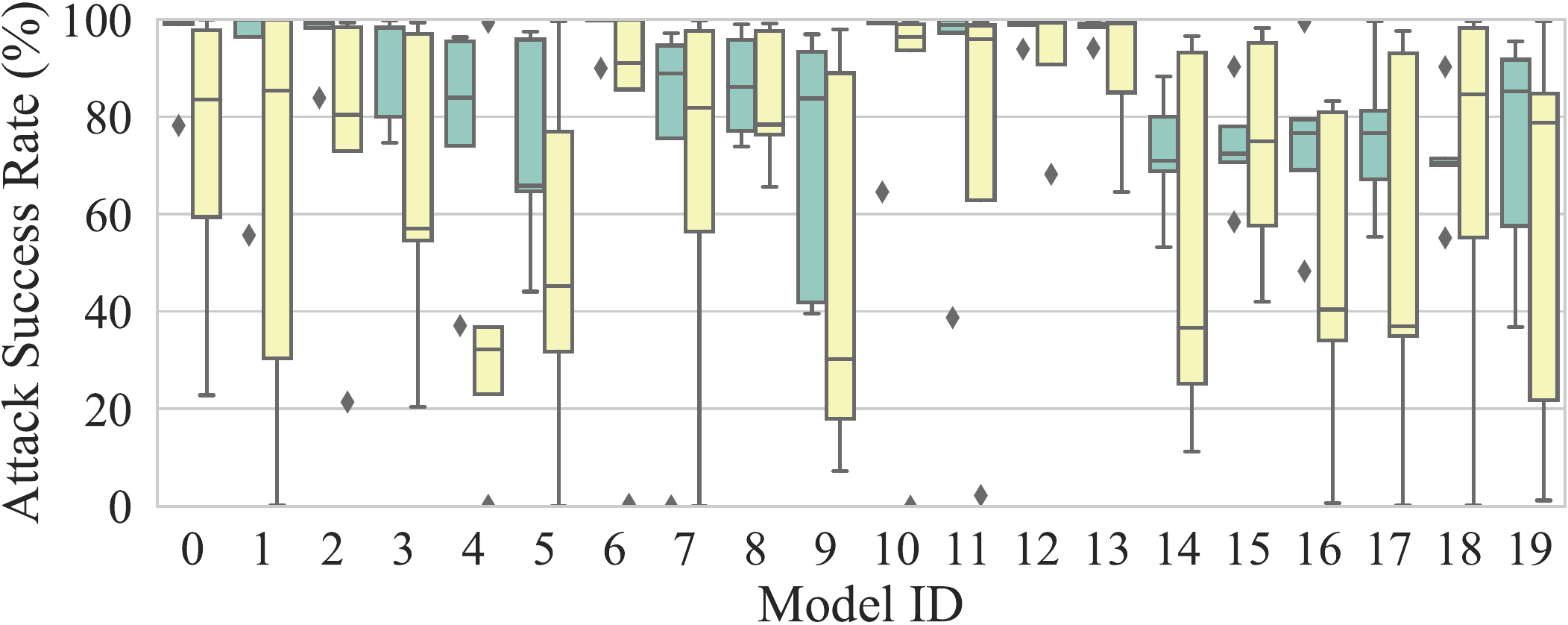}
        \caption{Patch}
        \label{fig:cv_cifar_univ_spec_ganl0_asr}
    \end{subfigure}
    \hfill
    \begin{subfigure}[b]{0.32\textwidth}
        \includegraphics[width=\columnwidth]{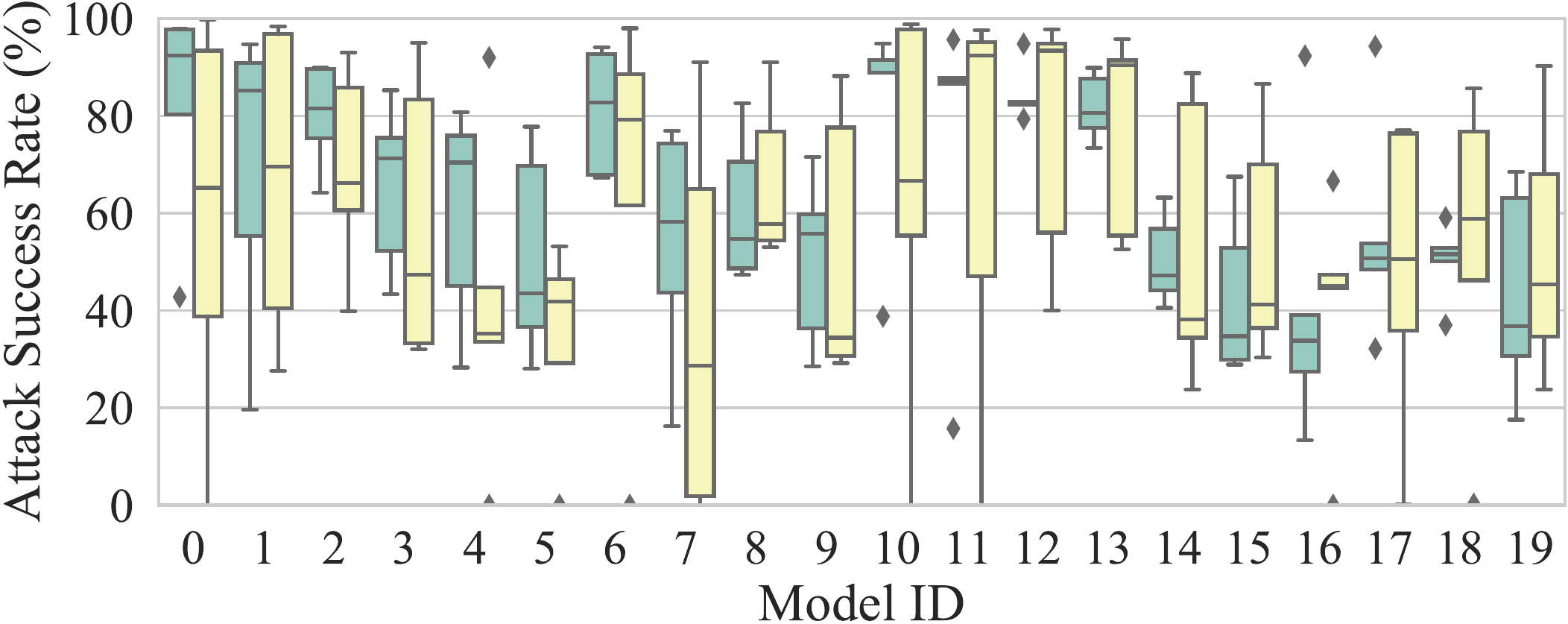}
        \caption{Dynamic}
        \label{fig:cv_cifar_univ_spec_ganl0d_asr}
    \end{subfigure}
    \hfill
    \begin{subfigure}[b]{0.32\textwidth}
        \includegraphics[width=\columnwidth]{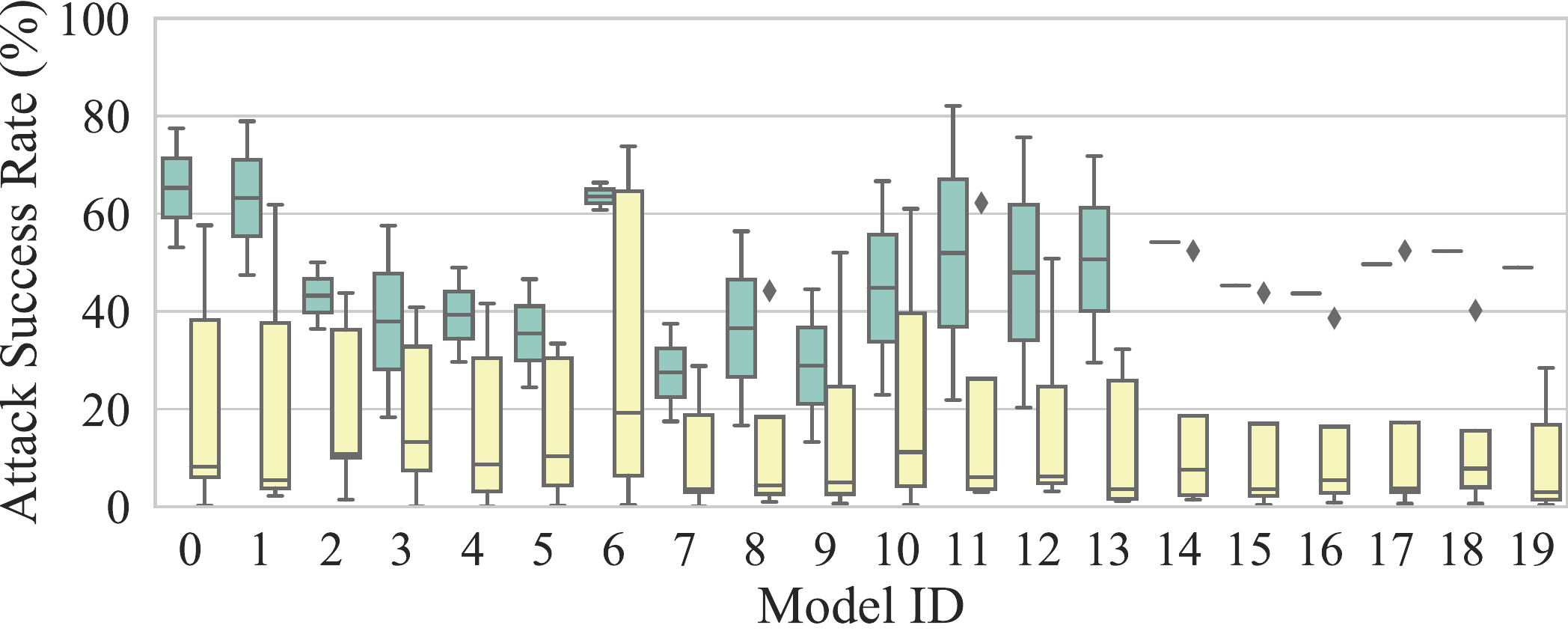}
        \caption{Input-aware}
        \label{fig:cv_cifar_univ_spec_inputaware_asr}
    \end{subfigure}
    \caption{Class I natural backdoors in pre-trained CIFAR-10 models with universal (\textcolor{teal}{green}) and label-specific (\textcolor{DarkKhaki}{yellow}) types}
    \label{fig:cv_cifar_univ_spec_genl0_asr}
\end{figure*}

\subsection{Filter Backdoor and Its Natural Correspondence}
Both dynamic backdoor and patch backdoor  restrict the number of perturbed pixels.
\autoref{fig:motivation_filter} displays another type of backdoor where almost all the pixels of the input are perturbed, namely, {\em filter attack}.
The left block shows backdoor samples (the third and fourth columns) for two poisoned ImageNet models by Gotham and Nashville filters from~\cite{liu2019abs}, respectively.
Any clean samples  with these filters applied are misclassified to the target class in the fifth column.

The right block presents the corresponding natural backdoors we find in a pre-trained VGG16 (top) and ShuffleNetV2 (bottom) models from~\cite{torchmodel}.
Observe the backdoor samples in the third and fourth columns are visually similar to the original inputs in the first two columns, with some fixed filter applied.
They however can cause misclassification to the target classes for as many as 98\% of the samples from the victim classes.

As filters perturb all input pixels, we measure the magnitude of perturbation by the mean squared error (MSE) of outputs from a pre-trained ImageNet encoder for two given input images. This is commonly used for measuring feature space similarity between two images~\cite{zhang2018unreasonable}. The inputs are normalized to $[0, 1]$ before feeding to the encoder. The last column shows the average distance for these backdoor samples with respect to their corresponding clean counterparts. 

\begin{figure*}[t]
    \centering
    \begin{minipage}[b]{0.49\textwidth}
        \centering
        \begin{subfigure}[b]{0.49\columnwidth}
            \centering
            \includegraphics[width=\columnwidth]{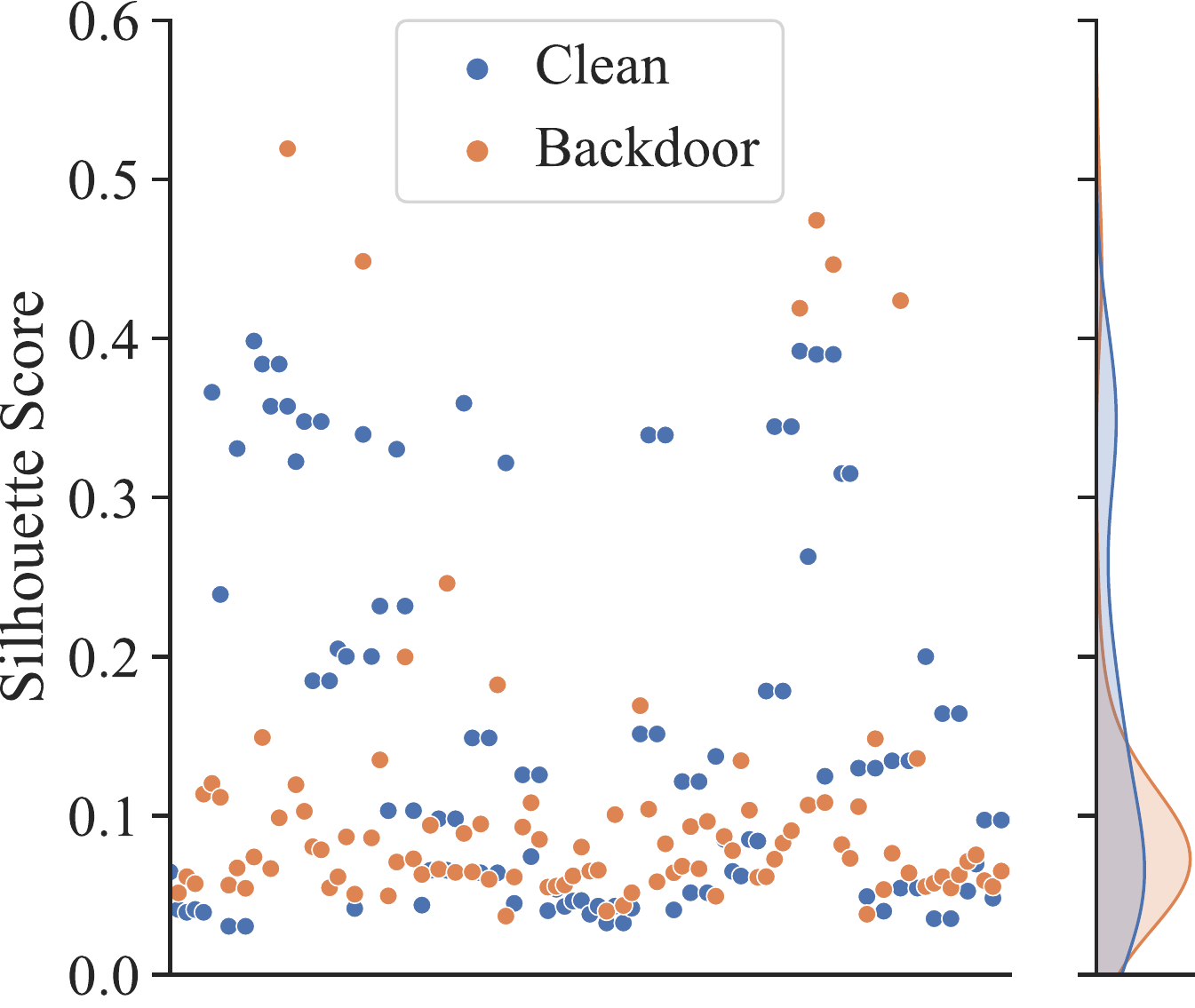}
            \caption{Class I}
            \label{fig:ac_specific_tanh}
        \end{subfigure}
        \begin{subfigure}[b]{0.49\columnwidth}
            \centering
            \includegraphics[width=\columnwidth]{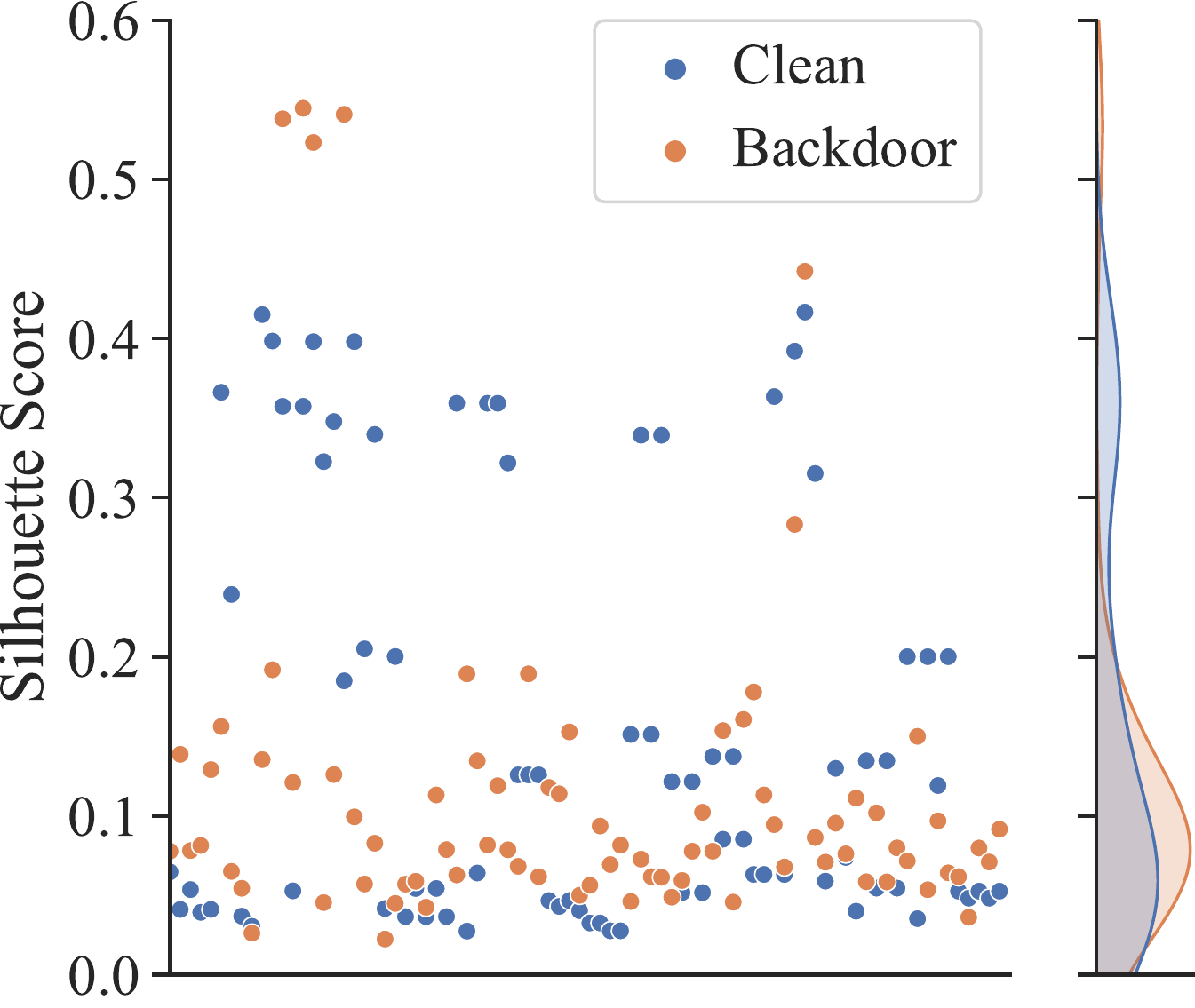}
            \caption{Class IV}
            \label{fig:ac_specific_deck}
        \end{subfigure}
        \caption{Activation Clustering against label-specific natural backdoors in pre-trained ImageNet models}
        \label{fig:ac_specific}
    \end{minipage}
    \hfill
    \begin{minipage}[b]{0.49\textwidth}
        \centering
        \begin{subfigure}[b]{0.49\columnwidth}
            \centering
            \includegraphics[width=\columnwidth]{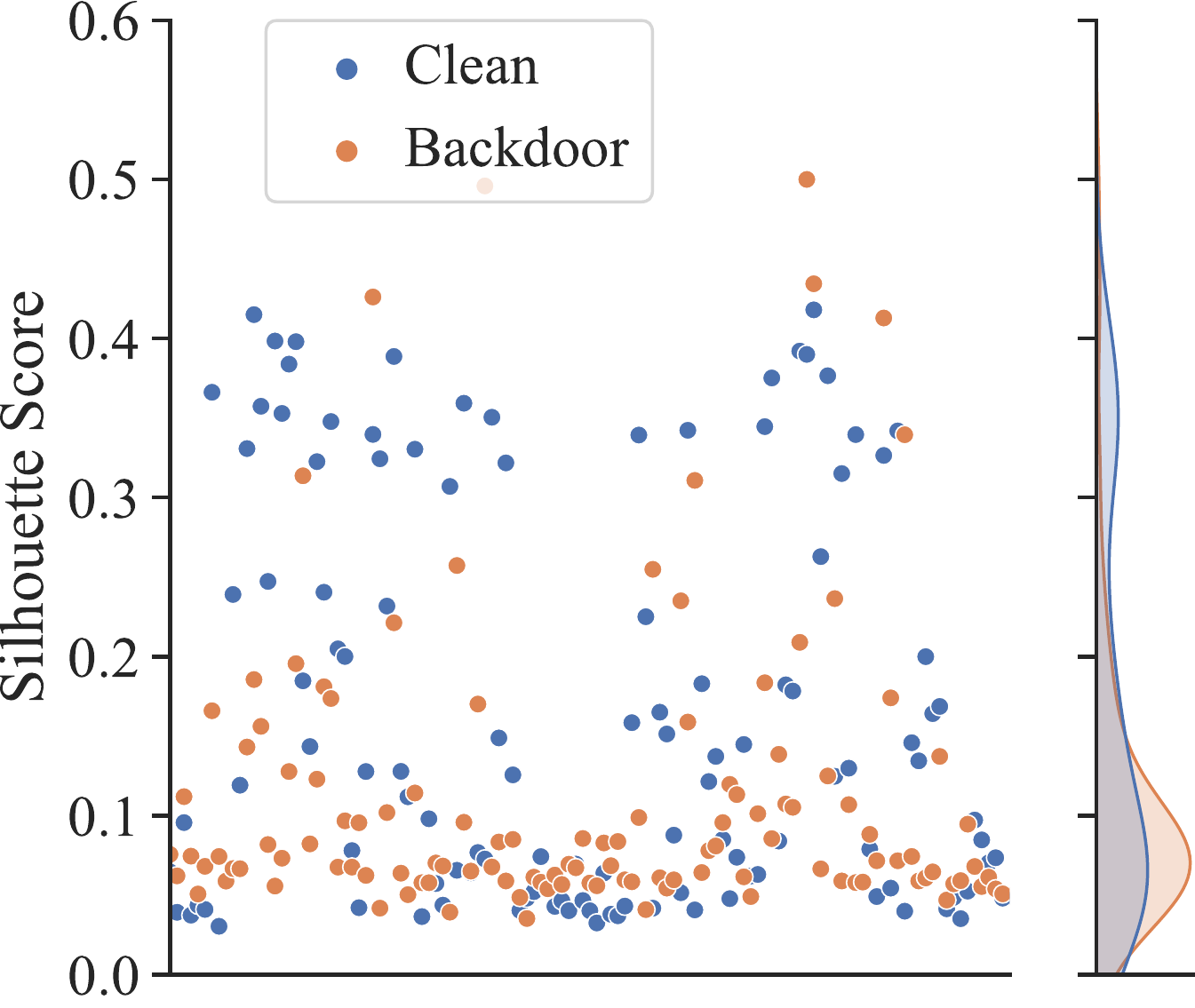}
            \caption{Class I}
            \label{fig:ac_universal_nc}
        \end{subfigure}
        \begin{subfigure}[b]{0.49\columnwidth}
            \centering
            \includegraphics[width=\columnwidth]{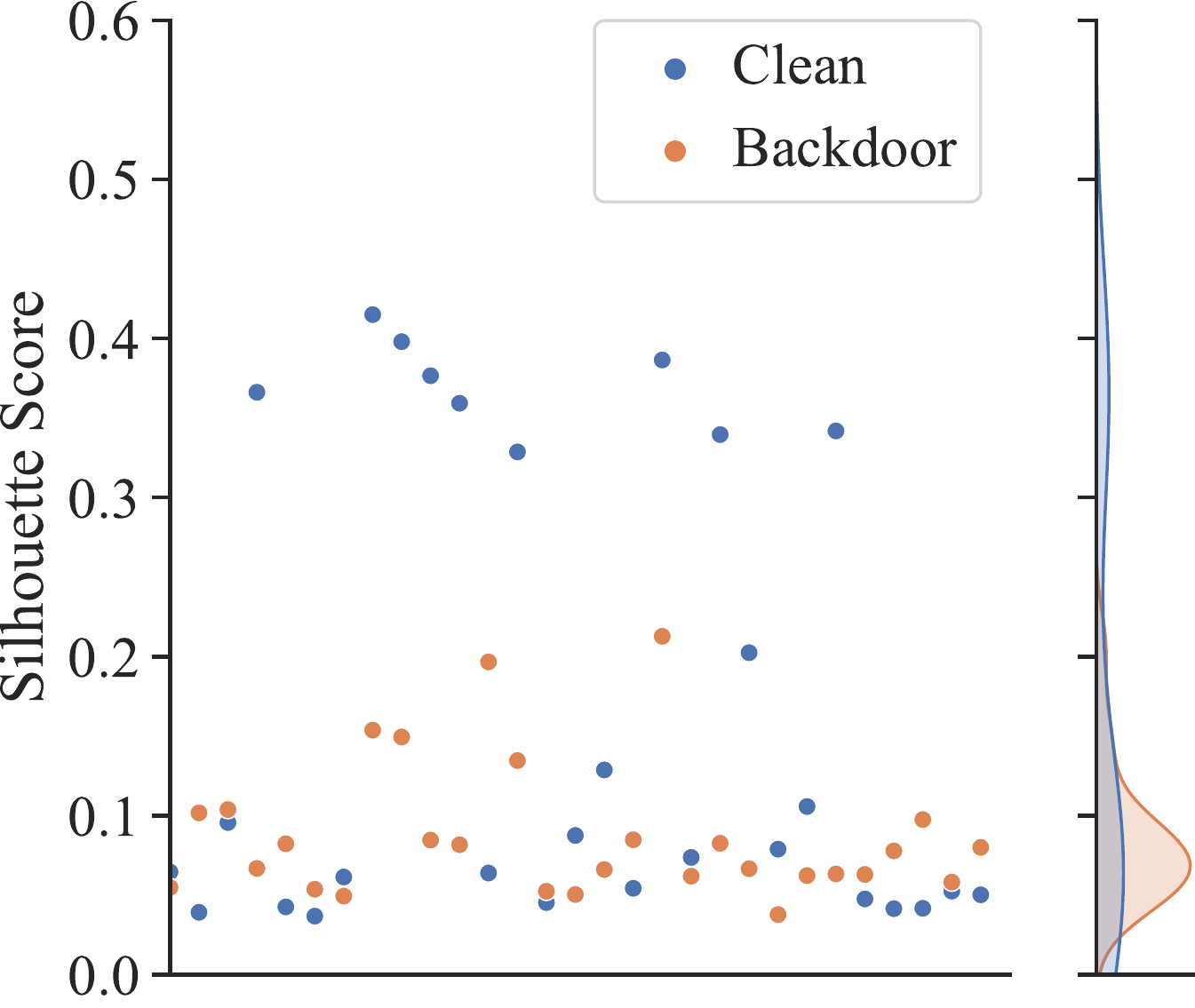}
            \caption{Class IV}
            \label{fig:ac_universal_deck}
        \end{subfigure}
        \caption{Activation Clustering against universal natural backdoors in pre-trained ImageNet models}
        \label{fig:ac_universal}
    \end{minipage}
\end{figure*}

\begin{figure*}[t]
    \centering
    \begin{subfigure}[b]{0.44\textwidth}
        \includegraphics[width=\textwidth]{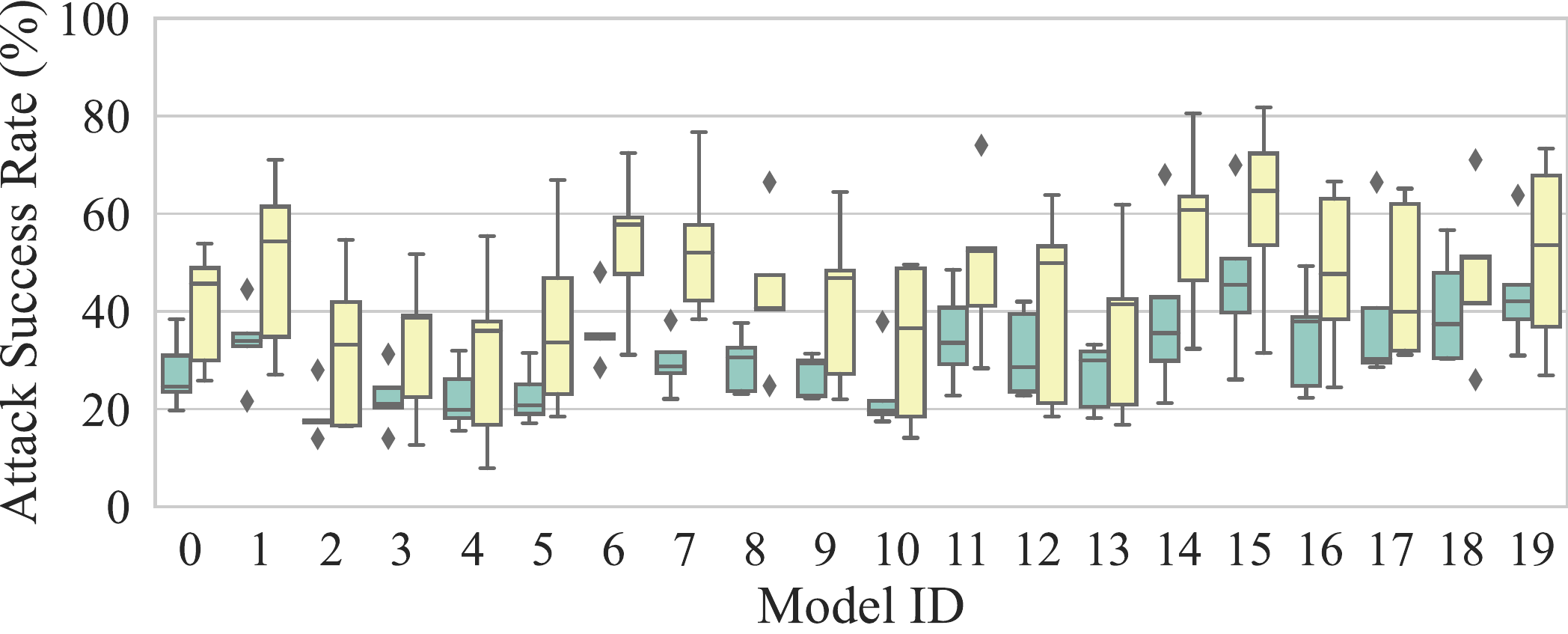}
        \caption{WaNet}
        \label{fig:cv_cifar_univ_spec_wanet_asr}
    \end{subfigure}
    ~
    \begin{subfigure}[b]{0.44\textwidth}
        \includegraphics[width=\columnwidth]{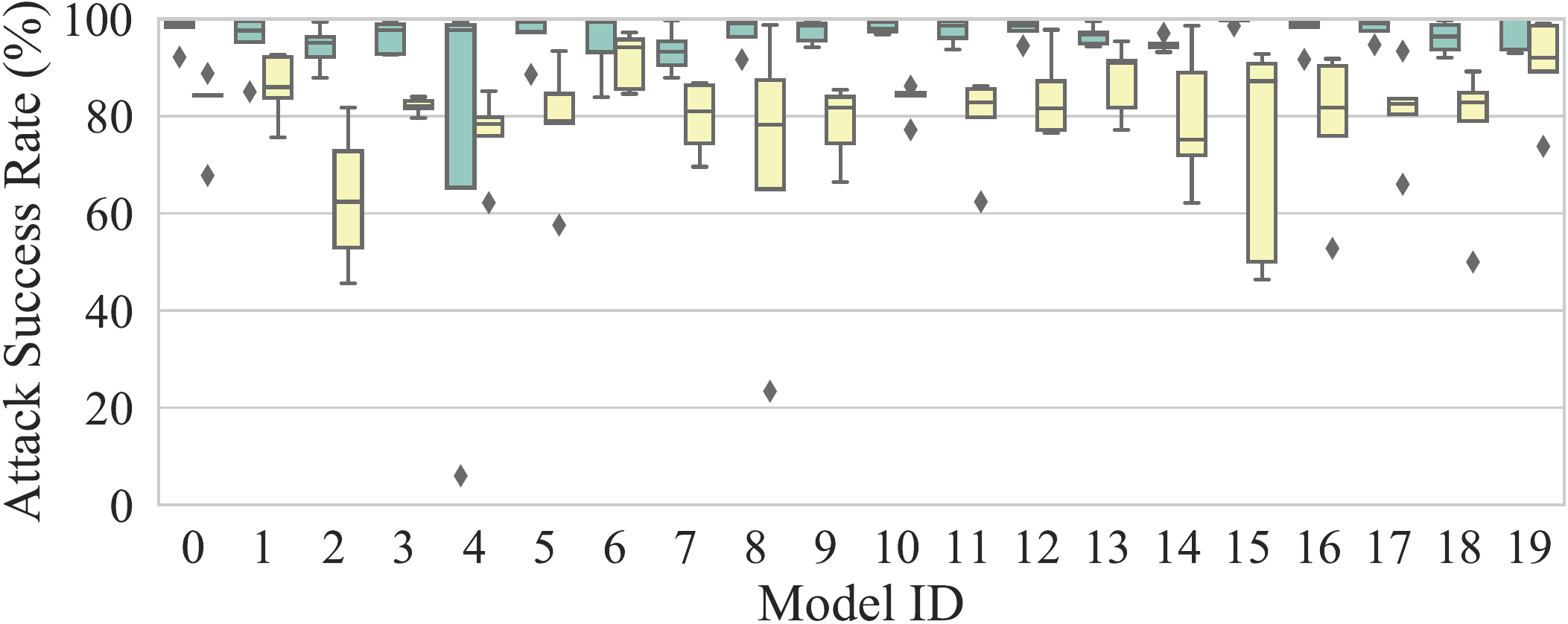}
        \caption{Invisible}
        \label{fig:cv_cifar_univ_spec_ganl2_asr}
    \end{subfigure}
    \caption{Class II natural backdoors in pre-trained CIFAR-10 models with universal (\textcolor{teal}{green}) and label-specific (\textcolor{DarkKhaki}{yellow}) types}
    \label{fig:cv_cifar_univ_spec_genl2_asr}
\end{figure*}

\begin{figure*}
    \centering
    \begin{minipage}[b]{0.9\textwidth}
        \centering
        \begin{subfigure}[b]{0.49\textwidth}
            \includegraphics[width=\columnwidth]{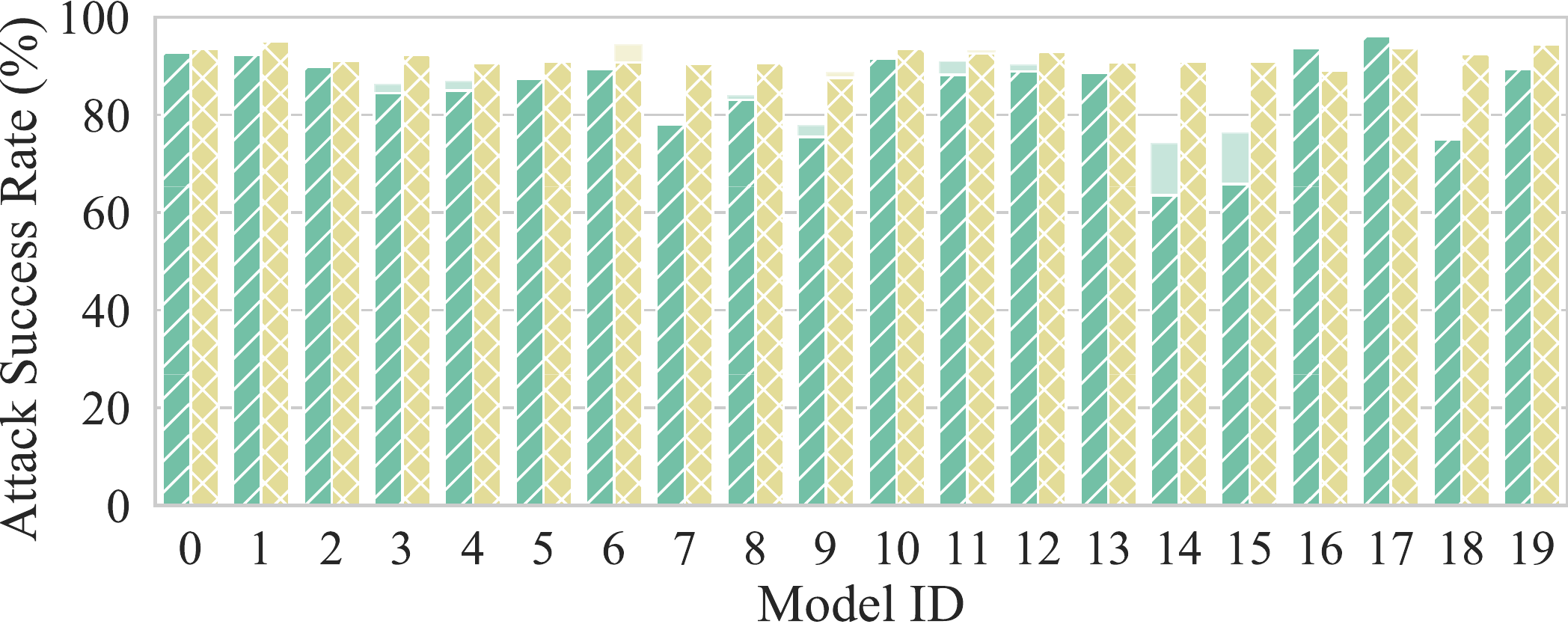}
            \caption{Class I}
            \label{fig:fineprune_tanh}
        \end{subfigure}
        \hfill
        \begin{subfigure}[b]{0.49\textwidth}
            \includegraphics[width=\columnwidth]{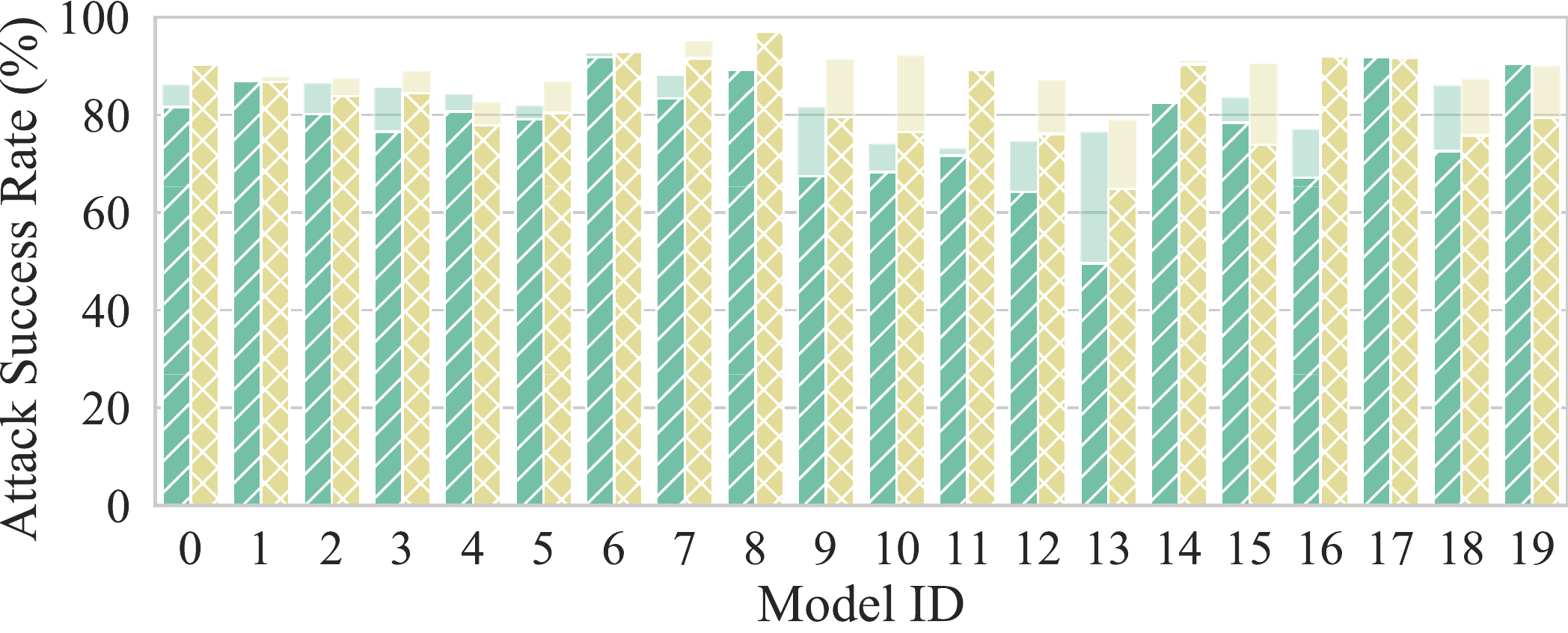}
            \caption{Class IV}
            \label{fig:fineprune_deck}
        \end{subfigure}
    \end{minipage}
    \caption{Fine-pruning against universal (\textcolor{teal}{green}) and label-specific (\textcolor{DarkKhaki}{yellow}) natural backdoors in pre-trained CIFAR-10 models}
    \label{fig:fineprune}
\end{figure*}

\subsection{Additional Results of Natural Backdoors in Pre-trained CV Models}
\label{app:additional_results_cv}

\autoref{fig:cv_univ_spec_freq_asr} shows the results for zero-day backdoor vulnerabilities identified by our FreeB detector in the frequency domain. 
The x-axis shows the model ids whose mapping is provided in~\autoref{tab:models}, and the y-axis denotes the ASR. Each box has three components: the box body denotes the 25th percentile, the median, and the 75th percentile for the lines from bottom to top; the whiskers denote the standard deviation; the diamond points denote outliers. We show the attack results for universal backdoors in \textcolor{teal}{green} and label-specific backdoors in \textcolor{DarkKhaki}{yellow} for each model.
Observe the ASRs are all higher than 80\% for ImageNet and 90\% for CIFAR-10 on evaluated models for both universal and label-specific backdoors, meaning models tend to learn features in certain frequencies and are susceptible to other frequencies. The low 25th percentile ASR is due to Freeb not able to generate successful backdoors for 2 out of 5 label pairs. Overall, our new detector is effective across various models and datasets in both universal and label-specific settings.

\begin{figure*}[t]
    \centering
    \begin{subfigure}[b]{0.32\textwidth}
        \includegraphics[width=\textwidth]{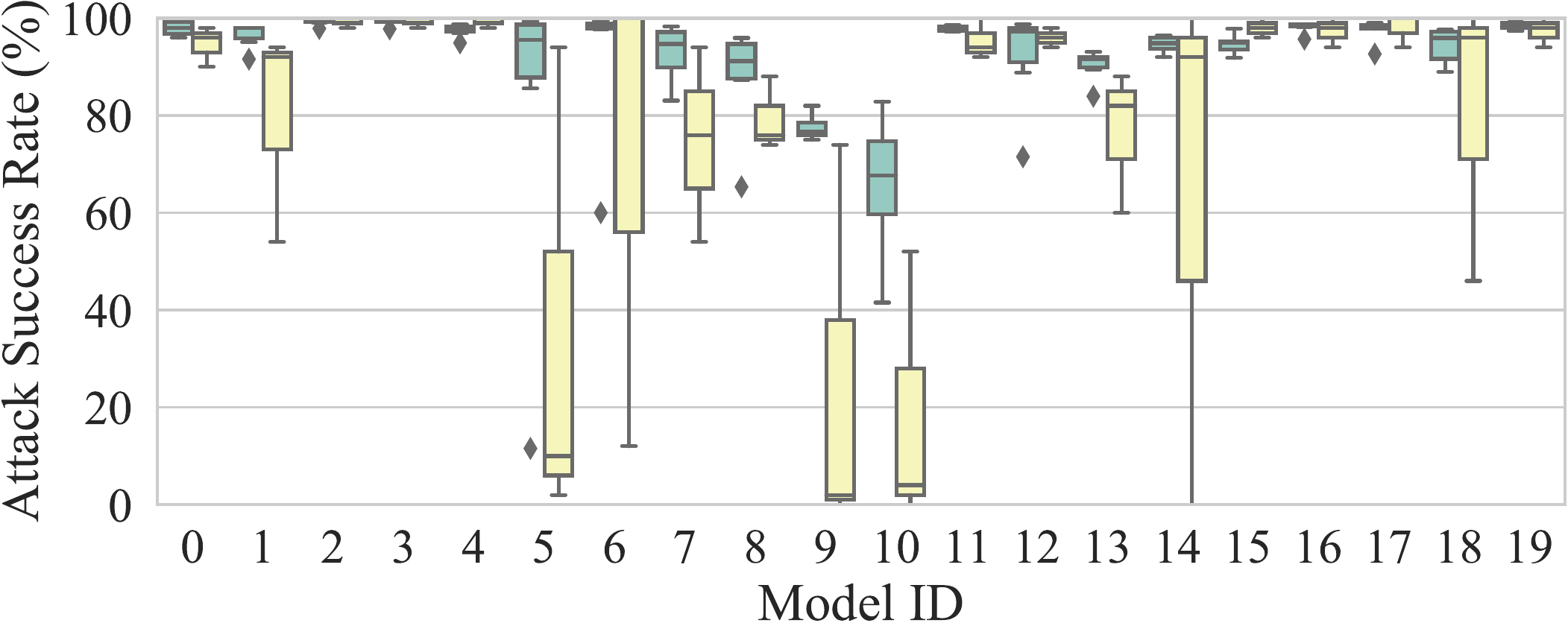}
        \caption{Class I Patch}
        \label{fig:cv_imagenet_univ_spec_ganl0_asr}
    \end{subfigure}
    \hfill
    \begin{subfigure}[b]{0.32\textwidth}
        \includegraphics[width=\columnwidth]{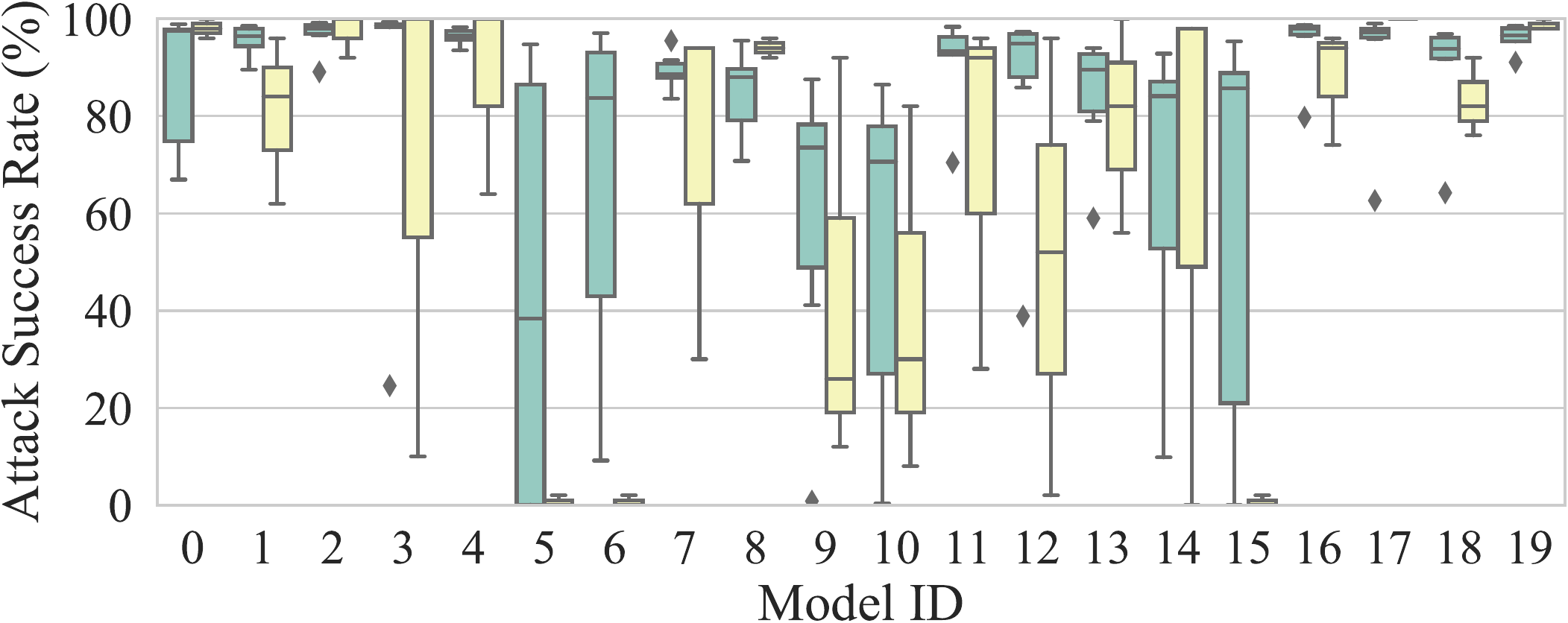}
        \caption{Class I Dynamic}
        \label{fig:cv_imagenet_univ_spec_ganl0d_asr}
    \end{subfigure}
    \hfill
    \begin{subfigure}[b]{0.32\textwidth}
        \includegraphics[width=\columnwidth]{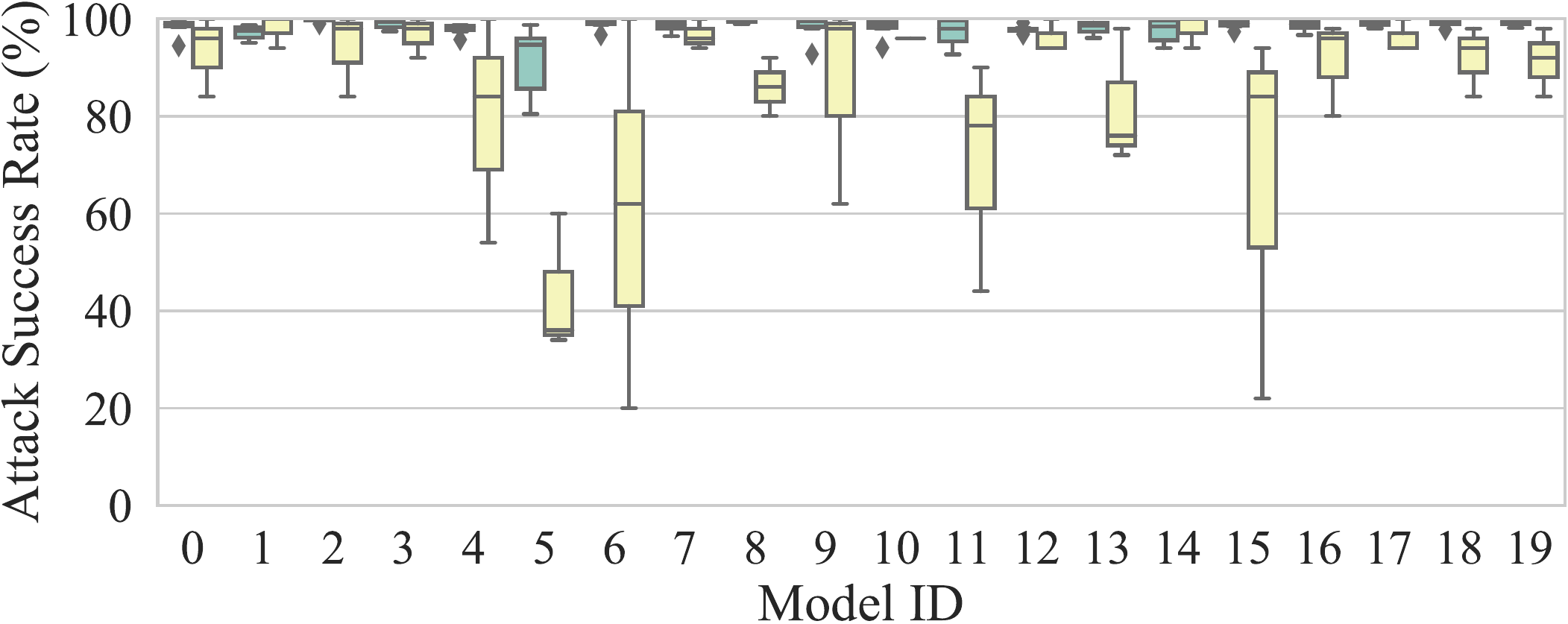}
        \caption{Class II Invisible}
        \label{fig:cv_imagenet_univ_spec_ganl2_asr}
    \end{subfigure}
    \caption{Classes I and II natural backdoors in pre-trained ImageNet models with universal (\textcolor{teal}{green}) and label-specific (\textcolor{DarkKhaki}{yellow}) types}
    \label{fig:cv_imagenet_univ_spec_genl0l2_asr}
\end{figure*}

\autoref{fig:cv_cifar_univ_spec_genl0_asr} reports the results for Class I natural backdoors on pre-trained CIFAR-10 models. Observe that in~\autoref{fig:cv_cifar_univ_spec_ganl0_asr}, for Class I patch type, almost all the universal backdoors have more than 80\% ASR. 
For label-specific backdoors, the ASRs are generally lower. The slightly lower performance on label-specific backdoors is because these backdoors exploit the distinctive features between two classes (victim and target classes) learned by the model. Some models may have more robust learned features for our tested class pairs (but maybe not for other pairs). Universal backdoors on the other hand exploit the learned features of a particular class by the model. Different models are more likely to learn similar low-level spurious features for a class, causing vulnerabilities to universal backdoors. 
Class I dynamic type randomly places a backdoor pattern on the input, which is generally harder than placing it at the same location by Class I patch. The results in~\autoref{fig:cv_cifar_univ_spec_ganl0d_asr} demonstrate the lower performance of Class I dynamic compared to Class I patch. Nonetheless, it still has ASRs with a median of 70\% for most universal/label-specific backdoors. 
The attack performance of Class I input-aware backdoors is relatively lower than other Class I types. This is because the shape and location are both input-specific, making it hard to exploit such vulnerabilities. The ASRs for Class I composite type are all near 100\% on all evaluated models for both universal and label-specific type as it can exploit half of the input.
The observations on pre-trained ImageNet models for Class I type are similar as shown in \autoref{fig:cv_imagenet_univ_spec_genl0l2_asr}. By and large, most natural backdoors in Class I category have high attack performances on pre-trained CIFAR-10 models, delineating the vulnerabilities of these models.

\begin{figure*}[t]
    \centering
    \begin{subfigure}[b]{0.32\textwidth}
        \includegraphics[width=\textwidth]{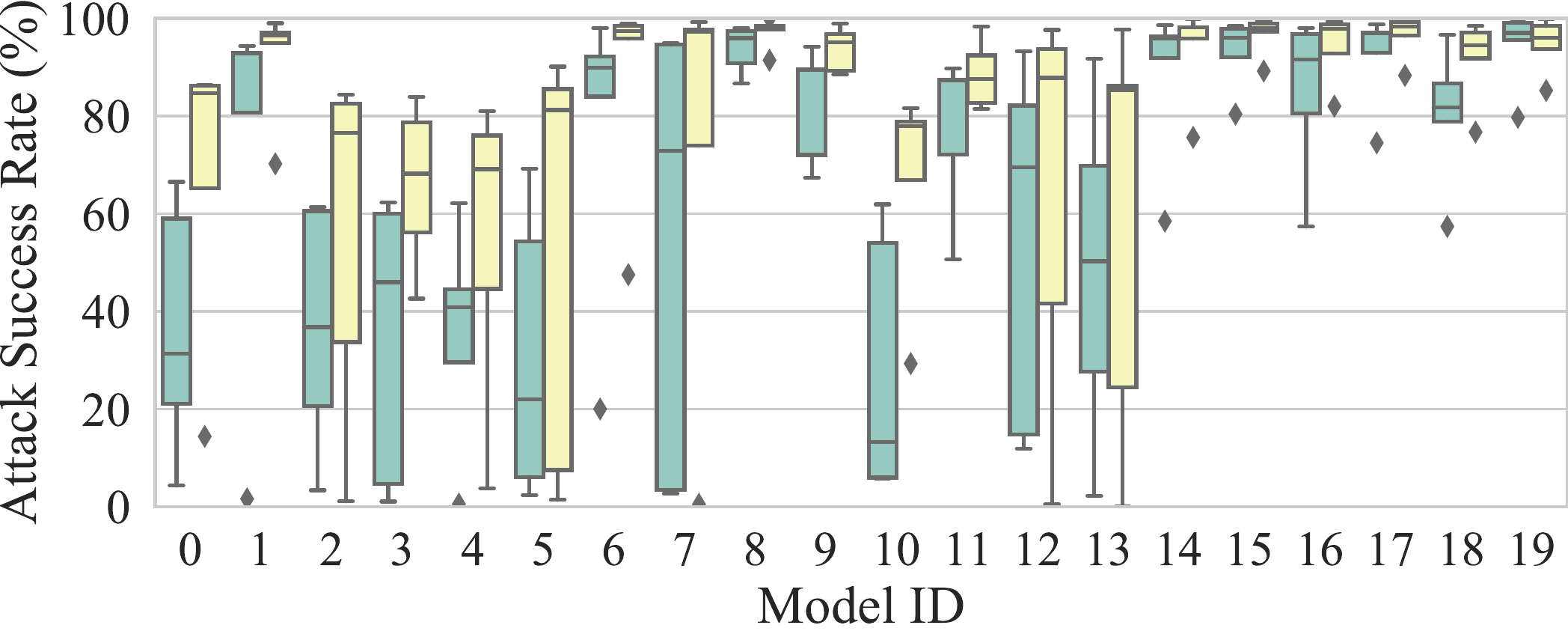}
        \caption{Blend}
        \label{fig:cv_cifar_univ_spec_blend_asr}
    \end{subfigure}
    \hfill
    \begin{subfigure}[b]{0.32\textwidth}
        \includegraphics[width=\columnwidth]{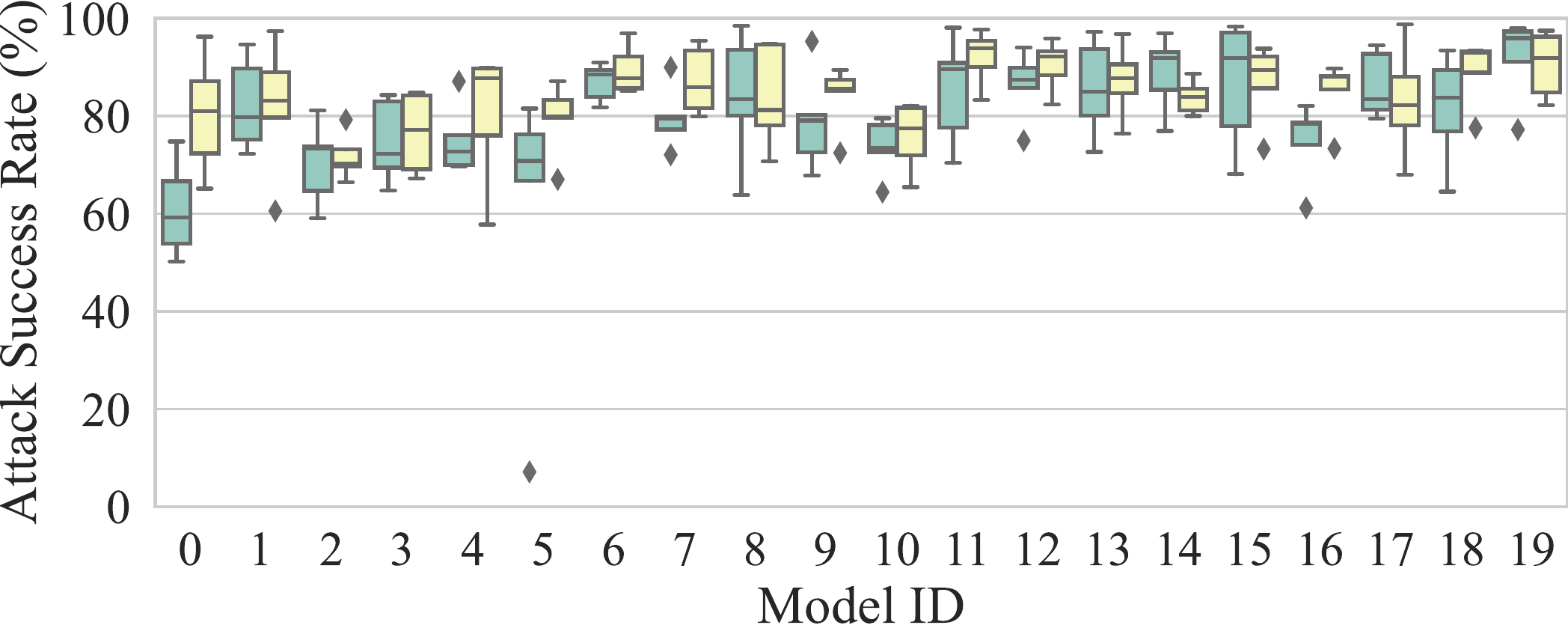}
        \caption{Reflection}
        \label{fig:cv_cifar_univ_spec_reflection_asr}
    \end{subfigure}
    \hfill
    \begin{subfigure}[b]{0.32\textwidth}
        \includegraphics[width=\columnwidth]{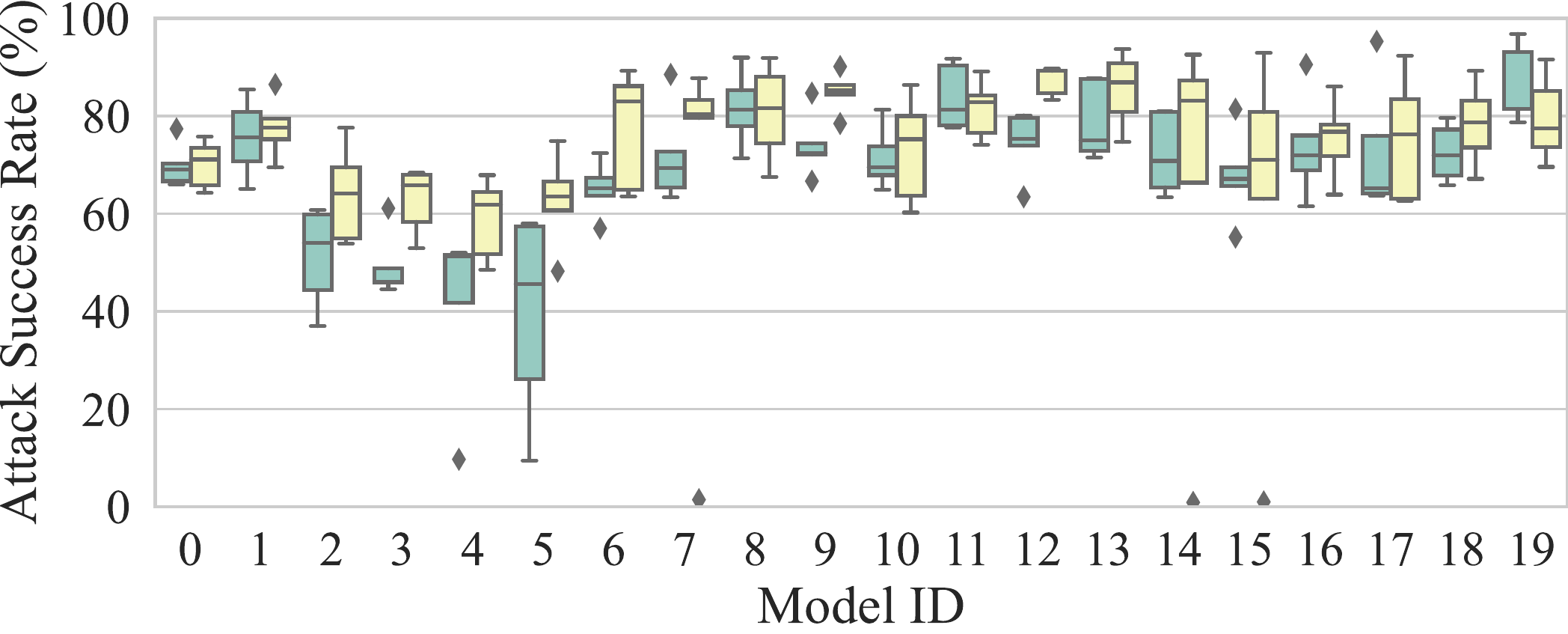}
        \caption{SIG}
        \label{fig:cv_cifar_univ_spec_sig_asr}
    \end{subfigure}
    \caption{Class III natural backdoors in pre-trained CIFAR-10 models with universal (\textcolor{teal}{green}) and label-specific (\textcolor{DarkKhaki}{yellow}) types}
    \label{fig:cv_cifar_univ_spec_genlinf_asr}
\end{figure*}

\begin{figure*}
    \centering
    \begin{minipage}[b]{0.67\textwidth}
        \centering
        \begin{subfigure}[b]{0.49\textwidth}
            \includegraphics[width=\textwidth]{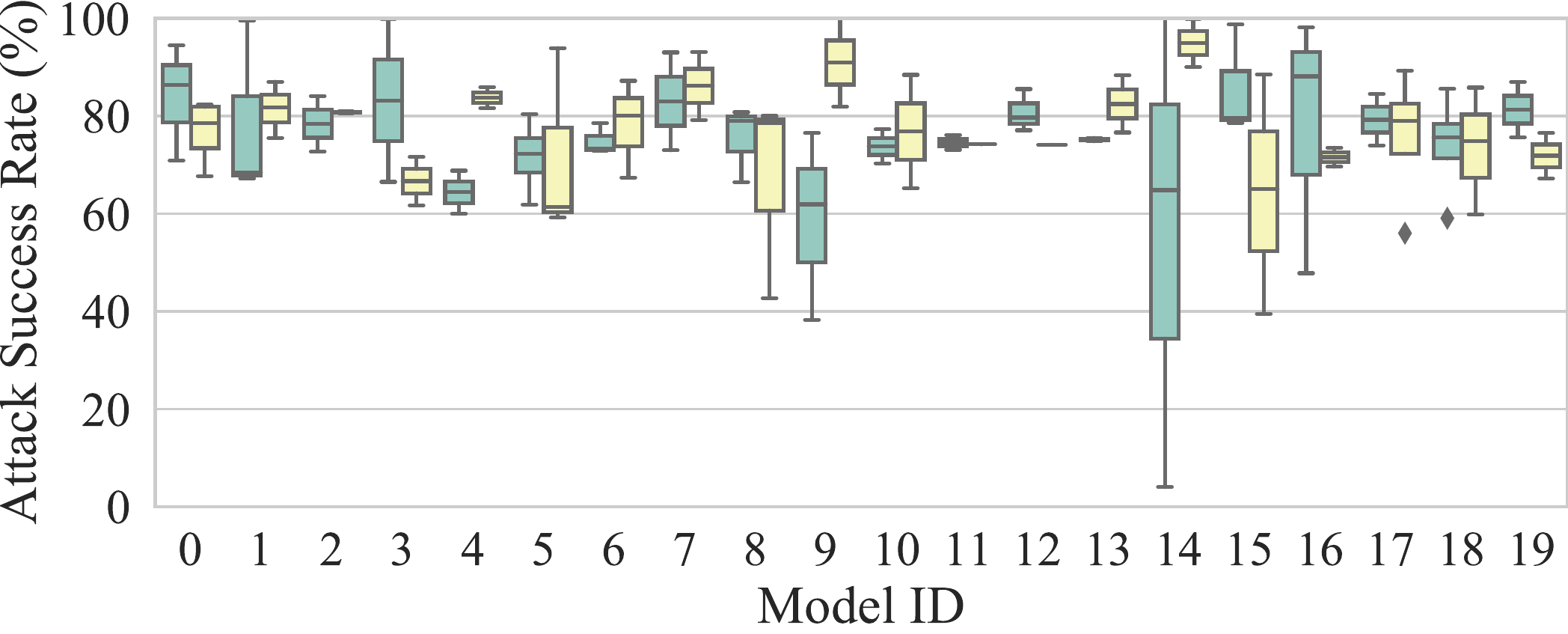}
            \caption{Filter}
            \label{fig:cv_cifar_univ_spec_filter_asr}
        \end{subfigure}
        \begin{subfigure}[b]{0.49\textwidth}
            \includegraphics[width=\columnwidth]{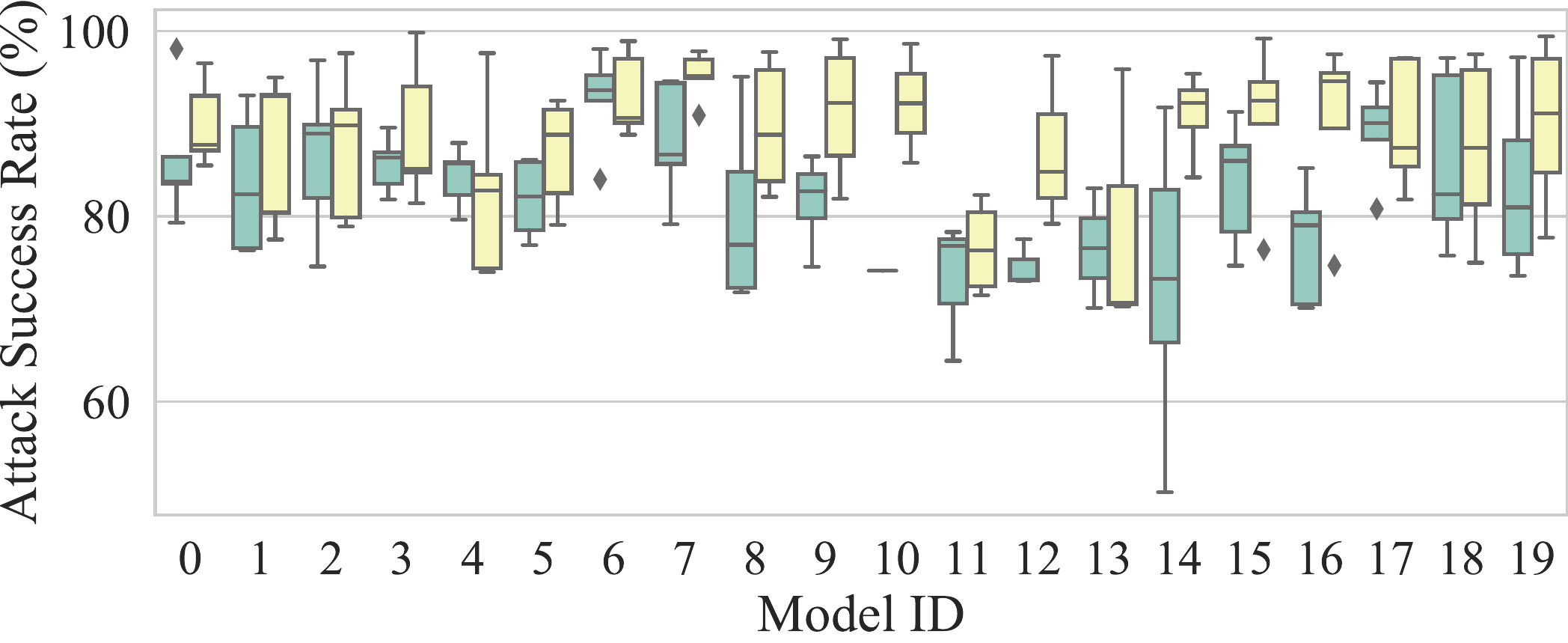}
            \caption{DFST}
            \label{fig:cv_cifar_univ_spec_deck_asr}
        \end{subfigure}
        \caption{Class IV natural backdoors in pre-trained CIFAR-10 models with universal (\textcolor{teal}{green}) and label-specific (\textcolor{DarkKhaki}{yellow}) types}
        \label{fig:cv_cifar_univ_spec_featurel2_asr}
    \end{minipage}
    \hfill
    \begin{minipage}[b]{0.32\textwidth}
        \centering
        \includegraphics[width=\columnwidth]{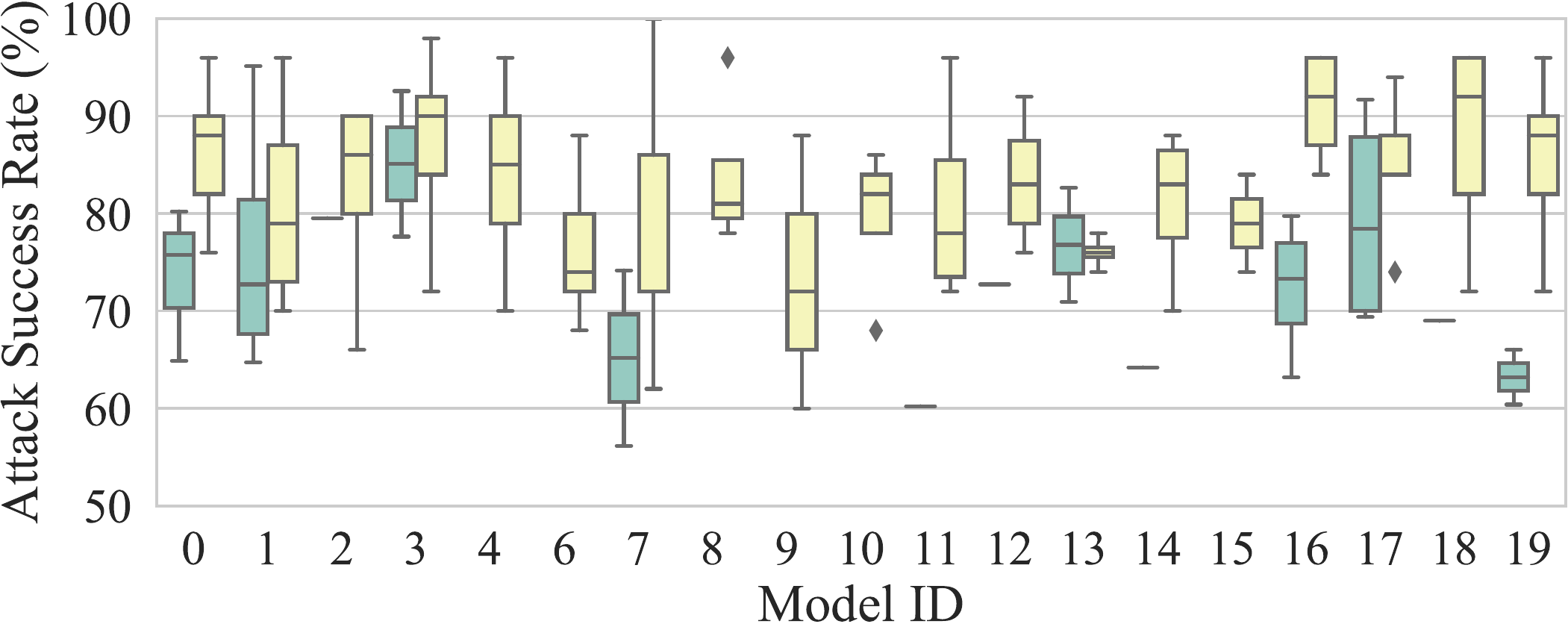}
        \caption{Class IV natural backdoors in pre-trained ImageNet models with universal (\textcolor{teal}{green}) and label-specific (\textcolor{DarkKhaki}{yellow}) types}
        \label{fig:cv_imagenet_univ_spec_deck_asr}
    \end{minipage}
\end{figure*}

\autoref{fig:cv_cifar_univ_spec_genl2_asr} reports the results of natural backdoors in Class II type on pre-trained CIFAR-10 models. Class II WaNet has low ASRs, meaning that there is no such a type of natural backdoors in the wild. The results of Class II invisible are much better. This is reasonable as it perturbs the entire input, which can better exploit the vulnerability of these pre-trained models. The observations on pre-trained ImageNet models for Class II type are similar as shown in \autoref{fig:cv_imagenet_univ_spec_ganl2_asr}.
\autoref{fig:cv_cifar_univ_spec_genlinf_asr} reports the results for Class III type. Observe natural backdoors of Class III blend have high ASRs on around half of the evaluated models. The results on the other half are slightly lower but still show the vulnerabilities of these models. The results are similar for Class III reflection and SIG. They both have very high ASRs on all the models.
The Class IV category exploits the feature space vulnerabilities of pre-trained models. The results in \autoref{fig:cv_cifar_univ_spec_featurel2_asr} and \autoref{fig:cv_imagenet_univ_spec_deck_asr} show such backdoor vulnerabilities are prevalent in pre-trained CIFAR-10 and ImageNet models.

\begin{figure*}[t]
    \centering
    \begin{subfigure}[b]{0.32\textwidth}
        \includegraphics[width=\columnwidth]{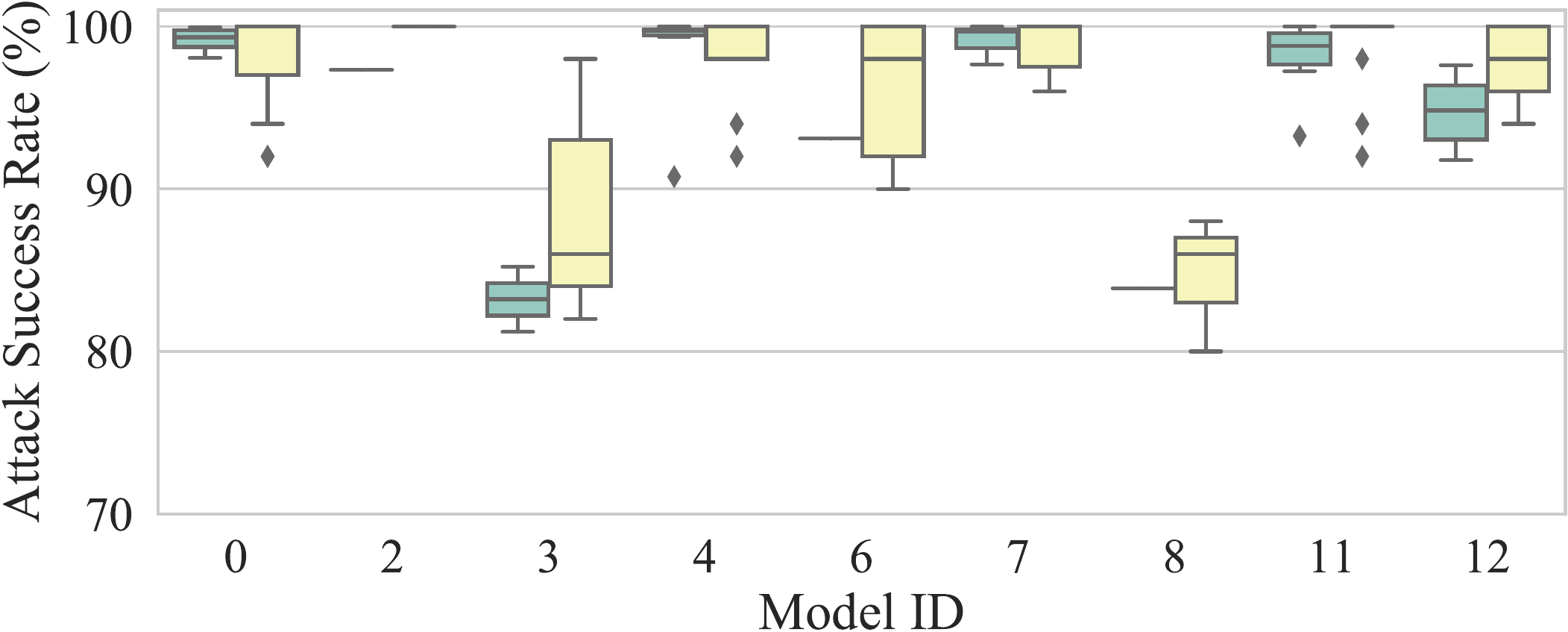}
        \caption{ABS}
        \label{fig:cv_imagenet_univ_spec_abs_asr}
    \end{subfigure}
    \hfill
    \begin{subfigure}[b]{0.32\textwidth}
        \includegraphics[width=\columnwidth]{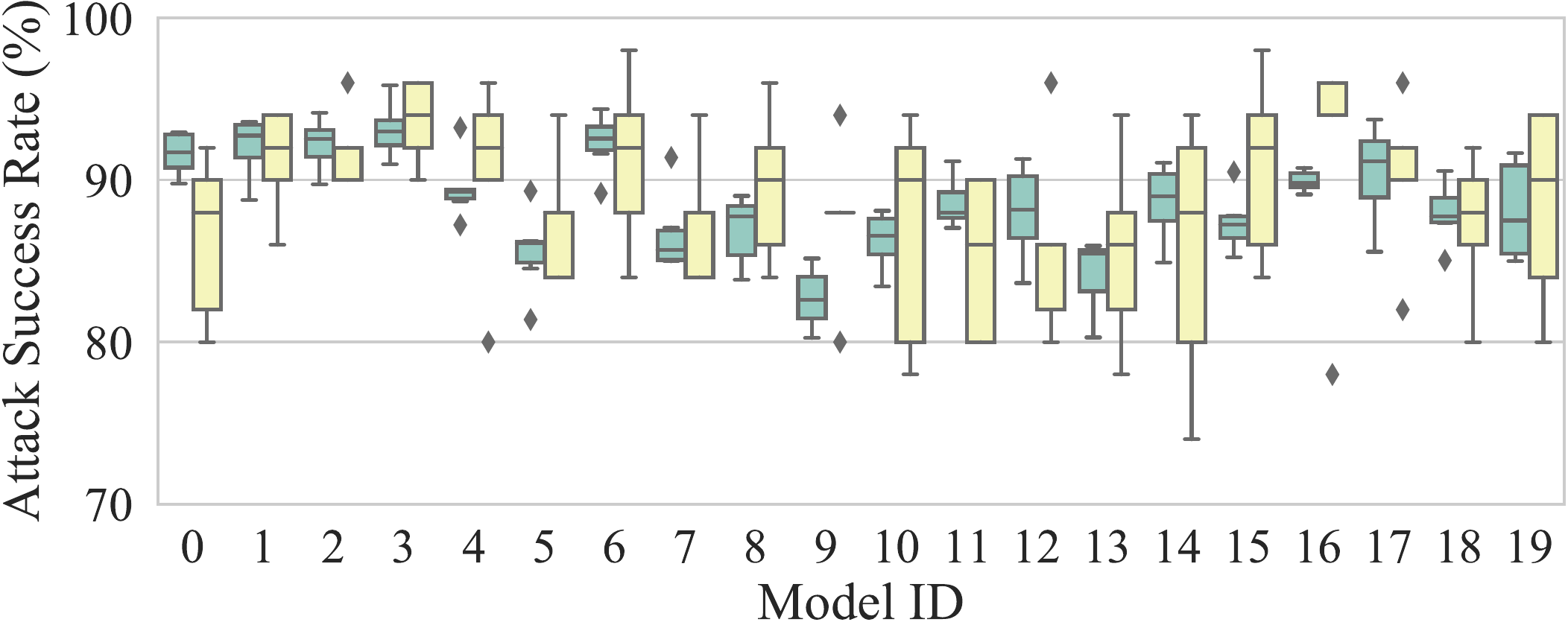}
        \caption{DualTanh}
        \label{fig:cv_imagenet_univ_spec_tanh_asr}
    \end{subfigure}
    \hfill
    \begin{subfigure}[b]{0.32\textwidth}
        \includegraphics[width=\columnwidth]{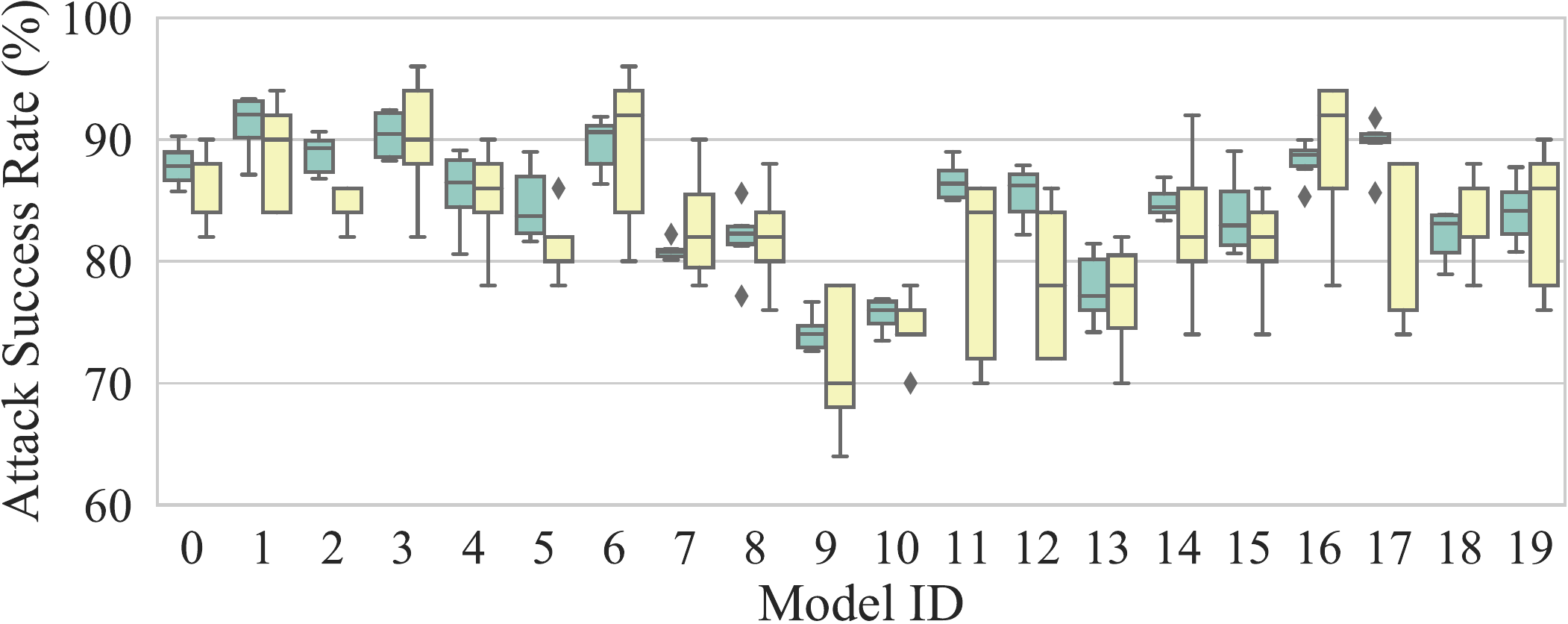}
        \caption{NC}
        \label{fig:cv_imagenet_univ_spec_nc_asr}
    \end{subfigure}
    \caption{Natural backdoors detected by existing scanners in pre-trained ImageNet models with universal (\textcolor{teal}{green}) and label-specific (\textcolor{DarkKhaki}{yellow}) types}
    \label{fig:cv_imagenet_univ_spec_atn_asr}
\end{figure*}

\begin{figure*}[t]
    \centering
    \begin{subfigure}[b]{0.32\textwidth}
        \includegraphics[width=\columnwidth]{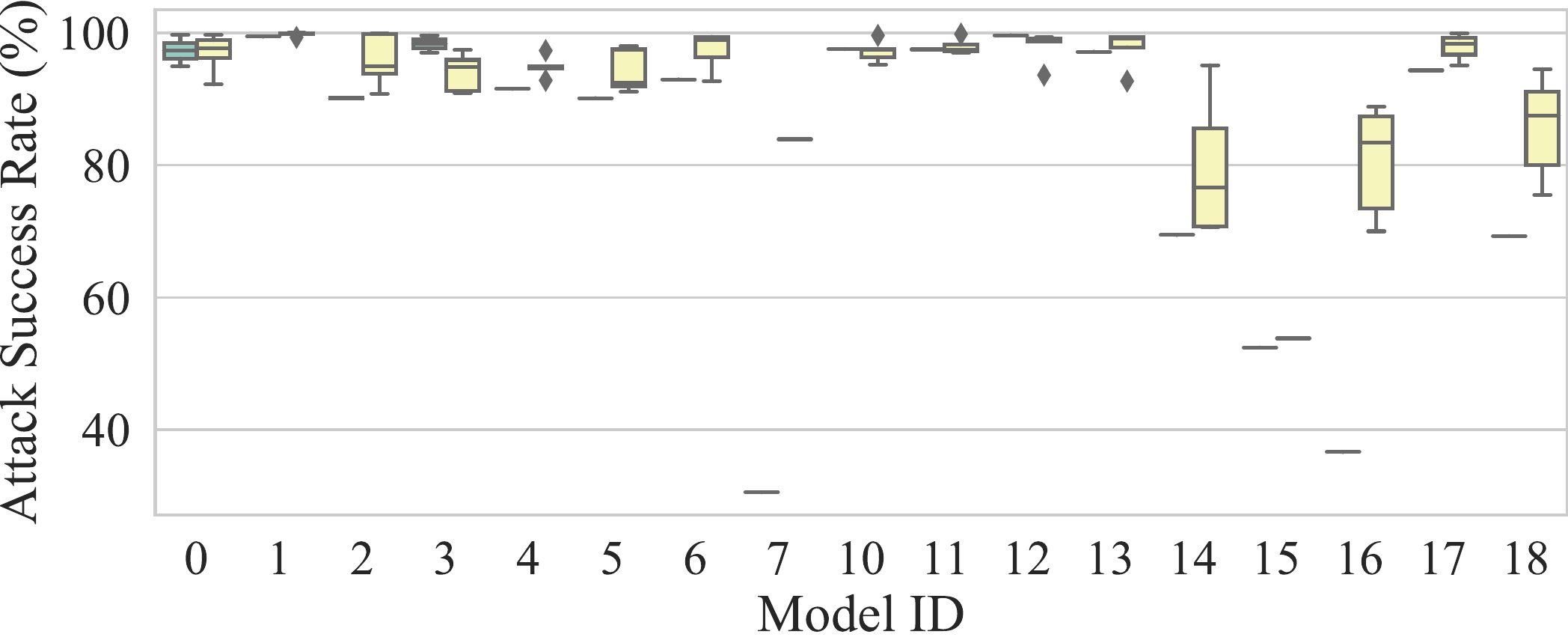}
        \caption{ABS}
        \label{fig:cv_cifar_univ_spec_abs_asr}
    \end{subfigure}
    \hfill
    \begin{subfigure}[b]{0.32\textwidth}
        \includegraphics[width=\columnwidth]{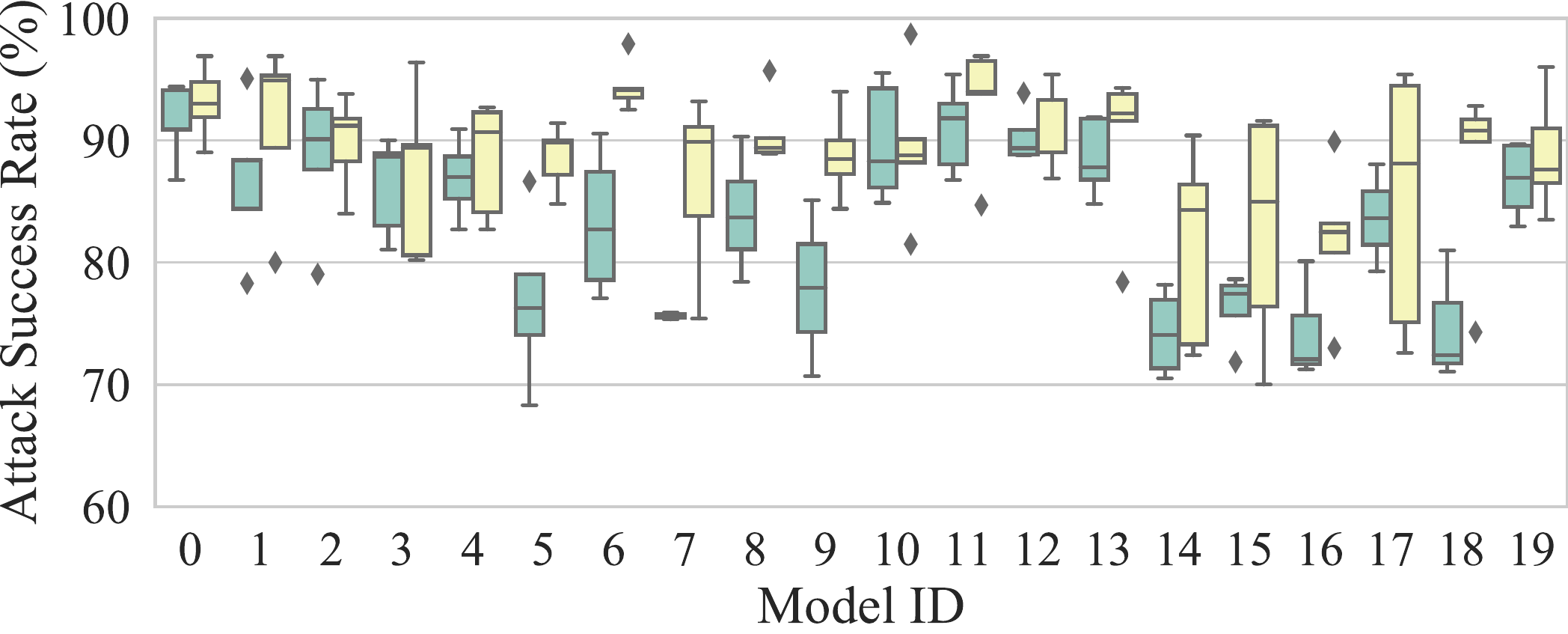}
        \caption{DualTanh}
        \label{fig:cv_cifar_univ_spec_tanh_asr}
    \end{subfigure}
    \hfill
    \begin{subfigure}[b]{0.32\textwidth}
        \includegraphics[width=\columnwidth]{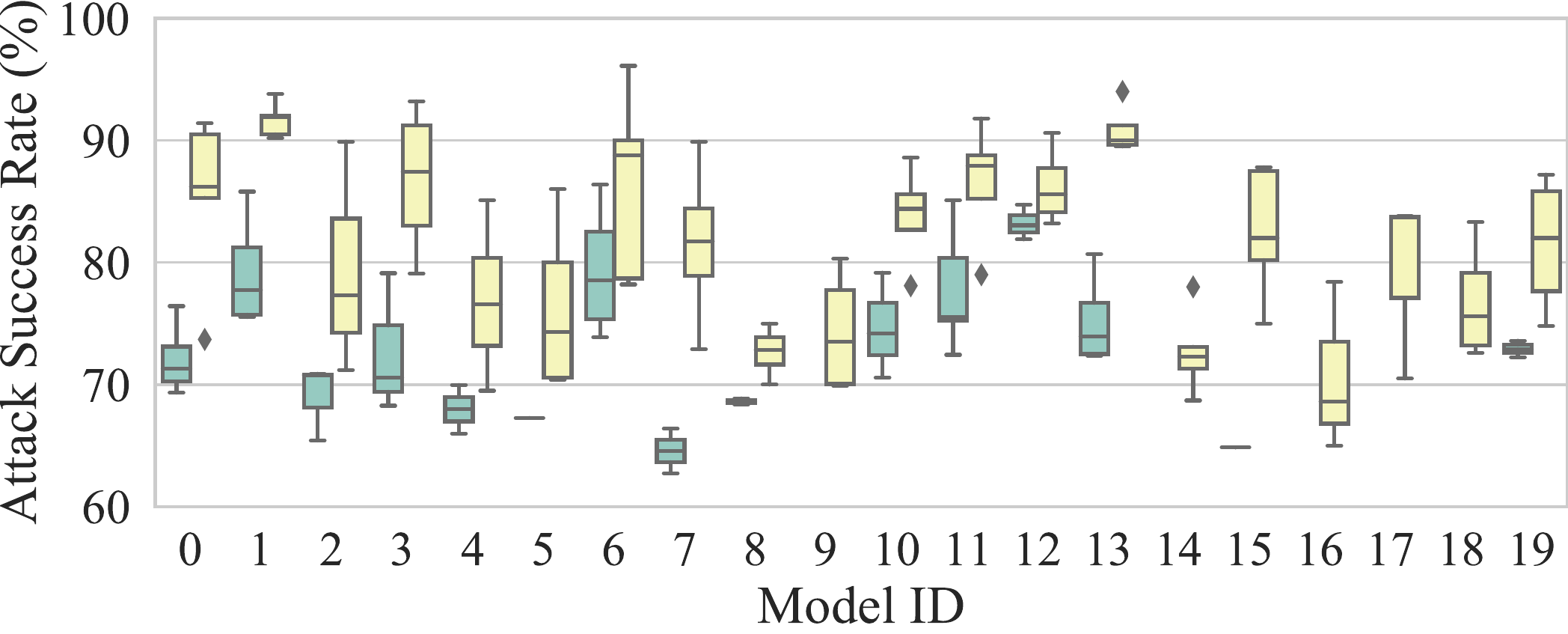}
        \caption{NC}
        \label{fig:cv_cifar_univ_spec_nc_asr}
    \end{subfigure}
    \caption{Natural backdoors detected by existing scanners in pre-trained CIFAR-10 models with universal (\textcolor{teal}{green}) and label-specific (\textcolor{DarkKhaki}{yellow}) types}
    \label{fig:cv_cifar_univ_spec_atn_asr}
\end{figure*}

We also construct natural backdoors using existing trigger inversion methods, such as NC~\cite{wang2019neural}, ABS~\cite{liu2019abs}, and DualTanh~\cite{tao2022better}, etc. We observe high ASRs of universal and label-specific natural backdoors across various models as well. Please see results in \autoref{fig:cv_imagenet_univ_spec_atn_asr} and \autoref{fig:cv_cifar_univ_spec_atn_asr} for pre-trained ImageNet and CIFAR-10 models.

\subsection{Additional Results of Defense}
\label{app:defense}

\subsubsection{Attack Instance Detection}
\label{app:instance_detection}

Activation Clustering~\cite{chen2018detecting} makes use of the activations from the last hidden layer of the model to distinguish backdoor samples from clean ones. Particularly, for each label, it utilizes clustering methods such as k-means~\cite{lloyd1982least} to separate a given set of samples into two clusters. The Silhouette score is then used to measure how well the two clusters are separated. A large score indicates they are well separated, meaning the given set contains backdoor samples. We use all the images in the validation set and Classes I and IV backdoors by our detectors GenL0 and FeatureL2 to conduct the experiments. The results are reported in~\autoref{fig:ac_specific} for label-specific backdoors. The y-axis denotes the computed Silhouette score. Each blue dot shows the score for the set with only clean images, while each orange dot for the set with both clean images and backdoor samples. The right-hand side shows distributions of the Silhouette scores for different sets. Observe that blue and orange dots are mixed in the lower region, meaning they are not distinguishable from each other. Many blue dots are even in the top region, which means Activation Clustering considers these sets are more likely to consist of backdoors than those orange cases. The observations are the same for two types of backdoors, and also universal backdoors shown in~\autoref{fig:ac_universal}. This is because natural backdoors exploit normal learned features, which are not distinguishable from clean samples as discussed previously.

\subsubsection{Backdoor Removal}
\label{app:removal}

Fine-pruning~\cite{liu2018fine} uses clean inputs to select neurons that have low activation values. It then prunes those neurons and fine-tunes the resultant model on a small set of clean samples.
We follow the original paper~\cite{liu2018fine} and ensure all the pruned models have less than 2\% accuracy degradation.
We generate two types of natural backdoors, Classes I and IV on the pruned models by Fine-pruning. The average attack results are shown in~\autoref{fig:fineprune}, where the x-axis denotes the model id and the y-axis the attack success rate (ASR). The ASRs on original models are bars in light color without the pattern.
For Class I backdoor, the ASRs on pruned models are no different from those of the original models, meaning Fine-pruning can hardly remove natural backdoors of Class I type.
For Class IV backdoor, the ASRs slightly drop after applying Fine-pruning. There is a relatively large ASR reduction on model id 13, which is a vgg19\_bn model according to~\autoref{tab:models}.
This model is less vulnerable to Class IV backdoor than other models as we can see from the original ASRs (in light background bars). Pruning neurons leads to lower ASR. Overall, Class IV backdoor still achieves high attack performance on almost all the pruned models.
We have similar observations on repaired models by other state-of-the-art defense such as ANP~\cite{wu2021adversarial} and NAD~\cite{li2021neural}. Detailed results are omitted due to space limit. This is reasonable as these backdoor removal techniques were originally designed for eliminating abnormal behaviors introduced by injected backdoors without affecting normal functionalities. Natural backdoors on the other hand are caused by low-level features learned by models, which are rooted in normal training data as discussed in Section~\ref{sec:root_causes}.

\subsection{More Related Works}
\label{app:related}
Existing study such as TrojanZoo~\cite{pang2020trojanzoo} focuses on evaluating existing backdoor attacks and defense in \textit{injected backdoor} scenario, which is orthogonal to our study of \textit{natural backdoor}. 
It characterizes attacks in four aspects: (1) architecture modifiability that means whether the model architecture is modified by the attack; (2) trigger optimizability that assesses whether the trigger is fixed or optimized during poisoning; (3) fine-tuning survivability that checks whether the backdoor remains effective when the model is fine-tuned; (4) defense adaptivity that evaluates whether the attack can evade possible defense. Our study, on the other hand, characterizes attacks based on how they transform the input, which is more important when exploring natural backdoors. This is a different characterization perspective from existing studies~\cite{pang2020trojanzoo,li2022backdoor,liu2020survey,goldblum2022dataset}. 
We also introduce a general definition that covers all the studied backdoor vulnerabilities. It provides a pragmatic definition that is actionable in practice when evaluating backdoor vulnerabilities, which was not studied in existing works~\cite{pang2020trojanzoo,li2022backdoor,liu2020survey,goldblum2022dataset}. 
Survey~\cite{li2022backdoor} categorizes backdoor triggers based on whether the trigger is optimized, whether it is input-specific, how many target labels it has, etc. It does not view the backdoor as a transformation function and summarize all existing attacks into a small number of formulas as in this paper.
Others~\cite{gao2020backdoor,li2022deep,guo2022overview} classify existing attacks based on adversary capabilities.
Study~\cite{kaviani2021defense} focuses on surveying defense techniques against injected backdoors, which is orthogonal to our study on evaluating defense against natural backdoors.

\begin{table*}[t]
    \centering
    \scriptsize
    \caption{Attack success rate of label-specific natural backdoors in pre-trained CIFAR-10 models}
    \begin{tabular}{l*{11}{r}}
        \toprule
        Model & Patch & Dynamic & Input-aware & Composite & WaNet & Invisible & Blend & Reflection & SIG & Filter & DFST \\
        \midrule
        vgg11\_bn\_1        & 96.90\% & 99.80\% & 57.60\% & 100.00\% & 40.84\% & 88.80\% & 86.30\% & 80.36\% & 70.71\% & 81.70\% & 96.50\% \\
        vgg13\_bn\_1        & 96.90\% & 98.40\% & 61.80\% & 99.96\% & 49.71\% & 92.60\% & 99.00\% & 81.93\% & 77.64\% & 87.00\% & 95.00\% \\
        resnet18            & 93.80\% & 93.00\% & 43.80\% & 99.98\% & 32.60\% & 81.80\% & 84.30\% & 71.73\% & 63.98\% & 80.60\% & 97.60\% \\
        resnet34            & 96.40\% & 95.00\% & 40.80\% & 99.87\% & 32.91\% & 84.00\% & 83.90\% & 76.49\% & 62.67\% & 71.60\% & 99.80\% \\
        resnet50            & 92.70\% & 92.00\% & 41.60\% & 99.96\% & 30.80\% & 85.20\% & 81.00\% & 80.20\% & 58.89\% & 85.90\% & 97.60\% \\
        densenet169         & 91.40\% & 53.20\% & 33.40\% & 99.89\% & 37.78\% & 93.40\% & 90.10\% & 79.40\% & 62.78\% & 93.90\% & 92.50\% \\
        googlenet           & 97.90\% & 98.00\% & 73.80\% & 100.00\% & 53.64\% & 97.20\% & 98.90\% & 89.51\% & 77.36\% & 87.20\% & 98.90\% \\
        inception\_v3       & 93.20\% & 91.00\% & 28.80\% & 99.98\% & 53.38\% & 86.80\% & 99.20\% & 87.22\% & 66.49\% & 79.20\% & 97.80\% \\
        resnet20            & 95.70\% & 91.00\% & 44.20\% & 99.98\% & 43.96\% & 98.80\% & 100.00\% & 83.89\% & 80.69\% & 78.60\% & 97.70\% \\
        resnet32            & 94.00\% & 88.20\% & 52.00\% & 100.00\% & 41.76\% & 85.40\% & 98.90\% & 84.00\% & 84.91\% & 81.90\% & 99.10\% \\
        vgg11\_bn\_2        & 98.70\% & 98.80\% & 61.00\% & 99.96\% & 33.49\% & 86.20\% & 81.60\% & 75.71\% & 73.09\% & 88.40\% & 98.60\% \\
        vgg13\_bn\_2        & 96.90\% & 97.60\% & 62.20\% & 99.91\% & 49.73\% & 86.20\% & 98.30\% & 92.04\% & 81.36\% & 74.20\% & 82.30\% \\
        vgg16\_bn           & 95.40\% & 97.80\% & 50.80\% & 99.93\% & 41.33\% & 97.80\% & 97.60\% & 90.38\% & 87.22\% & 88.30\% & 97.30\% \\
        vgg19\_bn           & 94.30\% & 95.80\% & 32.20\% & 99.82\% & 36.71\% & 95.40\% & 97.70\% & 87.22\% & 85.36\% & 76.60\% & 95.90\% \\
        mobilenetv2\_x0\_75 & 90.40\% & 88.80\% & 52.40\% & 100.00\% & 56.71\% & 98.60\% & 99.90\% & 83.84\% & 66.02\% & 90.00\% & 95.40\% \\
        mobilenetv2\_x1\_4  & 91.60\% & 86.60\% & 43.80\% & 99.91\% & 60.76\% & 92.80\% & 99.60\% & 86.84\% & 61.73\% & 88.50\% & 99.20\% \\
        shufflenetv2\_x1\_0 & 89.90\% & 66.60\% & 38.60\% & 100.00\% & 48.00\% & 91.80\% & 99.30\% & 84.96\% & 75.33\% & 73.50\% & 97.50\% \\
        shufflenetv2\_x1\_5 & 95.40\% & 77.00\% & 52.40\% & 100.00\% & 46.02\% & 93.40\% & 99.70\% & 83.02\% & 75.53\% & 74.80\% & 97.10\% \\
        shufflenetv2\_x2\_0 & 92.80\% & 85.60\% & 40.20\% & 100.00\% & 48.18\% & 89.20\% & 98.40\% & 89.20\% & 78.33\% & 74.80\% & 97.50\% \\
        repvgg\_a2          & 96.00\% & 90.20\% & 28.40\% & 100.00\% & 51.69\% & 99.00\% & 99.80\% & 90.51\% & 79.47\% & 67.20\% & 99.40\% \\
        
        \midrule
        Average             & 94.52\% & 89.22\% & 46.99\% & 99.96\% & 44.50\% & 91.22\% & 94.67\% & 83.92\% & 73.45\% & 68.31\% & 96.64\% \\
        
        \bottomrule
    \end{tabular}
    \label{tab:cifar_spec_asr}
\end{table*}

\end{document}